\documentclass[7pt, twocolumn]{extarticle}
\usepackage{amssymb}
\usepackage{amsmath}
\usepackage{amsthm}
\usepackage{color}
\usepackage{comment}
\usepackage{geometry}		% allows easy margin settings
\usepackage{graphicx}
\usepackage{array}
\usepackage{multirow}

\usepackage{natbib}
\usepackage{tcolorbox}
\usepackage[colorlinks=true, citecolor=blue]{hyperref}

\makeatletter
	\@addtoreset{equation}{section}
\makeatother

\DeclareFontFamily{OT1}{pzc}{}
\DeclareFontShape{OT1}{pzc}{m}{it}{<-> s * [1.10] pzcmi7t}{}
\DeclareMathAlphabet{\mathpzc}{OT1}{pzc}{m}{it}

\let\originalleft\left
\let\originalright\right
\renewcommand{\left}{\mathopen{}\mathclose\bgroup\originalleft}
\renewcommand{\right}{\aftergroup\egroup\originalright}

\theoremstyle{definition}

\title{Time series analysis of coupled slow-fast neuron models: From Hurst exponent to Granger causality}

\author{I.~Ghosh$^1$\footnote{Author to whom correspondence must be addressed.}, H.O.~Fatoyinbo$^2$, S.S.~Muni$^3$\\\\
$^1$School of Mathematical and Computational Sciences,\\
Massey University,\\
Palmerston North, 4410, New Zealand.\\
Email: \url{i.ghosh@massey.ac.nz}\\
$^2$School of Engineering, Computer and Mathematical Sciences,\\
Auckland University of Technology,\\
Auckland, 1142, New Zealand.\\
Email: \url{hammed.fatoyinbo@aut.ac.nz}\\
$^3$School of Digital Sciences,\\
Digital University Kerala,\\
Technopark Phase-IV campus, Mangalapuram - 695317, Kerala, India.\\
Email: \url{sishushankarmuni@gmail.com}
}

\begin{document}

% \maketitle

% % keywords: piecewise-linear; piecewise-smooth; border-collision bifurcation; chaotic attractor; invariant expanding cone
% % MSC codes: 37G35; 39A28

% \begin{abstract}

% Add later

% \end{abstract}

\twocolumn[
  \maketitle
  \begin{@twocolumnfalse}
    \begin{abstract}
      We perform time series analysis of small networks where every node is the slow-fast version of the denatured Morris--Lecar neuron proposed by Schaeffer and Cain. We choose popular coupling strategies from the literature and provide a detailed account of how varying their strength drives the dynamics of the small networks. Algorithms for time series analysis range from measuring their persistence (ability to remember past values), irregularity, chaos and quasiperiodicity, to synchronization between time series from every node within a network. Chaos is observed for inhibitory coupling strengths and for temperature higher than a reference temperature when the coupling is thermally sensitive. We observe quasi-periodicity when the coupling is very weak and synchronized bursting for highly excitatory coupling strength. In certain cases we also observe decay oscillations. Finally, a causality test is performed to detect whether the dynamics of one neuron is influencing the dynamics of the other in the coupled system.
    \end{abstract}
  \end{@twocolumnfalse}]

\section{Quotation}
Time series analysis of networks where each node is a dynamical system, is a hot topic in the current times. In this study, each node is realized by a reduced version of the Morris--Lecar neuron model, called the denatured Morris--Lecar model, put forward by Schaeffer and Cain. Additionally, these neurons are time-scale separated, and produce realistic bursting behavior. By leveraging different algorithms for time series analysis we try to explore whether we can quantify certain complex behaviors exhibited by small networks of these denatured Morris--Lecar neurons. We are able to use a combination of different measures to quantify chaos, quasiperiodicity, synchronized bursting, anti-phase oscillations, and decay oscillations. Furthermore, Granger's causality test gives us an intuition on whether the time series of one node causes the time series of another in a coupled system.

% \section{Probable Q1 Journals}
% \begin{enumerate}
%     \item Communications Physics (nature)
%     \item Computer Physics Communications
%     \item Chaos, Solitons \& Fractals
%     \item Applied Mathamatical Modelling
%     \item Journal of Computational physics
%     \item Chaos (AIP)
%     \item International Journal of Bifurcation and Chaos
% \end{enumerate}

\section{Introduction}

In this work we provide a unified approach to methods implemented to study the behavior of coupled neurons over time. A system that evolves with time can be mathematically modeled using {\em dynamical systems}. Neurons are the fundamental units of the nervous system, generally categorized as {\em excitable} cells exhibiting a plethora of highly complex dynamics. They act as suitable candidates for analysis using various tools and techniques from dynamical systems literature. In this work, we explore continuous-time models of neuron dynamics. We pick a handful of model equations for coupling strategies from the {\em neuronal dynamics} literature and investigate their behavior using a collection of methods that have been utilized recently by the research community. 

Schaeffer and Cain, in their book~\citep{ScCa18}, introduced a variant of the popular Morris--Lecar model~\citep{MoLe81}, which they called the {\em denatured} Morris--Lecar (dML) model, whose equations are given by
 \begin{equation}
\label{eq:dML_model}
\begin{aligned}
    \dot{x} &=  x^2(1-x) - y + I, \\
    \dot{y} &= A e^{\alpha x} - \gamma y.
\end{aligned}
\end{equation}
It can be noticed that~\eqref{eq:dML_model} looks structurally similar to the FitzHugh--Nagumo (FHN) model~\citep{Fi61} with a subtle difference in their $y$-nullclines. The $y$-nullcline in dML~\eqref{eq:dML_model} has an exponential growth term, whereas in FHN the $y$-nullcline is linear. The two variables in~\eqref{eq:dML_model} are the voltage-like variable ($x$) with cubic nonlinearity and the corresponding recovery variable ($y$), governing the dynamics of a typical excitable cell like neuron. The variable $x$ is self-reinforced as demonstrated by the positive feedback to the neuron via the nonlinearity, leading to its firing mechanism. The variable $y$ on the other hand, demonstrates a negative feedback via the exponential term, leading to the dynamics of the refractory period after the neuron fires. This has been detailed by Izhikevich and FitzHugh~\citep{IzFi06} for the case of the FitzHugh--Nagumo model. Parameter $\gamma$ represents the excitability of the neuron and together with the parameter $A$ determines the dynamics of the recovery variable $y$. Parameter $\alpha$ controls the exponential growth rate of $y$. Furthermore, $I$ is the external stimulus current that depolarizes the neuron, leading to the triggering of an action potential when the voltage crosses a critical value. All these parameters are kept positive for~\eqref{eq:dML_model}. We urge the readers to refer to Izhikevich's book~\citep{Iz07} to learn more about mathematical modeling of neuron dynamics. The original Morris--Lecar model~\citep{MoLe81} has sigmoidal activation functions which is biophysically inspired. The model~\eqref{eq:dML_model} although looks structurally similar to FitzHugh--Nagumo type model, has the exponential component, which pushes its dynamics more towards the sigmoidal nonlinearity seen in the original Morris--Lecar system. That is why~\eqref{eq:dML_model} is closer to a Morris--Lecar model rather than a FitzHugh--Nagumo model. Also, because~\eqref{eq:dML_model} drops its biophysical details, it is called a ``denatured'' Morris--Lecar system. In a previous work, Fatoyinbo {\em et al.}~\citep{FaMu22} has performed an extensive bifurcation analysis in terms of codimension-one and -two bifurcation plots for~\eqref{eq:dML_model}.

\begin{figure}
    \includegraphics[width=1\linewidth]{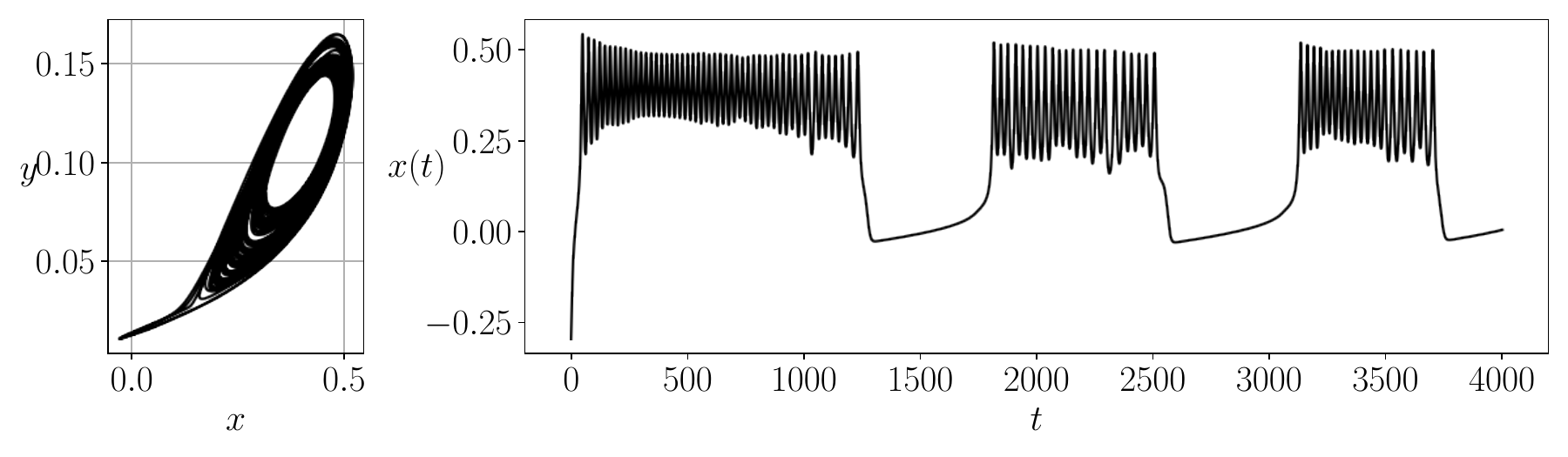}
    \caption{Typical time series and a phase portrait of~\eqref{eq:model_single} exhibiting bursting. Parameters considered are $A=0.0041$, $\alpha = 5.276$, $\gamma = 0.315$, and $\varepsilon = 0.0005$. Initial condition $x(0)$ is sampled randomly from the uniform distribution $[-1, 1]$, and $(y(0), I(0)) = (0.1, 0.019)$.}
    \label{fig:single}
\end{figure}

This two-dimensional model, however, does not portray repeated periodic bursting. Bursting with quiet intervals can be incorporated if the external current $I$ were to vary very slowly with time, rather than being a constant parameter. Schaeffer and Cain~\citep{ScCa18} discussed that the dynamics would then exhibit bistability between a stable periodic solution and a stable equilibrium point, which would lead to bursting. The system is now modified to a three-dimensional slow-fast variant given by
\begin{equation}
\label{eq:model_single}
\begin{aligned}
    \dot{x} &= f(x, y, I) = x^2(1-x) - y +I, \\
    \dot{y} &= g(x, y, I) = A e^{\alpha x} - \gamma y,\\
    \dot{I} & = h(x, y, I) =  \varepsilon\bigg[\frac{1}{60}\bigg\{1+\tanh\bigg(\frac{0.05-x}{0.001}\bigg)\bigg\} - I\bigg],
\end{aligned}
\end{equation}
where $\varepsilon$ is the {\em timescale parameter} that separates the system into two different time scales~\citep{Ku15}. A time series and a phase portrait of~\eqref{eq:model_single} is shown in Fig.~\ref{fig:single}. Here we have fixed $A = 0.0041$, $\alpha = 5.276$, $\gamma = 0.315$, and $\varepsilon = 0.0005$. We set the parameters to these values throughout the manuscript. The initial value $x(0)$ is sampled randomly from a uniform distribution with a closed interval $[-1, 1]$. The other initial conditions are $(y(0), I(0)) = (0.1, 0.019)$. We observe a {\em fold/homoclinic} type bursting where the resting state transitions to the spiking limit cycle via a saddle-node (fold) bifurcation and from the spiking limit cycle to the resting state via a saddle homoclinic orbit bifurcation. For a detailed discussion about Fold/homoclinic type bursting, please refer to Izhikevich~\citep{Iz07}. A detailed account of the qualitative analysis of the single-celled model~\eqref{eq:model_single} can be found in Ghosh {\em et al.}~\citep{GhFa25}. 

The purpose of this manuscript is twofold: (i) popularize the dML model, and (ii) provide a tutorial to perform time series analysis on data generated from various coupled systems of dML neurons. We extensively report how varying the coupling strength for different strategies gives rise to a battery of complex dynamics in the coupled systems. In some cases we observe chaos for inhibitory coupling. Chaos is mitigated as the coupling strength becomes weaker and a quasi-periodic regime arises close to zero coupling strength. The behavior of the time series can be distinctly quantified using Hurst exponent and sample entropy, showing a higher value close to chaos. The $0-1$ test quantifies chaos by allowing us to plot phase portraits of some translated variables from the dynamical variables, which show random Brownian motion. Synchronization and phase coherence are computed using a cross-correlation coefficient and Kuramoto's order parameter which help us quantify the collective dynamics of the coupled system. Intuitively, the nodes are asynchronized and out of phase when the systems behave chaotically. As coupling becomes more excitatory and positive, the nodes either display a synchronized bursting behavior or they decay into a symmetric equilibrium point. When the nodes are exhibiting bursting, they are mostly found to be phase-locked. In certain coupling strategies, we also observe anti-phase decay when the coupling strength is excitatory. Furthermore, temperature plays an important role in thermally sensitive coupling, where chaos is observed in the inhibitory coupling regime when the temperature is above a threshold. Below this threshold, the nodes always exhibit bursting, however, they oscillate in anti-phase for high inhibitory coupling, and settles to in-phase oscillation for excitatory coupling.

The manuscript is organized as follows: In section~\ref{sec:couplings} we introduce different coupling strategies implemented in this paper. They are gap junction, thermally sensitive gap junction, chemical, Josephson junction, memristive, higher-order gap junction, and random chemical couplings. Then in section~\ref{sec:methods} we review a combination of algorithms used to perform time series analysis on the data generated from the coupled systems. These are the Hurst exponent, sample entropy, $0-1$ test, Pearson's correlation coefficient, and Kuramoto's order parameter. We report all numerical analysis in section~\ref{sec:numerics}, and develop an account of how the above concepts work together to tell a story about the coupled dML neurons. In section~\ref{sec:granger} we provide a note on how Granger causality can be utilized to argue how one neuron drives the behavior of the other in the coupled system. Finally, in section~\ref{sec:conclusion} we detail concluding remarks and future steps.

\section{Coupling strategies}
\label{sec:couplings}
Neuronal interactions are usually modeled using a network-theoretic framework by the applied mathematics and computational neuroscience community. A network consists of nodes and edges connected via some coupling strategies that govern the dynamics of the network over time. A node can either represent a single neuron (in the microscopic scale) or a brain region (in the mesoscopic scale). Thus, nodes form the interacting unit within a network. An edge on the other hand represents how two neurons or regions are connected, and can be categorized into three types of connectivities: (i) {\em structural} connectivity for physical and anatomical links, (ii) {\em functional} connectivity for undirected statistical dependencies, and (iii) {\em effective} connectivity for directed causal relationships. 

In this work we will mathematically establish coupling strategies in small network models of {\em identical} neurons that will emulate these connectivities in one way or another. By ``identical", we mean the local parameters of each neuron are kept the same. For detailed discussion on these connectivities, refer to the review article by Park and Friston~\citep{PaFr13}. Our goal is to study how the properties of these couplings drive the dynamics of a small network as a whole. Furthermore, these couplings will have a scalar value associated with them, which will dictate their strength. This value can be either positive (excitatory coupling), negative (inhibitory coupling), or zero (absence of coupling). Positive coupling represents those consisting of asymmetric junctions alongside a dense protein complex associated with the postsynaptic membranes of the excitatory synapses~\citep{KaTa18}, and negative coupling represents those consisting of symmetrical synaptic junctions alongside lacking the dense protein complex~\citep{StPo14}.

\subsection{Gap junction coupling}
\begin{figure}
    \centering
    \includegraphics[width=0.5\linewidth]{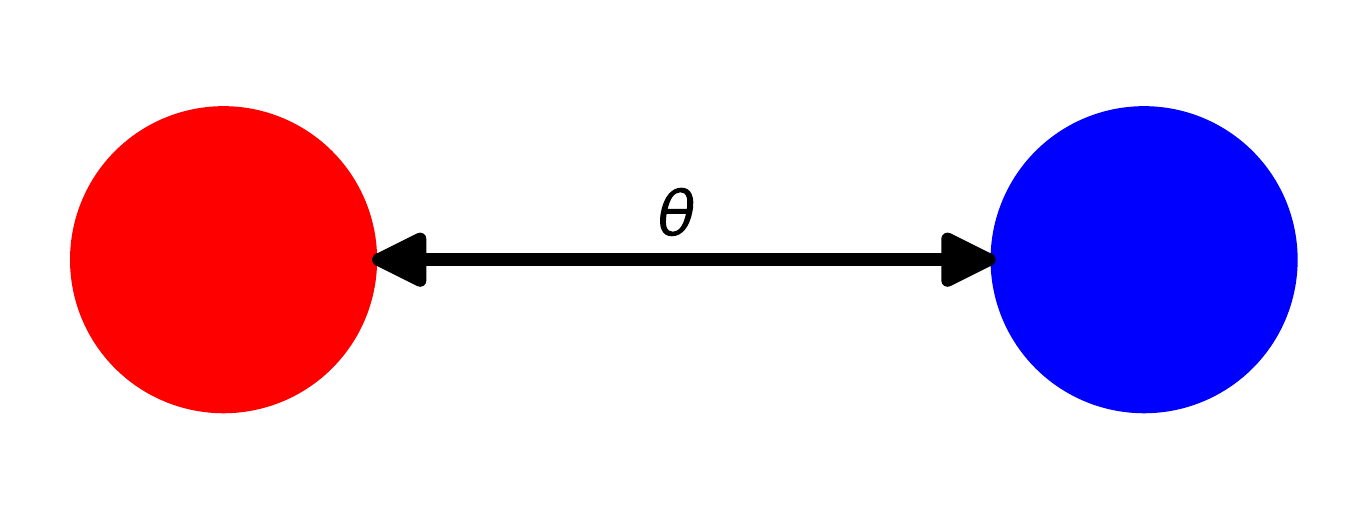}
    \caption{Schematic of a gap junction coupling between two neurons, with coupling strength $\theta$. The coupling is bidirectional.}
\label{fig:gap_schem}
\end{figure}

Gap junctions are intricate intercellular channels that provide a direct communication pathway between the cytoplasm of two coupled neurons via ions. It is the {\em synapse} that controls the connectivity between two neurons. A gap junction coupling can transmit signals equally in both directions. See Hormuzdi {\em et al.}~\citep{HoFi04} for an extensive report on electrical synapses. This straightforward strategy is given by
\begin{equation}
\label{eq:gapJunction}
\begin{aligned}
    \dot{x}_i &= f(x_i, y_i, I_i) + \sum_{j \in B(i)}\theta (x_j - x_i),\\
    \dot{y}_i &= g(x_i, y_i, I_i), \\
    \dot{I}_i & = h(x_i, y_i, I_i),
\end{aligned}
\end{equation}
with $B(i)$ representing the neighboring nodes of the node $i$. For a two coupled system $i, j = 1, 2$. Note that the gap junction coupling $\theta$ is considered bidirectional; see Fig.~\ref{fig:gap_schem}. The flow from node $j$ to $i$ is given by $x_j - x_i$. Gap junction coupling is by far the most extensively studied coupling strategy found in the computational neurodynamics literature~\citep{WaLu08, HoMe12, CaSt16, MaPe17, SiLo21, WeLi25, GoLa25, GhFa25}. 

\subsection{Thermally sensitive gap junction coupling}
One can modify the gap junction coupling to incorporate temperature effects into it. This kind of coupling has been looked into by Feudel {\em et al.}~\citep{FeNe00} and Yoshioka~\citep{Yo05} for Hodgkin--Huxley~\citep{HoHu52} type coupled neurons. More recently Follman {\em et al.}~\citep{FoJa24} explored the effects of temperature on neuronal synchronization in a network of Huber--Braun neurons and how it regulates seizures. Motivated by these, we have also modified~\eqref{eq:gapJunction} to include thermal effects, 
\begin{equation}
\label{eq:temperature}
\begin{aligned}
    \dot{x}_i &= f(x_i, y_i, I_i) + \sum_{j \in B(i)}\theta \delta(T) (x_j - x_i),\\
    \dot{y}_i &= g(x_i, y_i, I_i), \\
    \dot{I}_i & = h(x_i, y_i, I_i),
\end{aligned}
\end{equation}
where $\delta(T)$ is the Arrhenius function given by
\begin{align}
\label{eq:Arrhenius}
\delta(T) = \delta_0^{\frac{T - T_{\rm ref}}{10^{\circ}C}},
\end{align}
which is a temperature-dependent scaling factor. The coupling strength $\theta$ and temperature $T$ are the main bifurcation parameters, with $\delta_0 = 1.3$, and $T_{\rm ref} = 20^{\circ}C$ fixed. Yoshioka~\citep{Yo05} showed that temperature dependence drives chaotic firing in a modified Hodgkin--Huxley system of coupled neurons. Temperature fluctuations regulate the rate at which ions flow in and out of the neurons. A high temperature can lead to a neuron firing with higher frequencies, which needs to be mitigated via a higher excitatory coupling strength, such that synchrony is restored~\citep{Va20, FoJa24}.

\subsection{Chemical coupling}
\begin{figure}[h]
    \centering
    \includegraphics[width=0.5\linewidth]{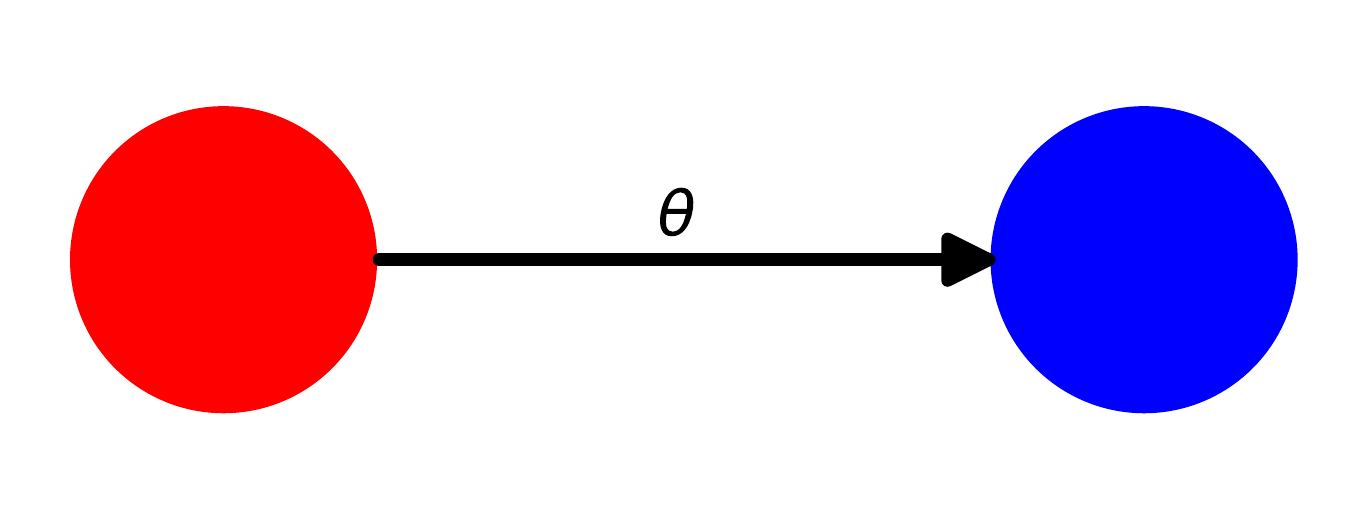}
    \caption{Schematic of a chemical coupling between two neurons, with coupling strength $\theta$. The coupling is unidirectional.}
\label{fig:chem_schem}
\end{figure}

Next, we move to chemical coupling, which has a non-local configuration. There is no continuity between the cytoplasm of two neurons, and the synaptic cleft is wider in comparison to the synaptic cleft in gap junction coupling to accommodate for neurotransmitter release and distribution~\citep{HoFi04}. Furthermore, chemical coupling is unidirectional, unlike electrical coupling, which is bidirectional, see Fig.~\ref{fig:chem_schem}. We model the coupling strategy for two dML neurons following the work of Belykh {\em et al.}~\citep{BeDe05} fetching us
\begin{equation}
\label{eq:dML_chemical}
\begin{aligned}
    \dot{x}_1 &= f(x_1, y_1, I_1),\\
    \dot{x}_2 &= f(x_2, y_2, I_2) +\theta\frac{v_s - x_2}{1+\exp\{-\lambda(x_1 - q)\}},\\
    \dot{y}_i &= g(x_i, y_i, I_i), \\
    \dot{I}_i & = h(x_i, y_i, I_i),
\end{aligned}
\end{equation} 
where $\theta$ is again the coupling strength and $v_s$ is the reversal potential. In the sigmoidal function $\frac{1}{1+\exp\{-\lambda(x - q)\}}$, the parameter $\lambda >0$ is the slope and $q$ is the synaptic threshold. Somers and Kopell~\citep{SoKo93} call this coupling scheme a {\em fast threshold modulation}.
Like gap junction coupling, researchers have shown interests in studying chemical coupling~\citep{MaPe17, MaBe19, SiLo21, WeLi25, ArSu25}. Throughout the manuscript we kept $v_s = 2$, $\lambda = 10$, and $q=-0.25$.

\subsection{Josephson Junction coupling}
\begin{figure}[h]
    \centering
    \includegraphics[width=0.5\linewidth]{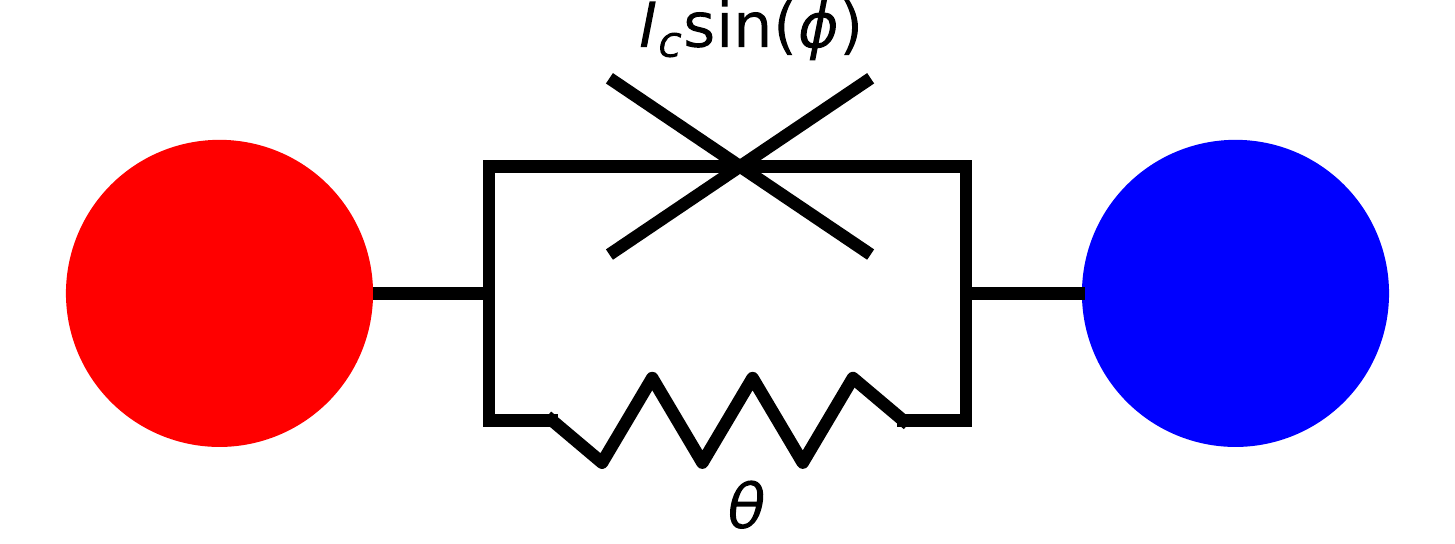}
    \caption{Schematic of a Josephson junction coupling, with coupling strength $\theta$ and the periodic junction current $I_c \sin(\phi)$.}
\label{fig:JJ_schem}
\end{figure}

Next, we move on to a more complicated coupling scheme that models a hybrid mechanism incorporating both electrical and chemical synapses. This was popularized by Njitacke {\em et al.}~\citep{NjRa22}, where they built a hybrid synapse by using a {\em Josephson junction} and a linear resistor. The coupling was incorporated between a two-dimensional Hindmarsh--Rose and a two-dimensional FitzHugh--Nagumo model. We employ this coupling scheme between two identical dML neurons (see Fig.~\ref{fig:JJ_schem}), which gives us
\begin{equation}
\label{eq:JJ}
\begin{aligned}
    \dot{x}_1 &= f(x_1, y_1, I_1) -I_c\sin(\phi) + \theta(x_2-x_1), \\
    \dot{x}_2 &= f(x_2, y_2, I_2) +I_c\sin(\phi) + \theta(x_1-x_2), \\
    \dot{y}_i &= g(x_i, y_i, I_i), \\
    \dot{I}_i & = h(x_i, y_i, I_i), \qquad i = 1,2, \\
    \dot{\phi} &= \mu(x_1 - x_2),
\end{aligned}
\end{equation}
where $I_c \sin (\phi)$ is the junction current across the Josephson junction that acts as an oscillatory modulation of the coupling strength consisting of two components, a phase difference variable $\phi \propto (x_1 - x_2)$ and the amplitude of the current $I_c$ that regulates the neurotransmitter release. The phase difference $\phi$ is time varying and evolves according to $x_1-x_2$ which incorporates feedback. The constant of proportionality $\mu$ is the rate at which the phase difference $\phi$ evolves. We have kept $I_c = 3$ and $\mu = 3$ in our manuscript.

The readers are also encouraged to look at Zhang {\em et al.}~\citep{ZhXu20} where the authors have built and studied the novel properties of a neural circuit including memristor and Josephson junction in parallel. Segall {\em et al.}~\citep{SeKe14} studied the synchronization between two coupled Josephson junction neurons, and Mishra {\em et al.}~\citep{MiGh21} found neuron-like firing and bursting behaviors in superconducting Josephson junctions. 

\subsection{Memristive coupling}
\begin{figure}[h]
    \centering
    \includegraphics[width=0.3\linewidth]{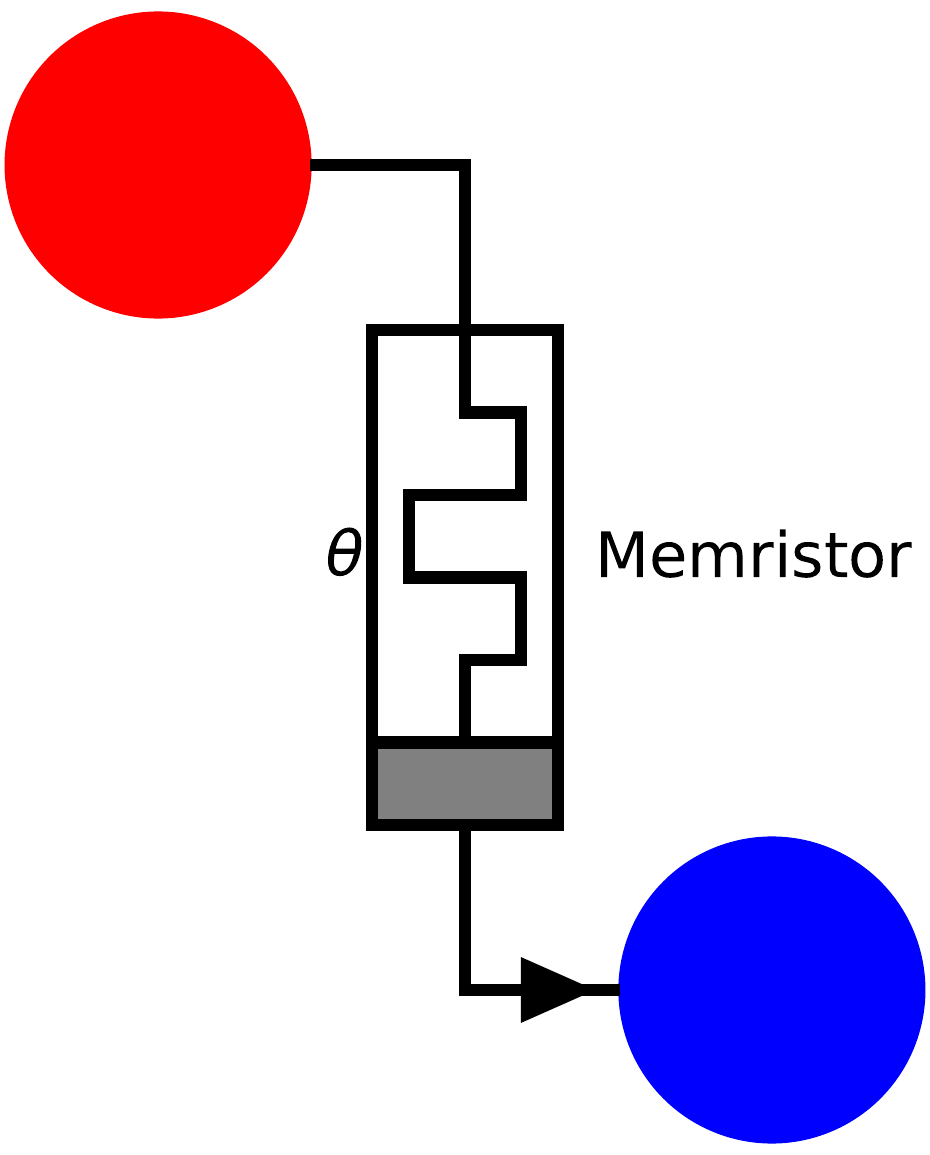}
    \caption{Schematic of a memristive coupling between two neurons, with $\theta$ as the coupling strength.}
    \label{fig:mem_schem}
\end{figure}

Another family of coupling strategies that recently garnered some attention is the memristive coupling. Considered as the fourth fundamental circuit element besides the resistor, capacitor, and the inductor, a {\em memristor}~\citep{Ch71, Ma08} consists of two terminals whose resistance fluctuates dynamically as a function of charge that has flowed through it already. It can be used to model a coupling between the membrane potential of the neuron and an electromagnetic flux, as was done by Xu {\em et al.}~\citep{XuJi17} where the authors coupled two FitzHugh--Nagumo neurons via a memristor. We follow the same approach to couple two dML neurons (see Fig.~\ref{fig:mem_schem}), giving us
\begin{equation}
\label{eq:mem_coupling}
\begin{aligned}
    \dot{x}_1 &= f(x_1, y_1, I_1) + \theta\rho(\phi)(x_2-x_1), \\
    \dot{x}_2 &= f(x_2, y_2, I_2) + \theta\rho(\phi)(x_1-x_2), \\
    \dot{y}_i &= g(x_i, y_i, I_i), \\
    \dot{I}_i & = h(x_i, y_i, I_i), \qquad i = 1,2, \\
    \dot{\phi} &= \theta(x_1 - x_2)
\end{aligned}
\end{equation} 
where $\theta$ is the induction coefficient (the coupling strength that we vary), and $\rho(\phi) (x_i - x_j)$ denotes the current induced by neuron $i$ on $j$. The function $\rho(\phi) = \kappa + 3 \beta \phi^2$ represents the conductance of the memristor. In our manuscript, we fix $\kappa = 10$ and $\beta = 5$. There has also been an extensive study on memristor-based couplings between neurons in recent years~\citep{KoSe21, YoTi23, KoRa24, ShCa24}.

\subsection{Higher-order gap junction coupling in the smallest ring-star network}
\begin{figure}[h]
    \centering
    \includegraphics[width=.4\linewidth]{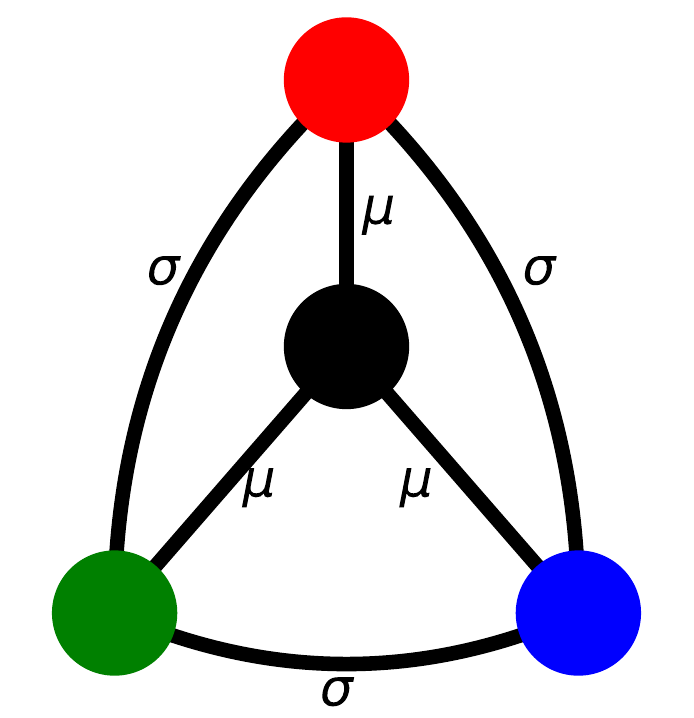}
    \caption{Schematic of a small ring star network with gap junction couplings between neurons. This diagram depicts the pairwise couplings $\sigma$ between two peripheral nodes and $\mu$ between the central node and a peripheral node.}
    \label{fig:ringStar}
\end{figure}

Next, we take a bit of a digression to study a {\em small} network of dML neurons which not only involves pairwise (dyadic) interactions but also {\em higher-order} (polyadic) interactions. By ``higher-order" we mean nonlinear interactions involving more than two nodes within a network. Higher-order interactions in networks are mathematically modeled using {\em hypergraphs} and {\em simplicial complexes}. A hypergraph is a generalization of a classical graph which consists of hyperedges connecting multiple nodes instead of just two. Standard references to learn more about hypergraphs and simplicial complexes are Bick {\em et al.}~\citep{BiGr23} and Berge~\citep{Be84}. In this work, we consider the smallest ring-star network of dML neurons, having one node at the center and three nodes surrounding this central node in the periphery, see Fig.~\ref{fig:ringStar}. Nair {\em et al.}~\citep{NaGh24} considered the smallest ring-star network of neurons with higher-order interactions where the dynamics of each node was governed by Chialvo's model~\citep{Ch95} (which is discrete in time unlike dML which is continuous).

\begin{figure}[h]
\begin{center}
\begin{tabular}{cc}
  \includegraphics[width=0.35\linewidth]{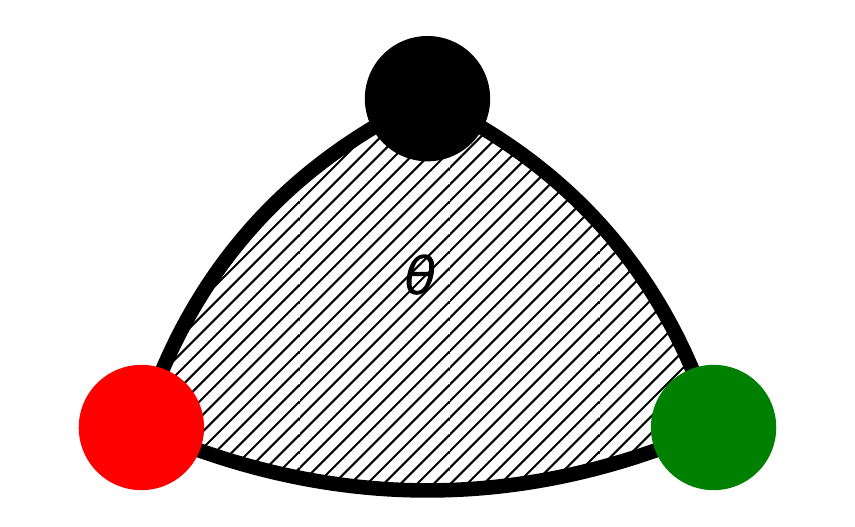} &  \includegraphics[width=0.35\linewidth]{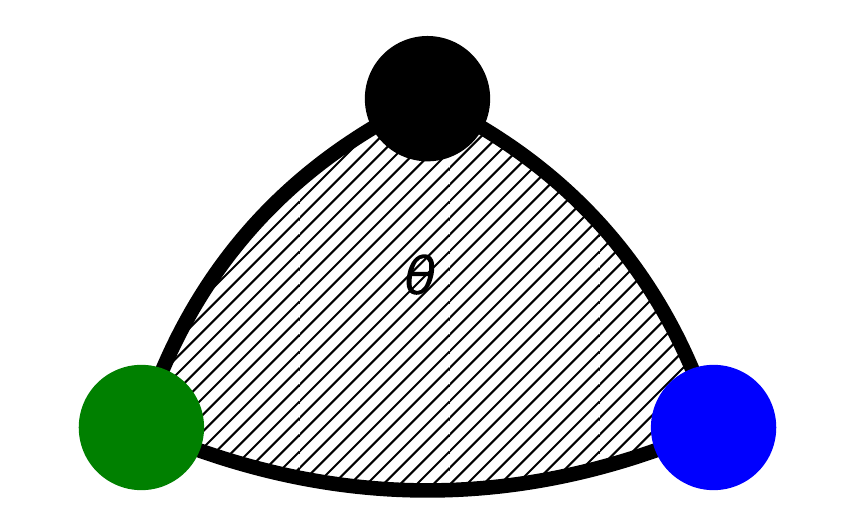}\\
  (a) & (b) \\
  \includegraphics[width=0.35\linewidth]{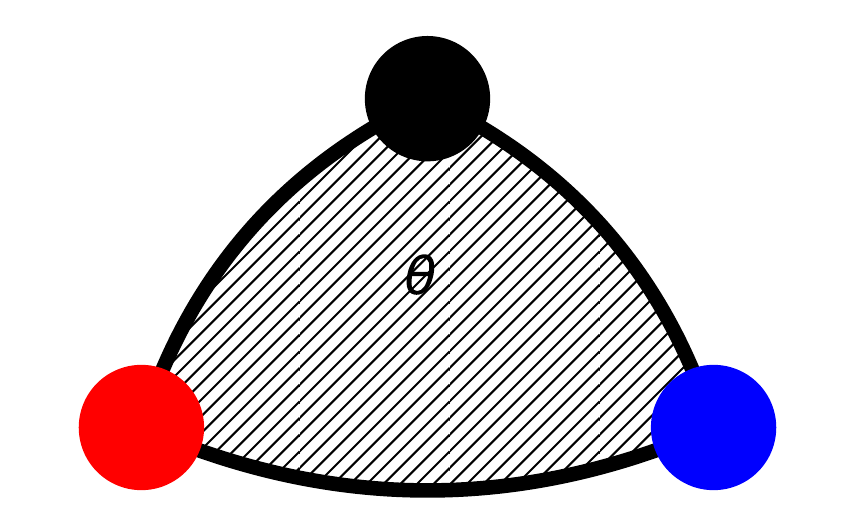} &  \includegraphics[width=0.35\linewidth]{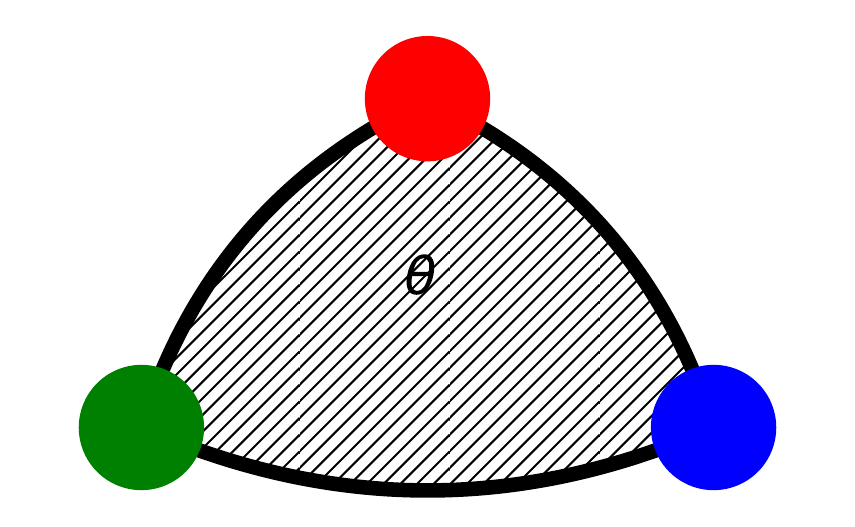}\\
  (c) & (d)
\end{tabular}
\end{center}
\caption{Schematics of higher-order connections within a triangle of neurons represented by two-simplicial complexes.}
\label{fig:higherOrder}
\end{figure}

Our network model is schematically represented by Fig.~\ref{fig:ringStar} depicting the smallest complete ring-star network. The central node is connected to the peripheral nodes with coupling strength $\mu$ and the peripheral modes are connected to each other with coupling strength $\sigma$. The nodes are numbered as (1) for the central node and (2, 3, 4) for the peripheral nodes. We consider bidirectional electrical couplings, thus omitting the directionality of the edges in the schematic. An edge is assumed to be bidirectional. Now, for allowing higher-order couplings, we incorporate two-simplicial complexes in the network. This means the highest-dimensional simplex is a triangle. There are four possible triangles \{1, 2, 3\}, \{1, 3, 4\}, \{1, 2, 4\}, and \{2, 3, 4\}. The higher-order coupling strength is denoted as $\theta$, which is the primary bifurcation parameter in our network model, see Fig.~\ref{fig:higherOrder}.
The model equations are given by
\begin{equation}
\label{eq:higher_order}
\begin{aligned}
    \dot{x}_1 &= f(x_1, y_1, I_1) + (\mu + 2\theta)(x_2+x_3+x_4 - 3x_1), \\
    \dot{x}_2 &= f(x_2, y_2, I_2) + \mu(x_1 - x_2) + \sigma(x_3+x_4-2x_2) \\ &+ 2\theta(x_1+x_3+x_4 - 3x_2),\\
    \dot{x}_3 &= f(x_3, y_3, I_3) + \mu(x_1 - x_3) + \sigma(x_2+x_4-2x_3) \\
    &+ 2\theta(x_1+x_2+x_4 - 3x_3),\\
    \dot{x}_4 &= f(x_4, y_4, I_4) + \mu(x_1 - x_4) + \sigma(x_2+x_3-2x_4) \\ 
    &+ 2\theta(x_1+x_2+x_3 - 3x_4),\\
    \dot{y}_i &= g(x_i, y_i, I_i), \\
    \dot{I}_i & = h(x_i, y_i, I_i), \qquad i = 1,\ldots, 4.
\end{aligned}
\end{equation}
We have fixed $\mu=\sigma = 0.01$ for the smallest higher-order ring-star network of dML neurons in this manuscript. We have closely followed the modeling of Nair {\em et al.}~\citep{NaGh24}. Ring-star network topology of the Chua circuit was studied by Muni {\em et al.}~\citep{MuPr20}. Since then, ring-star networks of neuron models~\citep{MuFa22, GhMu23, BoBe24, NaGh24} and other dynamical systems~\citep{ViMu25} have been widely studied. The research community has recently developed a widespread interest in higher-order network models of neuron dynamics~\citep{TlLe19, NoGa20, PaMe22, MiMe22, WaCh24, HuWu25}.

\subsection{A small random network with autapse and chemical couplings.}
\begin{figure}[h]
    \centering
    \includegraphics[width=0.3\linewidth]{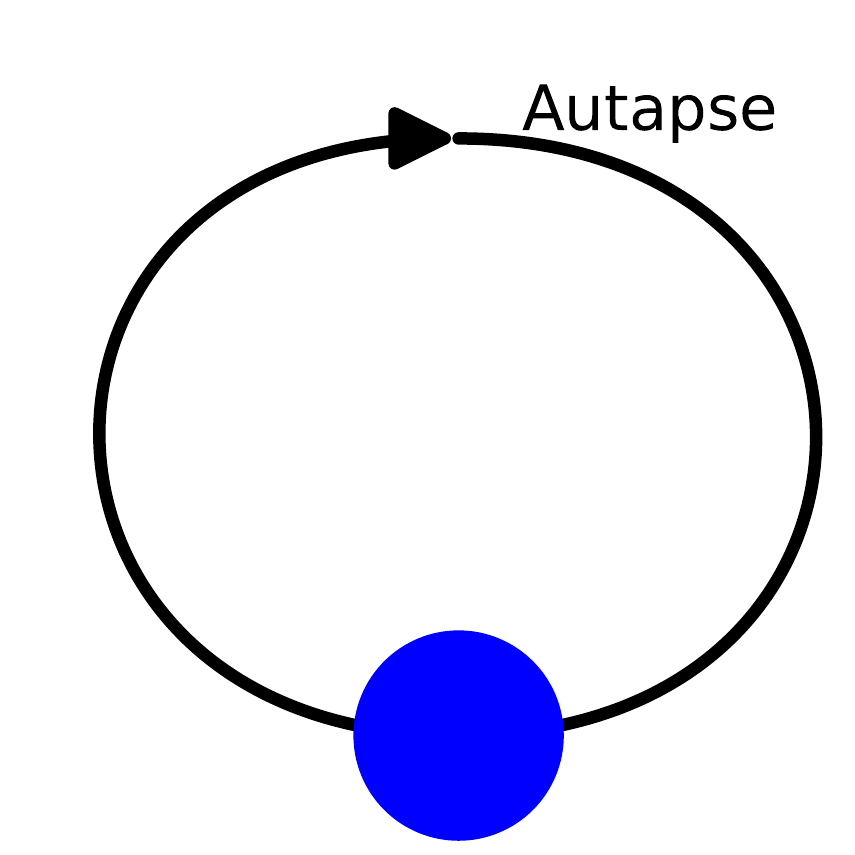}
    \caption{Schematic of an autapse. Note that the coupling considered is chemical}
    \label{fig:autapse_schem}
\end{figure}

\begin{figure}[h]
    \centering
    \includegraphics[width=.4\linewidth]{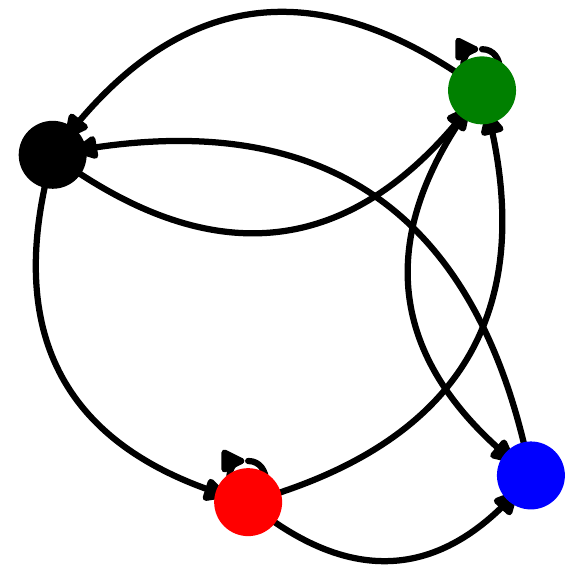}
    \caption{Schematic of a random network of four networks connected by chemical couplings.}
    \label{fig:random_schem}
\end{figure}

Finally, we introduce a small random network made of four nodes (dML neurons) which are connected via chemical couplings (which are unidirectional in structure), see Fig.~\ref{fig:random_schem} for example. In this network we add an extra level of complexity in this small network by allowing autapses, however, ignore any higher-order interactions. An {\em autapse} is a self-loop which biologically represents a ``single neuron with a synapse onto itself"~\citep{SeLe00}. In the brain network, an autapse can be of two categories: excitatory (glutamate-releasing) and inhibitory (GABA-releasing)~\citep{IkBe06}. Thus it makes sense to consider chemical coupling as the coupling strategy. A schematic of an autapse is shown in Fig.~\ref{fig:autapse_schem}. The model equations are given by
\begin{equation}
\label{eq:random}
\begin{aligned}
    \dot{x}_i &= f(x_i, y_i, I_i) + \sum_{j \in B(i)}\theta A_{i, j}\frac{v_s - x_i}{1+e^{-\lambda(x_j - q)}},\\
    \dot{y}_i &= g(x_i, y_i, I_i), \\
    \dot{I}_i & = h(x_i, y_i, I_i),
\end{aligned}
\end{equation}
with $B(i)$ representing the neighboring nodes of $i$. We consider four nodes $i, j = 1, \ldots, 4$. Also, $A_{i, j}$ indicates $\{i, j\}^{th}$ element of the adjacency matrix $A$ of the small network with binary values. This means that there exists a directional coupling from node $i$ to $j$ if $A_{i, j} = 1,$ else $A_{i, j} = 0$. The asymmetric structure of $A_{i, j}$ indicates directional couplings. As this network allows autapse, the diagonal of matrix $A$ can have values $1$, meaning a node coupling to itself.

\section{Methods}
\label{sec:methods}
In this section, we review a list of tools and techniques from non-linear dynamics literature that allow us to analyze the time series data generated from our dynamically simulated small network models. We discuss Hurst exponent, sample entropy, $0-1$ test, Pearson's cross-correlation coefficient, and Kuramoto's order parameter. We report how these measures indicate different dynamical properties of the time series data we have at hand. 

\subsection{Hurst exponent: measuring persistence in time series}
We employ the classical Hurst exponent~\citep{Hu51} computed using the rescaled-range (R/S) analysis following Qian and Rasheed~\citep{QiRa04}. Although the above paper explores the dynamics of financial time series data using Hurst exponent, this statistical measure has been explored to classify models like Lorenz's chaotic model~\citep{FrOs15}. Given a time series data $\{x(t),t=1, \ldots, N\}$ with length $N$, the steps to compute the Hurst exponent using the R/S analysis starts by computing the mean
$$
\overline{x} = \frac{1}{N} \sum_{t=1}^N x(t).
$$
Then the time series is shifted by the mean to calculate the adjusted series
$$
x^{\rm adj}(t) = x(t) - \overline{x}, \qquad t = 1, \ldots, N.
$$
From this adjusted series, one can next compute the cumulative deviate series
$$
Q(t) = \sum_{j = 1}^tx^{\rm adj}(j), \qquad t = 1, \ldots, N.
$$
We now utilize $Q(t)$ to devise the last two important series: the range series
\begin{align}
  R(t) &= \max(Q(1), Q(2), \ldots, Q(t)) \nonumber\\ 
  &-\min(Q(1), Q(2), \ldots, Q(t)), \qquad t = 1, \ldots, N, \nonumber  
\end{align}

and the standard deviation series
$$
S(t) = \sqrt{\frac{1}{t} \sum_{j=1}^t (x(j) - z)^2}, \qquad t = 1, \ldots, N,
$$
where $z$ is the mean given by $z = \frac{1}{t}\sum_{k=1}^tx(k)$. Thus the rescaled range series is given by
$$
RS(t) = \frac{R(t)}{S(t)}, \qquad t = 1, \ldots N,
$$
which is an element-wise fraction. It should be noted that the series $RS(t)$ is averaged over all partial time series $\{x(1), x(t)\}$, $\{x(t+1), x(2t)\}$, till $\{x((m-1)t), x(mt)\}$. Here $m = \lfloor \frac{N}{t} \rfloor$. Hurst reported that 
$$
RS(t) = ct^H,
$$
where $c$ is a constant and $H$ is the exponent we are trying to evaluate. The simplest way to do it is to plot the series $RS(t)$ as a function of $t$ on a $\log-\log$ scale. The slope of the line will be the approximation of the exponent $H$. Note that the length of the time series should be $N>10$. Otherwise the approximation of $H$ would be inaccurate.

Note that the value of $H$ lies in the range $[0, 1]$. Depending on the value a time series can fall into three major categories: (i) an {\em anti-persistent} time series with $H \in [0, 0.5)$ meaning the time series exhibits some negative dependence on previous values, (ii) a {\em random} time series with $H \approx 0.5$ behaving like a random walk, and (iii) a {\em persistent} time series with $H \in (0.5, 1]$ meaning the time series exhibits some positive dependence on previous values. An $H<0.5$ indicates that if a time series has an upward trend, it is more likely to exhibit a downward trend next, and an $H>0.5$ indicates that if a time series exhibits an upward trend, it is more likely to go upward in the subsequent time steps. Persistence in a time series refers to its tendency to memorize its past values. As put forward by Mandelbrot~\citep{Ma85}, the Hurst exponent is related to the fractal dimension $D$ for a self-similar time series by the formula $D = 2-H$. A higher value of $H$ implies a smoother time series, having a lower fractal dimension.

\subsection{Sample Entropy: measuring complexity of time series}

Next, we discuss the implementation of the sample entropy on a time series data following Richman and Moorman~\citep{RiMo00}. For a time series $\{x(t), t = 1, \ldots, N\}$, let $p \le N$ be a non-negative integer, using which we can define $N-p +1$ vectors from the time series
$$
\tilde{x}_p(j) = \{x(j +k)\vert 0\le k \le p-1\},
$$
with $1 \le j \le N - p + 1$. This set consists of $p$ elements which enumerates from $x(j)$ to $x(j+p - 1)$. Now the Euclidean distance between two such vectors is given by
$$
\mathcal{E}(\tilde{x}_p(j), \tilde{x}_p(n)) = \max_{0 \le k \le p - 1} \{\vert x(j+k) - x(n+k) \}.
$$
For a positive tolerance $\varepsilon$, Richman and Moorman defined the term $B^p_j(\varepsilon)$ as $\frac{1}{N - p -1}$ times the number of vectors $x_p(j)$ satisfying $\mathcal{E}(x_p(j), x_p(n)) \le \varepsilon$. Note that $1\le j \le N - p$ with $j\ne i$ (this is to ensure that self matches are excluded). The weighted average is then given by
$$
B^p(\varepsilon) = \frac{1}{N-p}\sum_{j=1}^{N-p}B_j^p(\varepsilon),
$$
which represents the probability that two series will match for $p$ points. Similarly, the authors defined the term $A^p_j(\varepsilon)$ as $\frac{1}{N - p -1}$ times the number of vectors $x_{p+1}(j)$ satisfying $\mathcal{E}(x_{p+1}(j), x_{p+1}(n)) \le \varepsilon$. Note that again $1\le j \le N - p$ with $j\ne i$ to exclude self-matches. The weighted average is then given by
$$
A^p(\varepsilon) = \frac{1}{N-p}\sum_{j=1}^{N-p}A_j^p(\varepsilon),
$$
which represents the probability that two series will match for $p+1$ points. Then the sample entropy is given by
$$
{\rm SE}(p, \varepsilon, N) = \lim_{N \to \infty} \bigg(-\log_e \frac{A^p(\varepsilon)}{B^p(\varepsilon)} \bigg).
$$
A higher sample entropy value indicates higher irregularity in the time series, directly relating to a higher complexity and lesser self-similarity.

\subsection{$0-1$ test: measuring chaos from time series}
We employ the $0-1$ test to detect chaos from our time-series data following the works of Gottwald and Melbourne~\citep{GoMe09, GoMe09b, GoMe16}. From the time series $\{x(t), t = 1, \ldots, N\}$, the first step is compute two translation variables given by $\tilde{p}(t; e)$ and $\tilde{q}(t; e)$ given by
\begin{align}
    \tilde{p}(t; e) &= \sum_{k=1}^t x(k) \cos(ek), \nonumber \\
    \tilde{q}(t; e) &= \sum_{k=1}^t x(k) \sin(ek). \nonumber 
\end{align}
Here $e \in (0, 2\pi)$ is a small number. The plot of $\tilde{p}(t; e)$ vs $\tilde{q}(t; e)$ will generally be bounded for a regular dynamical variable from the system in question, or will approximate a two-dimensional Brownian motion with $\sqrt{n}$ evolution rate and a drift of $0$ for when the dynamical variable of the system behaves irregularly/chaotically. A mean-square displacement term can be computed
\begin{align}
m(t; e) &= \frac{1}{N}\sum_{j=1}^N \bigg[\bigg(\tilde{p}(j + t; e) - \tilde{p}(j; e) \bigg)^2 \nonumber \\
&+ \bigg(\tilde{q}(j + t; e) - \tilde{q}(j; e) \bigg)^2 \bigg]. \nonumber
\end{align}
For better numerical results with $N \to \infty$, $t$ must follow $t \ll N$. We also require to have a critical value $N_{\rm crit}$ such that $m(t; e)$ is computed only for an $t \le N_{\rm crit}$, with the extra requirement $N_{\rm crit} \ll N$. Gottwald and Melbourne recommend a value $N_{\rm crit} \le \frac{N}{10}$ for a better performing algorithm. The next step is adding a correction term to $m(t; e)$,
$$
\tilde{m}(t; e) = m(t; e) - \bigg(\lim_{N \to \infty} \frac{\sum_{j=1}^N x(j)}{N} \bigg)^2 \times \frac{1-\cos(et)}{1-\cos(e)}.
$$
This updated term $\tilde{m}(t; e)$ converges better than $m(t; e)$ and grows asymptotically at the same rate as $m(t; e)$. This rate can be computed by 
$$
K(e) = \lim_{t \to \infty} \frac{\log m(t; e)}{\log t}.
$$
Now this term can be computed in two different statistical ways: a {\em regression} approach or a {\em correlation} approach. For the regression approach, one deals with the term $\tilde{m}(t; e)$ because it exhibits a lesser variance than $m(t; e)$. As $\tilde{m}(t; e)$ can have negative values, it can be updated further to accommodate for the correction
$$
r(t; e) = \tilde{m}(t; e) - \min_{1 \le t \le N} \tilde{m}(t; e).
$$
One can then evaluate $K(e)$ numerically by plotting $r(t;e)$ against $t$ on a $\log-\log$ scale and using a regression method to approximate the growth rate. If the mean square displacement $m(t; e)$ is bounded then $K(e) \approx 0$ (indicates regular dynamics in the system), whereas if $m(t; e)$ grows linearly then $K(e) \approx 1$ (indicates chaos in the system). It is recommended to compute different $K(e)$ values for a range of $e$ values and taking the median of these $K(e)$ values to approximate an accurate value of the growth rate. 

Now to compute this growth rate using the correlation approach, we start by preparing a vector of time steps $s_1 = (1, 2, \ldots, N_{\rm crit})$ and a corresponding vector of $\tilde{m}(t; e)$ values given by $s_2 = (\tilde{m}(1; e), \tilde{m}(2; e), \ldots, \tilde{m}(N_{\rm crit}; e))$. The growth rate is then given by the correlation formulation between the two vectors
$$
K(e) = \frac{\overline{s_1^{\rm adj}s_2^{\rm adj}}}{\sqrt{\bigg(\overline{s_1^{\rm adj}}\bigg)^2 \bigg(\overline{s_2^{\rm adj}}\bigg)^2}},
$$
where $s_i^{\rm adj} = s_i - \overline{s_{i}}$ and $\overline{s}_i = \frac{1}{N_{\rm crit}}\sum_{j=1}^{N_{\rm crit}} s_i(j)$. Note that $K(e) \in [0, 1]$. Again, it is advisable to compute different $K(e)$ values for a range of $e$ values and take the median of these $K(e)$ values to approximate an accurate growth rate value.

\subsection{Pearson's correlation coefficient: measuring synchronization}
Nodes in a dynamical network tend to synchronize~\citep{Su06, St12} as a signature of complex collective behavior. Neurons synchronize when the voltage variables coincide on a spatio-temporal regime, leading to coordinated changes to the extracellular ionic activities. Neuron synchronization depends on gap junction, chemical, and other types of nonlinear synaptic interactions~\citep{TiBa12}. As a quantity to characterize this synchronization, researchers have widely employed Pearson's normalized correlation coefficient in both neuron dynamical systems~\citep{RyZa21, GhMu23, NaGh24, GhFa25}, and other kinds of dynamical systems~\citep{VaSt16, RyAn19, ShBu20, BuRy17}.

Given the time series data generated for two different nodes $m$ and $n$ from a network, the correlation coefficient is defined by
\begin{align}
    \Gamma_{m, n} = \frac{\overline{x_m^{\rm adj}(t) x_n^{\rm adj}(t)}}{\sqrt{\bigg(\overline{x_m^{\rm adj}(t)}\bigg)^2 \bigg(\overline{x_n^{\rm adj}(t)}\bigg)^2}}.
\end{align}
Now, for a network of $M$ nodes connected via some form of coupling strategies, the normalized correlation coefficient is given by
$$
\Gamma = \frac{1}{M-1}\sum_{n=1, n\ne m}^M \Gamma_{n, m}.
$$
When the nodes are synchronized with each other we will have $|\Gamma| = 1$. Asynchrony between the nodes is characterized by $|\Gamma|<1$. Moreover, $\Gamma = 1$ represents in-phase synchrony and $\Gamma = -1$ represents anti-phase synchrony.

\subsection{Kuramoto Order parameter: measuring synchronization}
Kuramoto and Battogtokh~\citep{KuBa02} introduced an order parameter to quantify coherence in phases of Kuramoto oscillators, that is, measure how stable the phase difference between these oscillators are. Since then, Kuramoto's order parameter has been utilized to study synchronization in neuron dynamics models like the Leaky Integrate-and-Fire neurons~\citep{AnPr22}, FitzHugh--Nagumo neurons~\citep{OmPr15, RyZa21}, Chialvo neurons~\citep{NaGh24}, and others.

In order to compute the order parameter, one starts by finding the phase $\zeta_m$ of an oscillator at time $t$ which is given by
\begin{align}
    \zeta_m = \tan^{-1}\bigg(\frac{y_m(t)}{x_m(t)}\bigg).
\end{align}
The complex-valued Kuramoto index $B$ is then given by
\begin{align}
    B_m(t) = \exp(i\zeta_m(t)), \qquad i = \sqrt{-1}.
\end{align}
The index at time $t$ is $B(t) = \bigg|\frac{1}{N}\sum_{m=1}^N B_m(t)\bigg|$. This notation represents the mean of all phases within the unit circle. Thus, the time-averaged value is given by $B = \langle B(t) \rangle_t$. When $B=1$, this means the nodes are all fully coherent and their phases are all locked. Any value $B <1$ represents incoherence, with the magnitude of incoherence increasing as $B \to 0$. More detailed explanations on Kuramoto's order parameter are provided in Kuramoto~\citep{Ku84}, Strogatz~\citep{St00}, Bick {\em et al.}~\citep{BiGo20}, among others.

\section{Numerics}
\label{sec:numerics}

In this section, we implement the methodologies from Sec.~\ref{sec:methods} to the models introduced in Sec.~\ref{sec:couplings}. A reminder to the readers on the parameter values are provided in Table~\ref{tab:params}. These parameter values were chosen after performing extensive numerical experiments and ensuring that the dynamics of the network do not diverge. For each model, we first provide a collection of phase portraits and time series by varying the primary bifurcation parameter. For every value of the bifurcation parameter, we also compute the metrics from time-series analysis and display them on top of the respective phase portrait-time series plot. This provides an overview of the qualitative nature of the dynamics of the coupled system, governed by the primary bifurcation parameter. Additionally, we show how these metrics behave over a parameter sweep, which allows us to compare how the different coupling strategies control the dynamics of the coupled system.

\begin{table}[h]
    \centering
    \begin{tabular}{|c|p{2.5cm}|p{2.6cm}|}
        \hline
         Model & Constants & Bifurcation range \\
        \hline\hline

        \multirow{4}{*}{\eqref{eq:model_single}} & $A = 0.0041$ & \multirow{4}{*}{N/A} \\
        \cline{2-2}
        & $\alpha = 5.276$ & \\
        \cline{2-2}
        & $\gamma = 0.315$ & \\
        \cline{2-2}
        & $\varepsilon = 0.0005$ & \\
        \hline

        \eqref{eq:gapJunction} & \eqref{eq:model_single} & $\theta \in [-10, 10]$\\
        \hline

        \multirow{3}{*}{\eqref{eq:temperature}, \eqref{eq:Arrhenius}} & \eqref{eq:model_single} & \multirow{3}{*}{} \\
        \cline{2-2}
        & $\delta_0 = 1.3$ & $\theta \in [-5, 5]$  \\
        \cline{2-3}
        & $T_{\rm ref} = 20^{\circ} C$ & $T \in [0, 40] ^\circ C$ \\
        \hline

        \multirow{4}{*}{\eqref{eq:dML_chemical}} & \eqref{eq:model_single} & \multirow{4}{*}{ $\theta \in [-0.005, 0.1]$} \\
        \cline{2-2}
        & $v_s =2$ & \\
        \cline{2-2}
        & $\lambda=10$ & \\
        \cline{2-2}
        & $q = -0.25$ & \\
        \hline

        \multirow{3}{*}{\eqref{eq:JJ}} & \eqref{eq:model_single} & \multirow{3}{*}{$\theta \in [-1, 1]$} \\
        \cline{2-2}
        & $I_c = 3$ & \\
        \cline{2-2}
        & $\mu=3$ & \\
        \hline

        \multirow{3}{*}{\eqref{eq:mem_coupling}} & \eqref{eq:model_single} & \multirow{3}{*}{$\theta \in [-0.02, 0.01]$} \\
        \cline{2-2}
        & $\kappa = 10$ & \\
        \cline{2-2}
        & $\beta=5$ & \\
        \hline

        \multirow{3}{*}{\eqref{eq:higher_order}} & \eqref{eq:model_single} & \multirow{3}{*}{$\theta \in [-0.1, 0.1]$} \\
        \cline{2-2}
        & $\mu = 0.01$ & \\
        \cline{2-2}
        & $\sigma=0.01$ & \\
        \hline

        \multirow{4}{*}{\eqref{eq:random}} & \eqref{eq:model_single} & \multirow{4}{*}{ $\theta \in [-0.01, 0.01]$} \\
        \cline{2-2}
        & $v_s =2$ & \\
        \cline{2-2}
        & $\lambda=10$ & \\
        \cline{2-2}
        & $q = -0.25$ & \\
        \hline

    \end{tabular}
    \caption{Parameter values considered for models introduced in Sec.~\ref{sec:couplings}.}
    \label{tab:params}
\end{table}

We now describe various numerical tools and techniques that we have incorporated to perform the time series analysis for our models. For every simulation we run it for $t \in [0, 4000]$ time span. We generally create a time series consisting of $50000$ data points that we utilise for most of our simulations except as otherwise stated. Note that our models are initial value problems and we solve them using the \texttt{solve\_ivp()} function from \texttt{Python scipy}'s \texttt{integrate} package. For the time integration of the differential equations, we use \texttt{method = `RK45'} which is the explicit Runge-Kutta method of order 5(4)~\citep{DoPr80}. 

Let us first talk about the \texttt{nolds}~\citep{Sc19} package written in \texttt{Python}, which stands for `NOnLinear measures for Dynamical Systems'. We utilize \texttt{nolds} to compute the Hurst exponent and sample entropy from a time series. The function to compute the Hurst exponent is \texttt{hurst\_rs()} that is built upon the rescaled-range approach that we talked about in Sec.~\ref{sec:methods}. In the package there exists flexibility to choose the number of partial time series $\{x(1), x(t)\}$, $\{x(t+1), x(2t)\}$, till $\{x((m-1)t), x(mt)\}$, with $m = \lfloor \frac{N}{t} \rfloor$. By default it says ``15 logarithmically spaced values in the medium 25\% of the logarithmic range". We keep the default value for our simulations. We report the averaged Hurst exponent in our simulations by first computing the Hurst exponent from time series of every node in the network and then taking the average. The function \texttt{sampen()} provides for evaluating the sample entropy of the nodes in our coupled network models. We then similarly report the averaged sample entropy of the system. This function is built on the algorithm by Richman and Moorman~\citep{RiMo00}. We have employed this package to analyze both continuous-time neuron dynamics~\citep{GhFa25} and map-based models of neuron dynamics~\citep{GhMu23, GhNa24, NaGh24}. The default values of $p$ and $\varepsilon$ in the function are set as $2$ and $0.2$ times the standard deviation of the input time series, respectively. 

For testing chaos in our time series, we implement the $0-1$ test. We refer to the open-source \texttt{Julia} package \texttt{01ChaosTest.jl}~\citep{amJulia} for this purpose. The \texttt{Julia} package is converted to \texttt{Python}\footnote{We thank Anjana S. Nair for helping with this conversion.} to keep the code base entirely homogeneous. For the computation of $K$ we divide the time range $t \in [0, 4000]$ to $10000$ data points instead of $50000$. For every model in~\ref{sec:couplings}, we keep $c=1.1$. Also $N_{\rm crit} = 20 \le 10^3$ except for~\eqref{eq:mem_coupling} and~\eqref{eq:higher_order} where $N_{\rm crit} = 50 \le 10^3$, and for~\eqref{eq:random} where $N_{\rm crit} = 80 \le 10^3$. We use both the correlation method and the regression method of $0-1$ test in our simulations. The $p$ vs $q$ plots generated from the time series are displayed besides the phase portraits of the coupled systems.

Finally, the computations of the cross-correlation coefficient $\Gamma$ and the Kuramoto order parameter $B$ are very straightforward, and we have implemented the formulas from Sec.~\ref{sec:methods} in \texttt{Python}. For $\Gamma$, we discard the first $5000$ data points from our time series, except for~\eqref{eq:random} in which case we discard the first $500$ data points. All data files and \texttt{Python} scripts are openly available for download and implementation from our GitHub repository: \url{https://github.com/indrag49/TS-SlowFast-dML}.

\begin{figure*}[h]
\begin{tabular}{ccc}
  \includegraphics[width=0.3\linewidth]{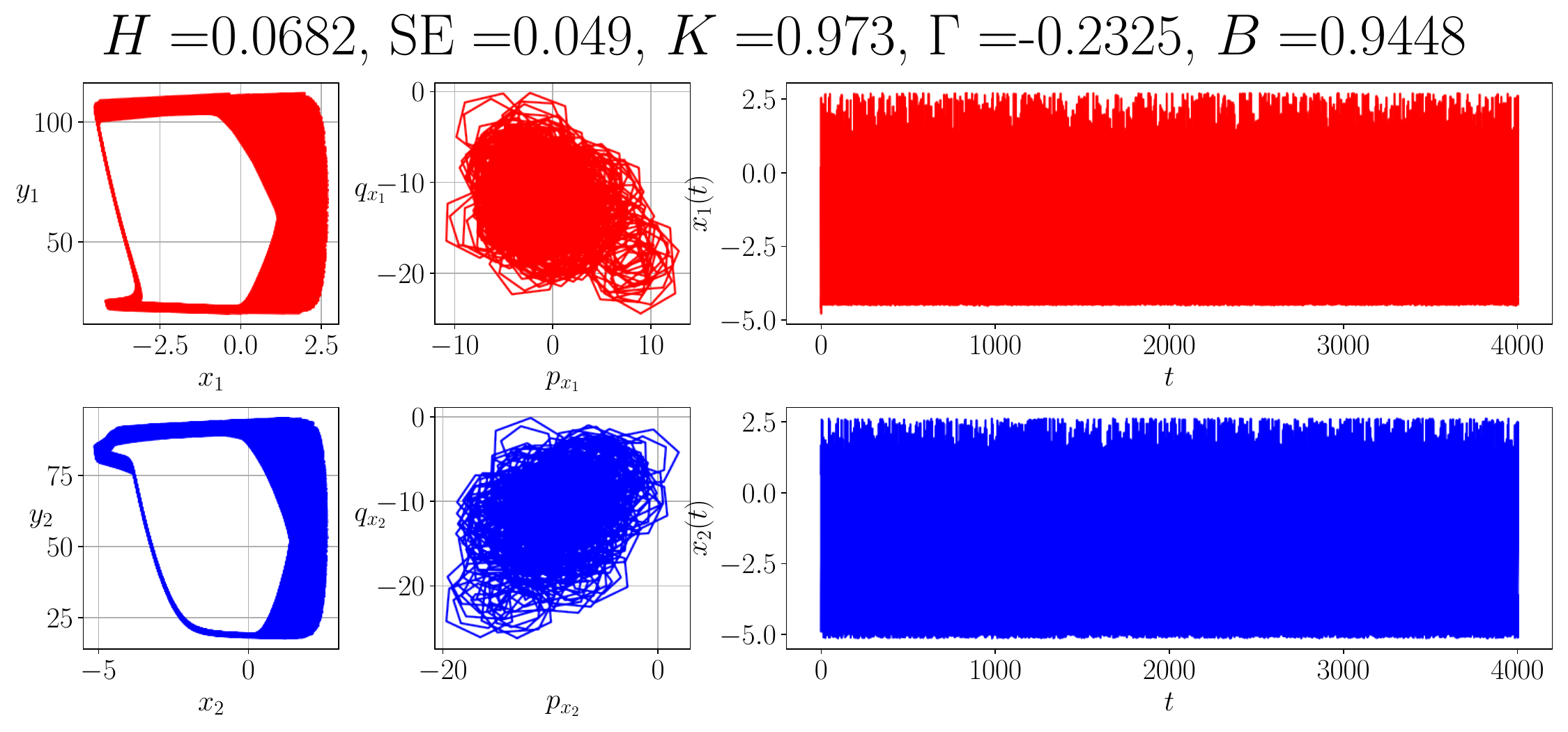} &  \includegraphics[width=0.3\linewidth]{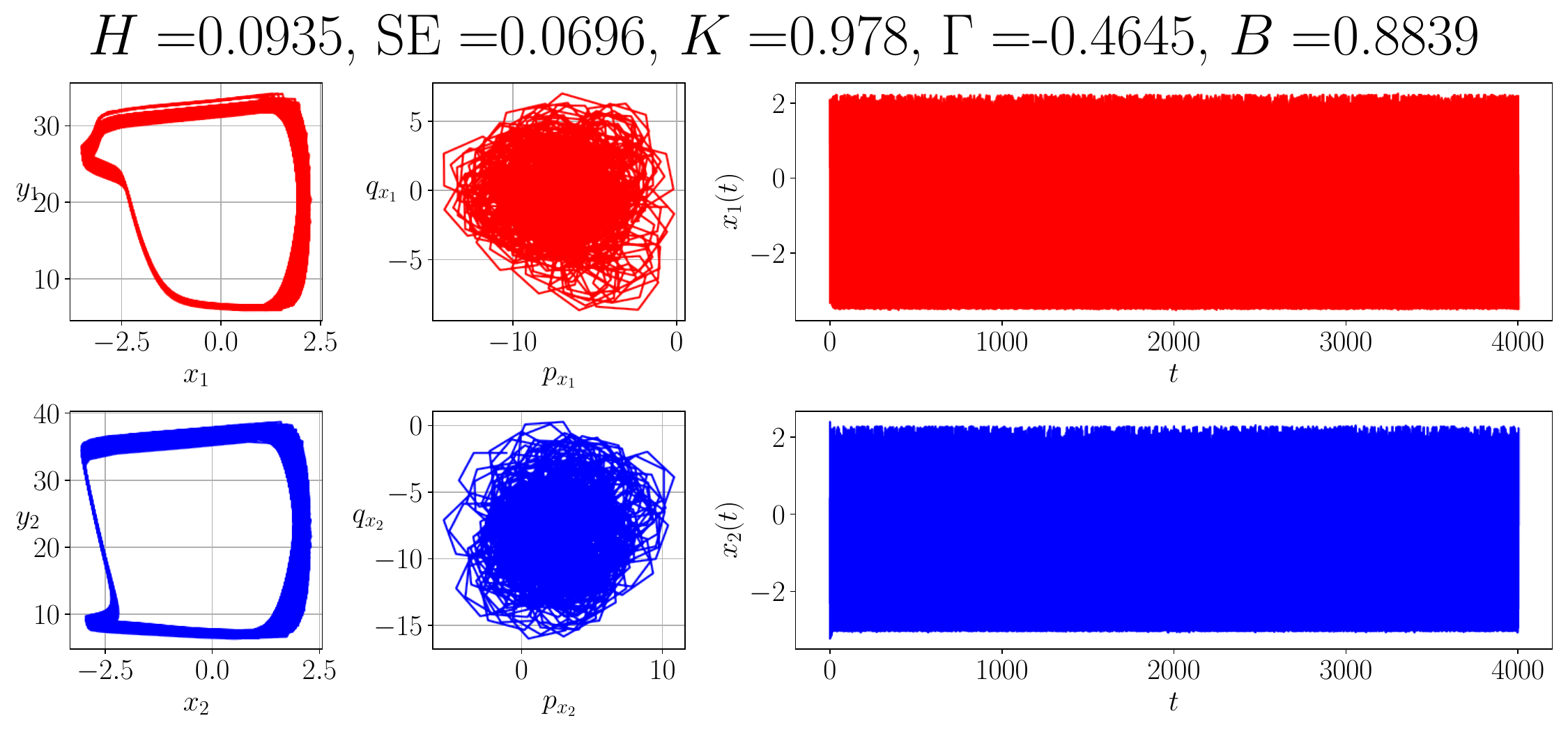} &  \includegraphics[width=0.3\linewidth]{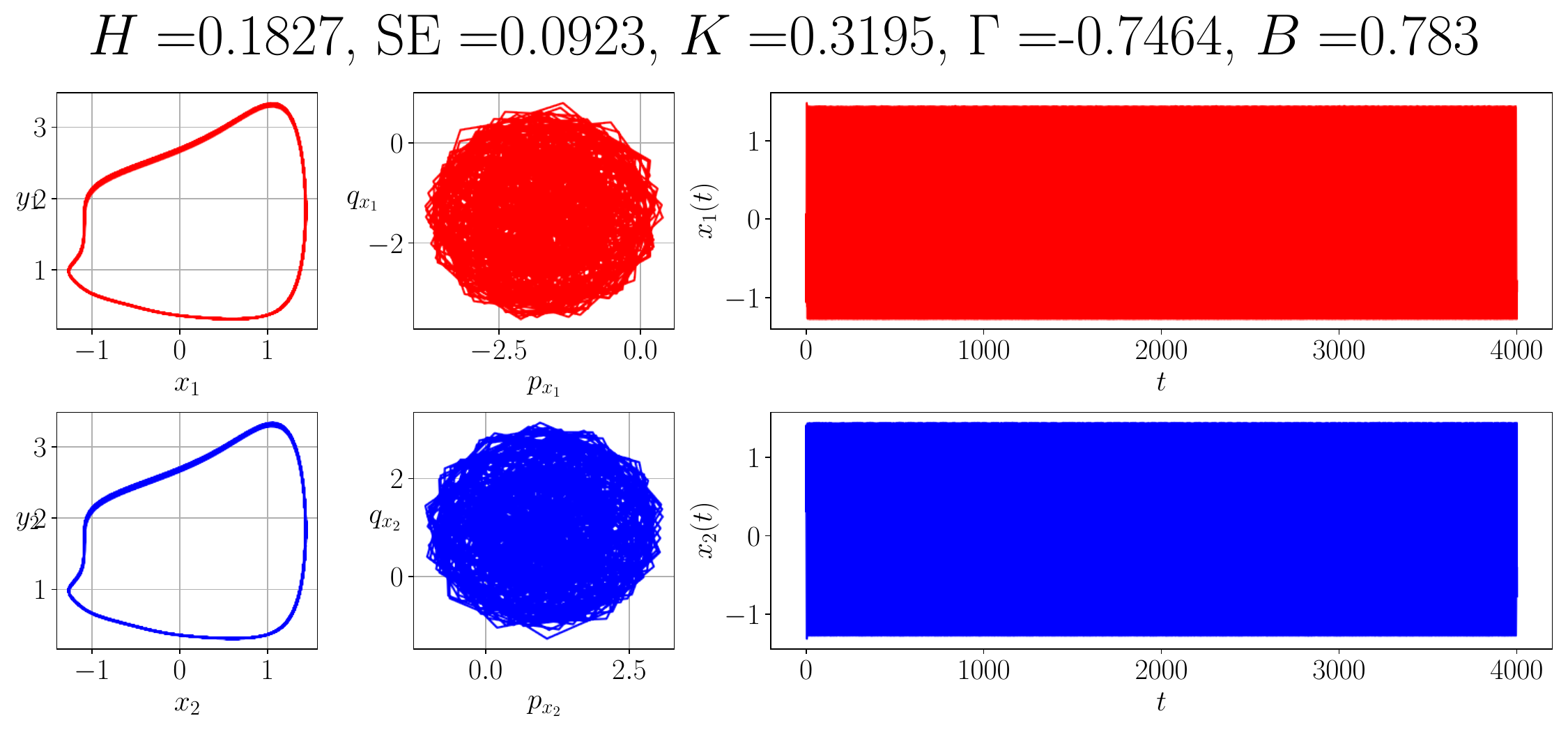} \\
(a) $\theta = -10$ & (b) $\theta = -5$ & (c) $\theta = -1$ \\
\includegraphics[width=0.3\linewidth]{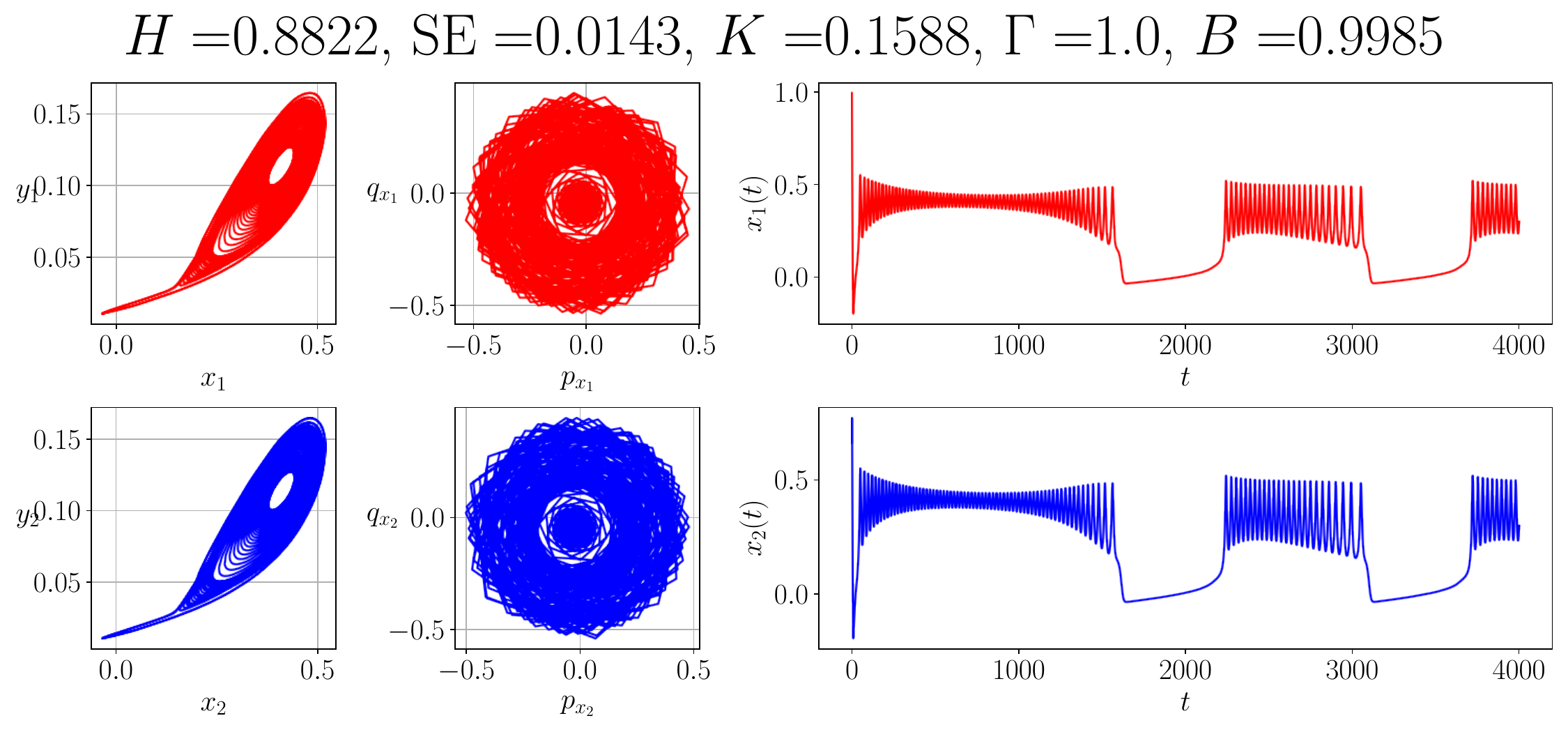} &  \includegraphics[width=0.3\linewidth]{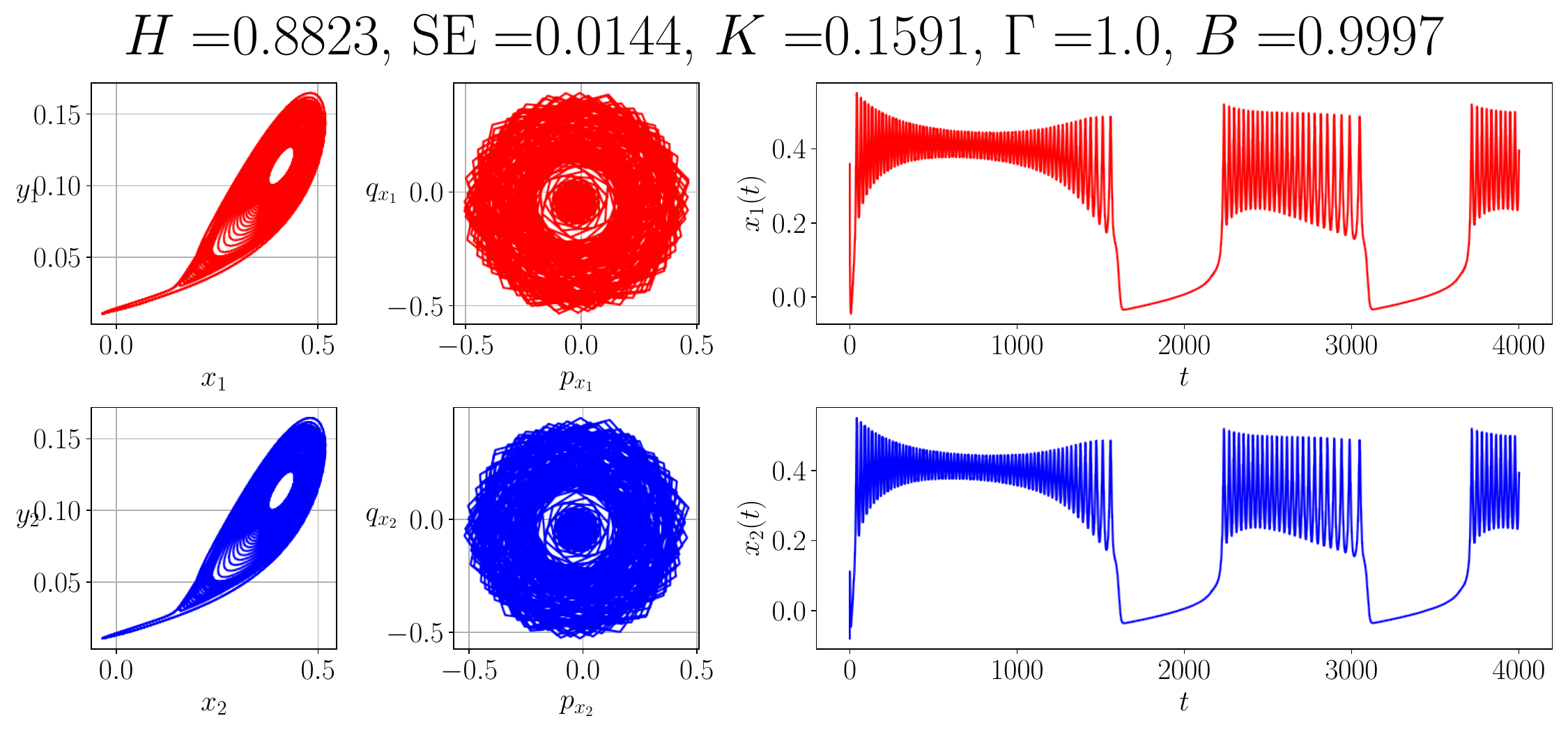} &  \includegraphics[width=0.3\linewidth]{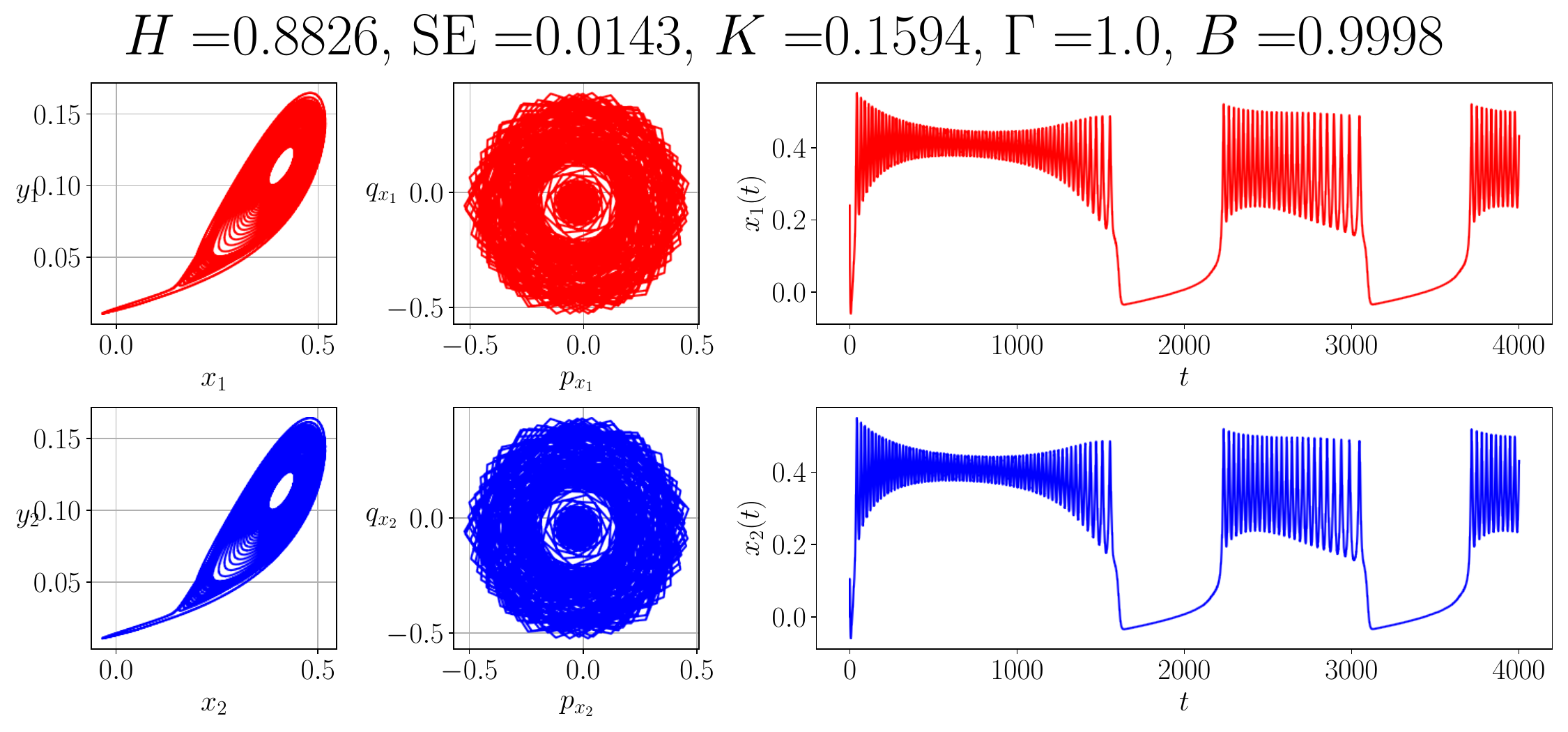} \\
(d) $\theta = 1$ & (e) $\theta = 5$ & (f) $\theta = 10$\\
\end{tabular}
\caption{Phase portraits, $p$ vs $q$ plots, and time series of~\eqref{eq:gapJunction} with varying $\theta \in [-10, 10]$. Other parameters are fixed as in table~\ref{tab:params}, and initial conditions $x_1(0)$ and $x_2(0)$ are sampled randomly from the continuous uniform distribution over the interval $[-1, 1]$. Other initial conditions are fixed as $y_1(0) = y_2(0)= 0.1$, $I_1(0) = 0.019$, and $ I_2(0)= 0.022$. We observe anti-persistence and chaos for a strong inhibitory coupling, which dies down to a more regular behavior exhibiting quasi-periodicity as $\theta$ approaches $0$. When $\theta$ becomes excitatory and more positive, the system exhibits persistence, with the two nodes portraying bursting and fully synchronized. The corresponding parameter sweep plots are shown in Fig.~\ref{fig:BifGap}.}
\label{fig:Gap_pp}
\end{figure*}

We start with simulating~\eqref{eq:gapJunction}, see Fig.~\ref{fig:Gap_pp}. Parameter $\theta$ is varied from $-10$ to $10$, and we show six instances in this range. The initial conditions $x_1(0)$ and $x_2(0)$ are sampled randomly from the continuous uniform distribution over the interval $[-1, 1]$. Other initial conditions are fixed as $y_1(0) = y_2(0)= 0.1$, $I_1(0) = 0.019$, and $ I_2(0)= 0.022$. For a highly inhibitory coupling strength $\theta = -10$, we notice an anti-persistent behavior in the time series with $H = 0.0682 \ll 0.5$ and a higher unpredictability with its sample entropy being a high value of ${\rm SE}\approx 0.05$. From the phase portrait, $p$ vs $q$ plot (exhibiting a random Brownian motion), and the time series itself, we observe a clear chaotic pattern (reflected by the Hurst exponent and the sample entropy already). This is further strengthened by $K =0.973$, which is very close to $1$. Both nodes are oscillating in anti-synchronous fashion with respect to each other, thus a value of $\Gamma = -0.2325$. The value of $B=0.9448$ does not tell a story yet, but will be more evident when we compare the behavior of the system's dynamics for a higher value of $\theta$. As $\theta$ increases to $\theta = -5$, the qualitative property of the system remains similar to that of $\theta = -10$. On increasing $\theta$ further to $\theta = -1$, we see the appearance of a quasi-periodic orbit. The time series becomes less anti-persistent with an increase in the Hurst exponent to $H = 0.1827$. There is even higher unpredictability with sample entropy increasing to ${\rm SE} = 0.0923$. The attractor exhibits a thinner trajectory, however still shows irregularity. The $p$ vs $q$ plots for both nodes have become more regular compared to a more negative value of $\theta$. Quasiperiodicity is supported by a smaller value of $K = 0.3195$. The nodes are still asynchronous with $\Gamma = -0.7464$. Note that $B = 0.783$ indicates asynchrony between the phases of both nodes. As soon as $\theta >0$, both nodes show fold/homocilinc bursting behavior. The time series for both nodes become highly persistent with $H >0.88$, with sample entropy dropping down to ${\rm SE} \approx 0.0143$. Note that the $p$ vs $q$ plots becomes more bounded with $K$ value dropping down to $K \approx 0.159$. Both nodes oscillate in synchrony as $\Gamma =1$ and they are phase-locked with $B \approx 1$, indicating a high coherence.

\begin{figure}[h]
    \centering
    \includegraphics[width=0.7\linewidth]{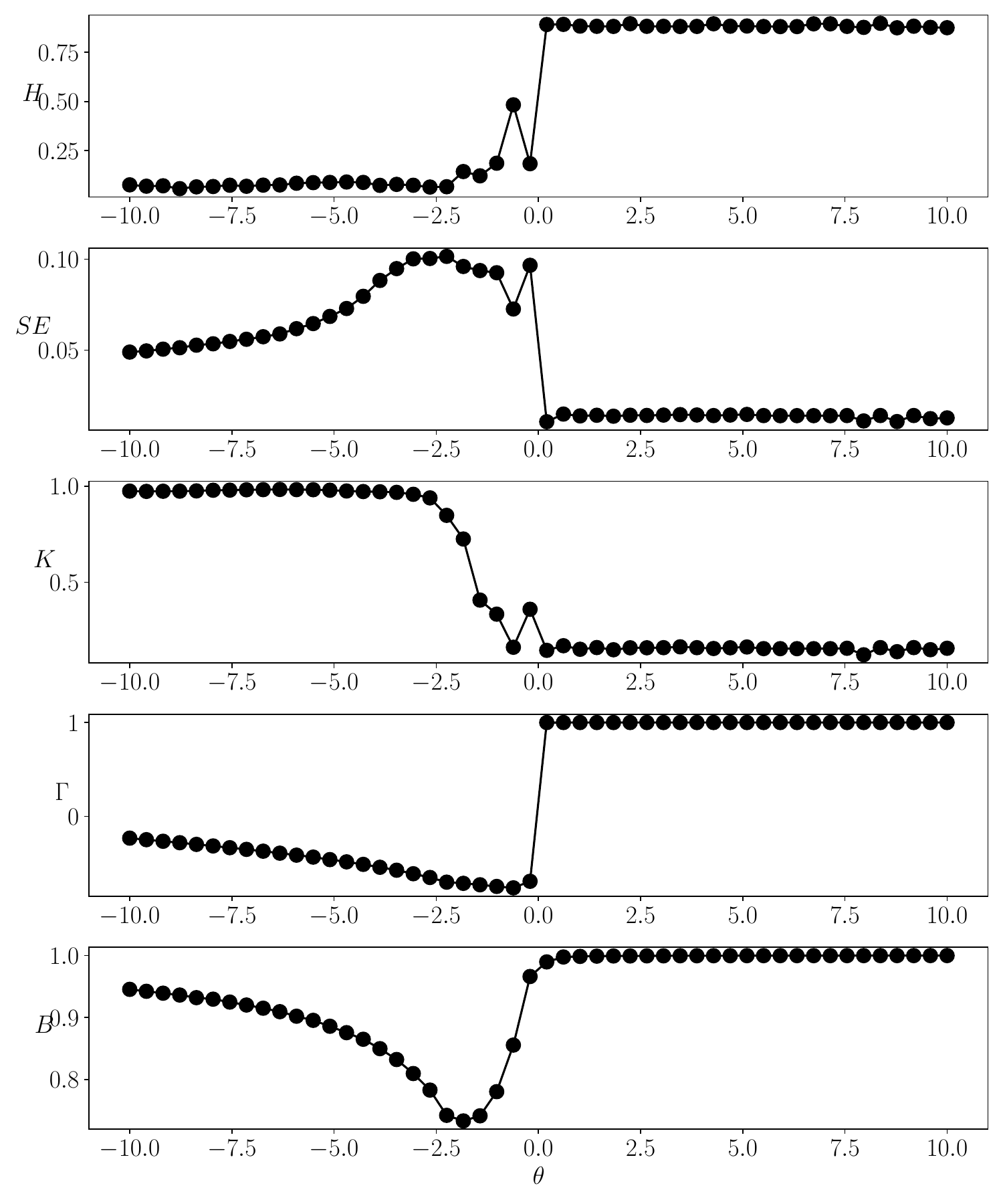}
    \caption{Bifurcation plots of different metrics performing a parameter sweep on $\theta \in [-10, 10]$ for model~\eqref{eq:gapJunction}.}
    \label{fig:BifGap}
\end{figure}

We then portray a collection of bifurcation plots of these metrics doing a parameter sweep on $\theta \in [-10, 10]$, see Fig.~\ref{fig:BifGap}. We do this by dividing the range of $\theta$ into fifty equally spaced values. When $\theta<0$, we observe an anti-persistent behavior in the time series with $H \ll 0.5$, and an unpredictable complexity with ${\rm SE}$ increasing monotonically from close to $0.05$ to approximately $0.1$ as $\theta$ approaches $0$. A maximum is reached around $\theta \approx -0.25$. In this range the system is chaotic with $K \approx 1$. Both nodes are asynchronous to each other with $-1<\Gamma <0$ and exhibits incoherence as indicated by $B$ decreasing monotonically from slightly above $0.9$ to beloved $0.8$. As $\theta$ increases and approaches the quasiperiodic regime close to $\theta \approx 0$, the system behaves in a more persistent manner with ${\rm SE}$ lowering down to $\approx 0.0144$. We see a sharp decrease in $K$ close to $\approx 0.32$, supporting a quasiperiodic behavior. We see a sharp jump of $\Gamma$ to $1$ indicating the nodes becoming totally synchronous to each other and a sharp increase in their coherence as $B$ approaches $1$ as well. As soon as $\theta>0$, $H$ stabilizes to $\approx 0.88>0.5$, showing a highly persistent behavior, and ${\rm SE}$ stabilizing to $\approx 0.0144$. Also, $K$ stabilizes to $\approx 0.159$ showing a regular behavior. Furthermore, both nodes show a total in-phase synchrony ($\Gamma = 1$, $B \approx 1$). A comprehensive analysis of a gap junction coupled dML system is provided recently by Ghosh {\em et al.}~\citep{GhFa25}.

\begin{figure*}[h]
\begin{tabular}{ccc}
  \includegraphics[width=0.3\linewidth]{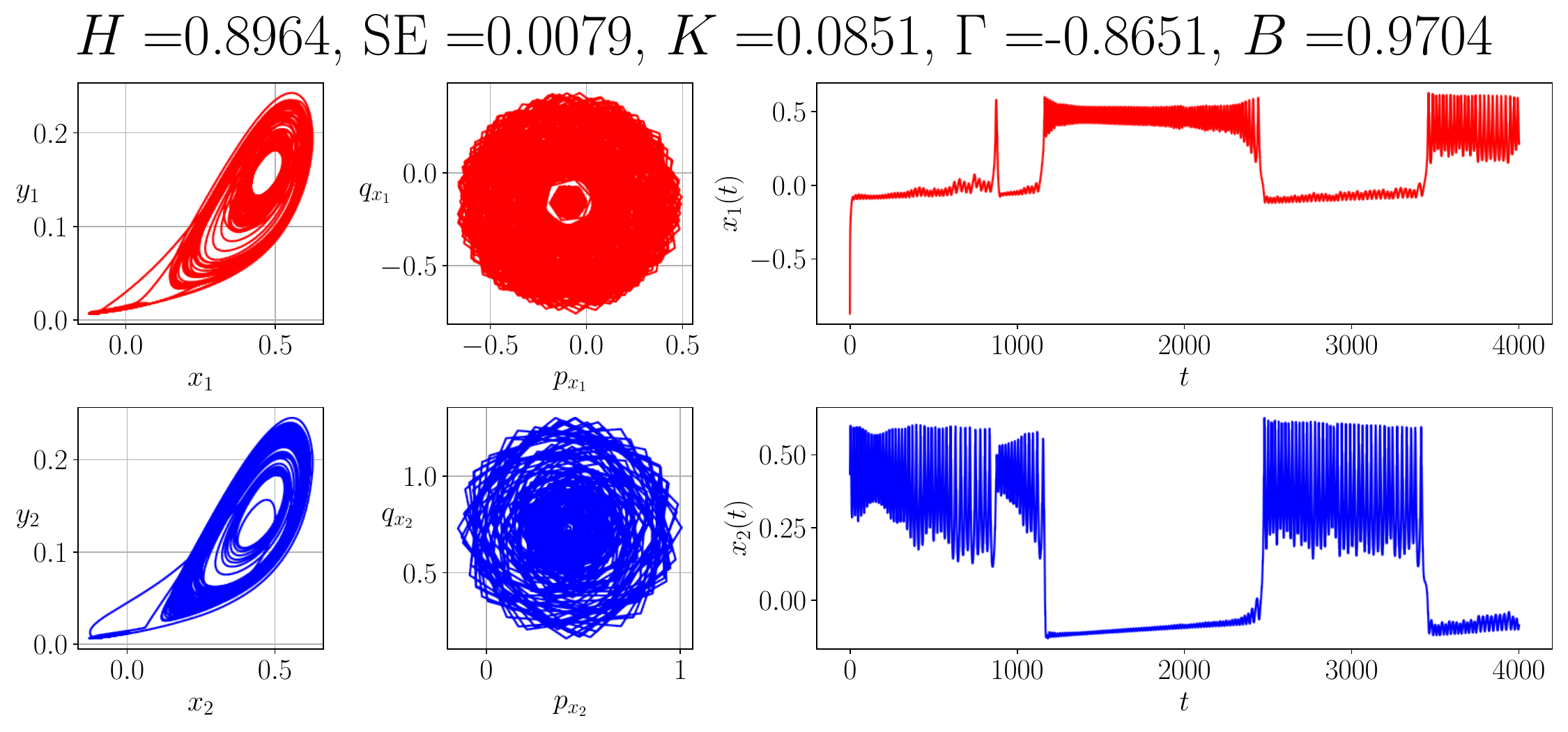} &  \includegraphics[width=0.3\linewidth]{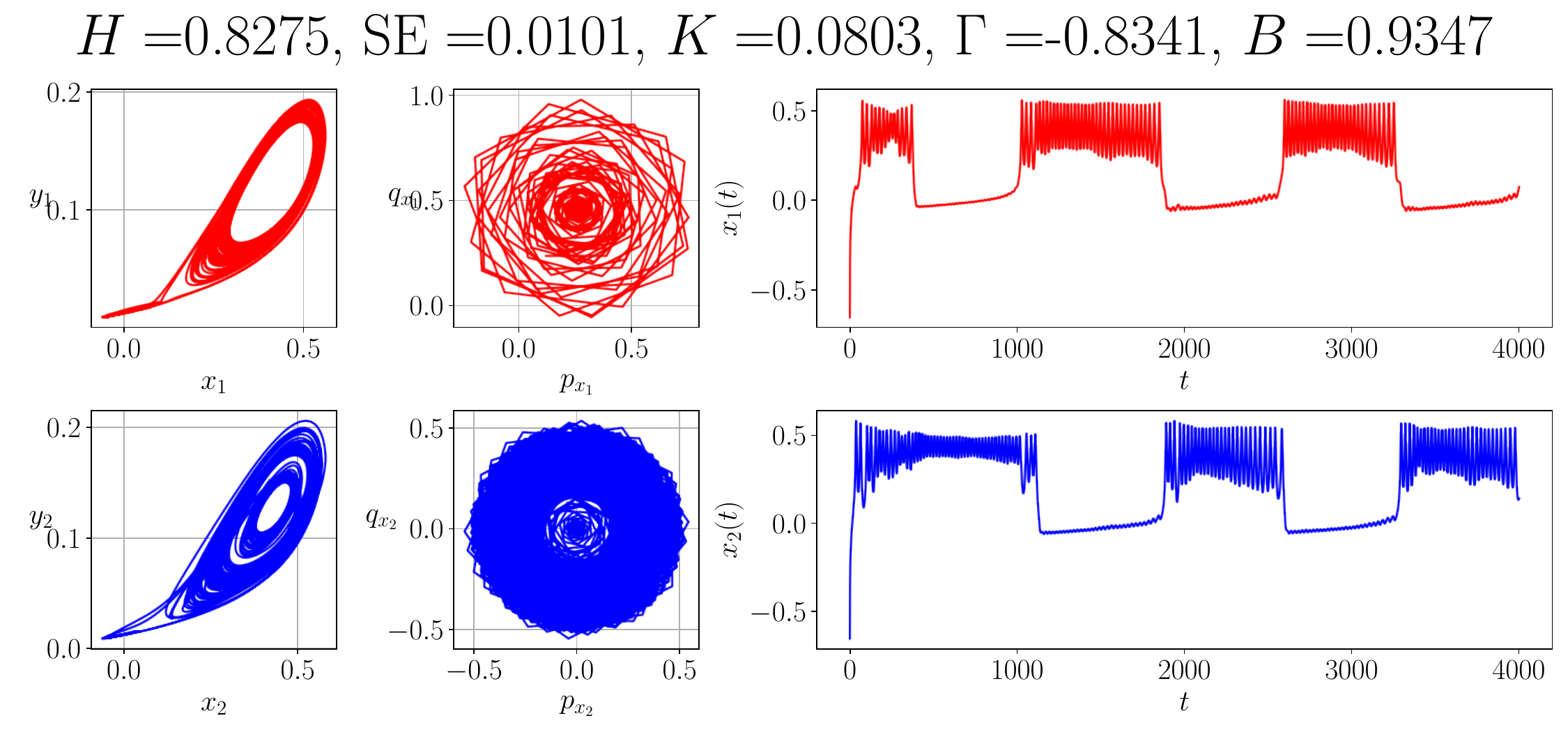} &  \includegraphics[width=0.3\linewidth]{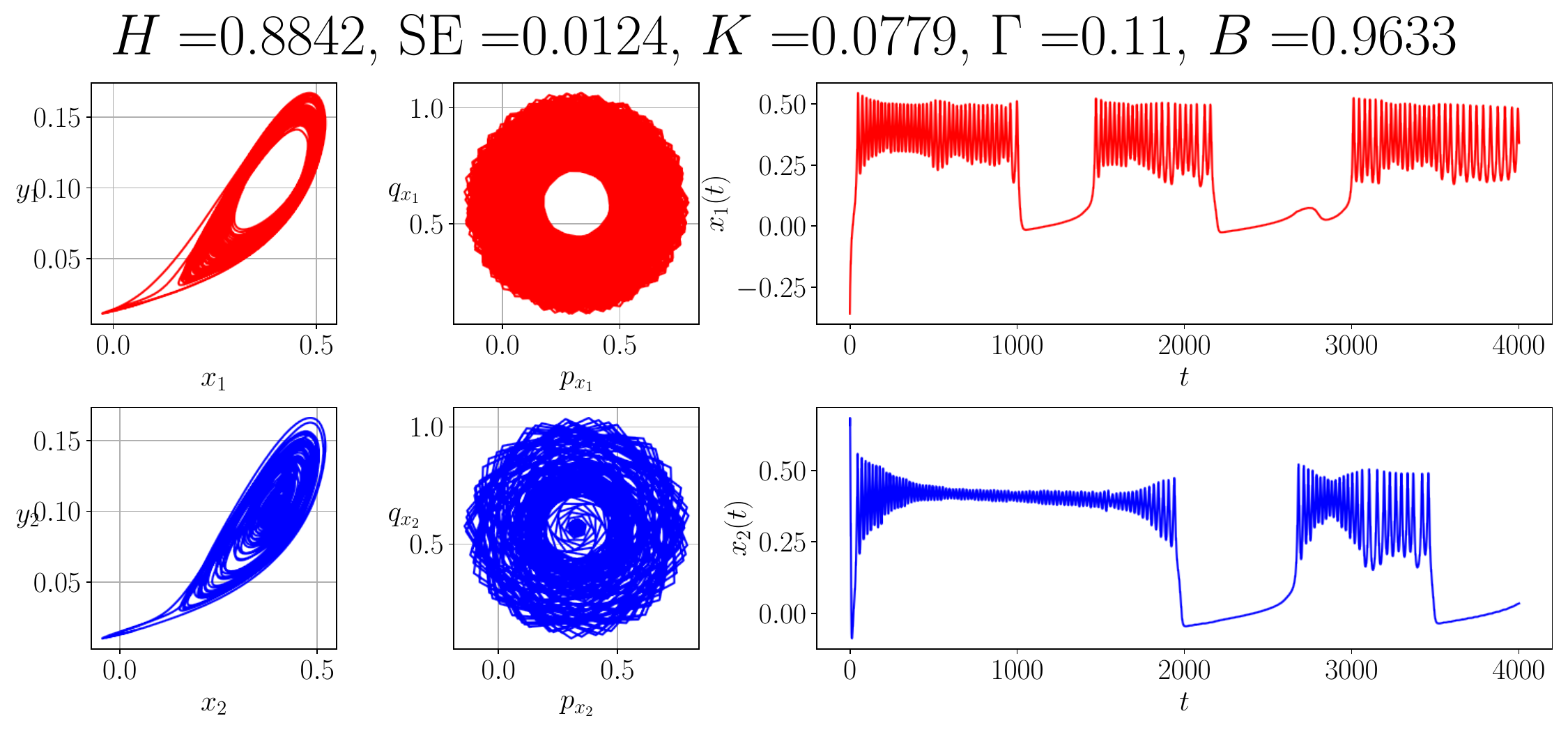} \\
(a) $\theta = -5$ & (b) $\theta = -2$ & (c) $\theta = -0.1$ \\
\includegraphics[width=0.3\linewidth]{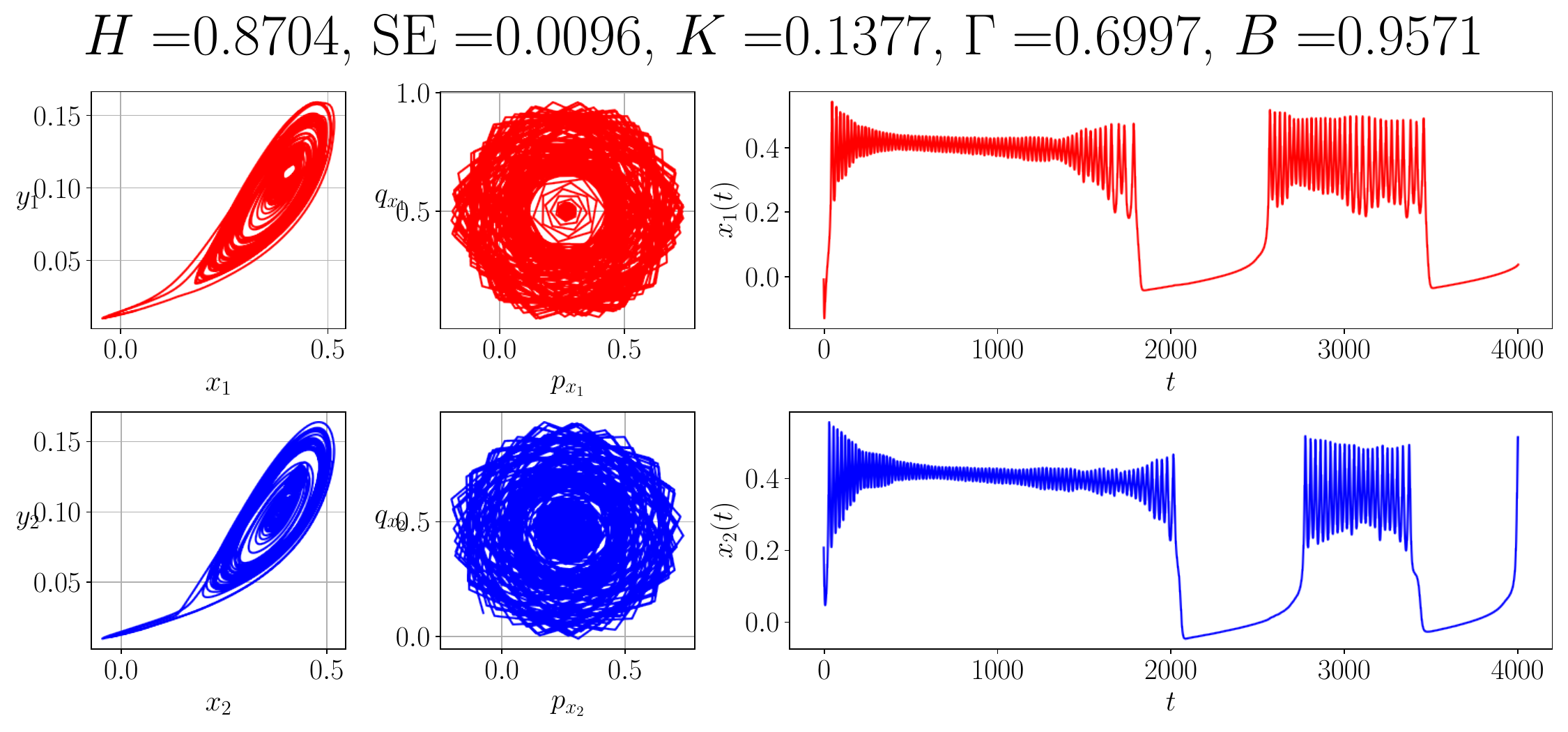} &  \includegraphics[width=0.3\linewidth]{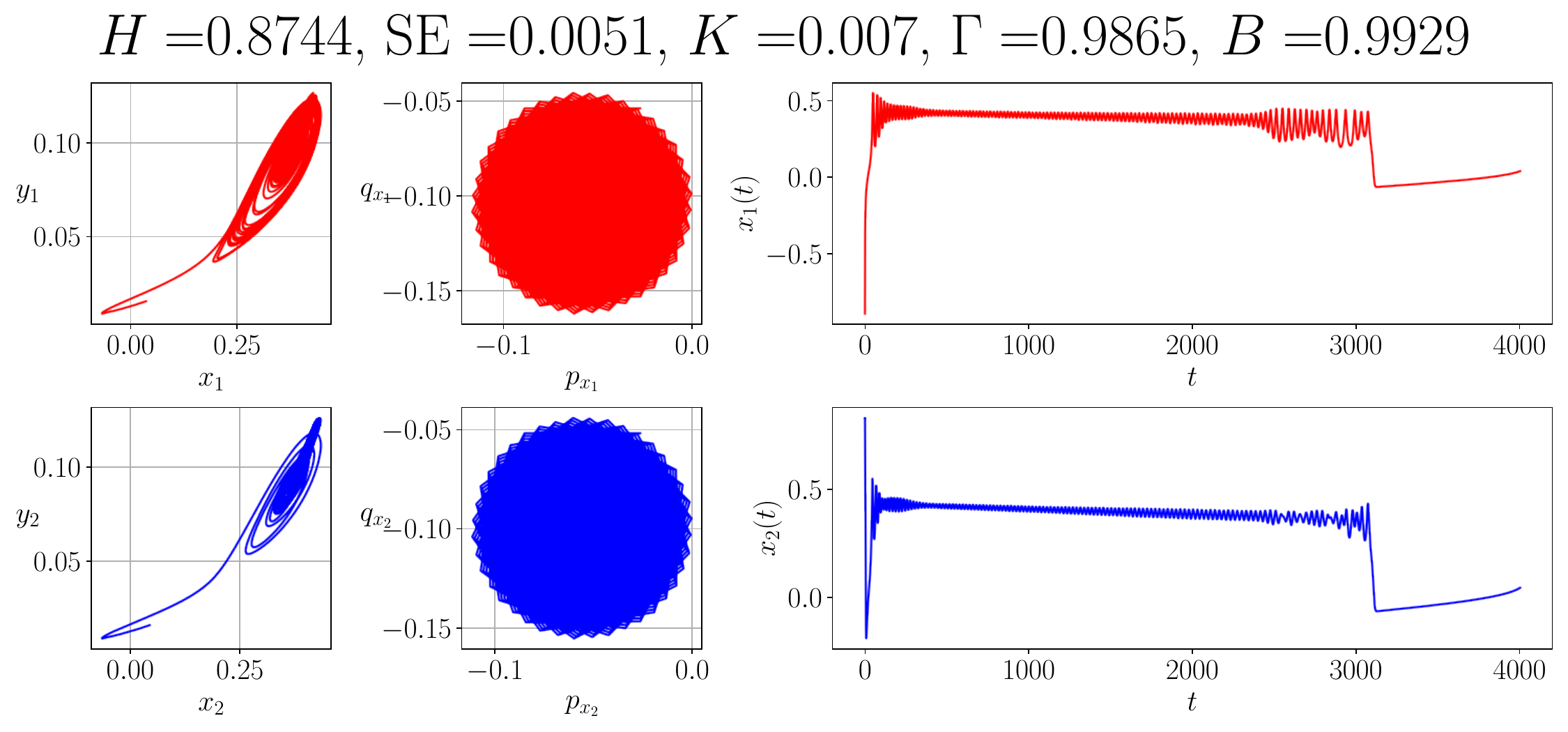} &  \includegraphics[width=0.3\linewidth]{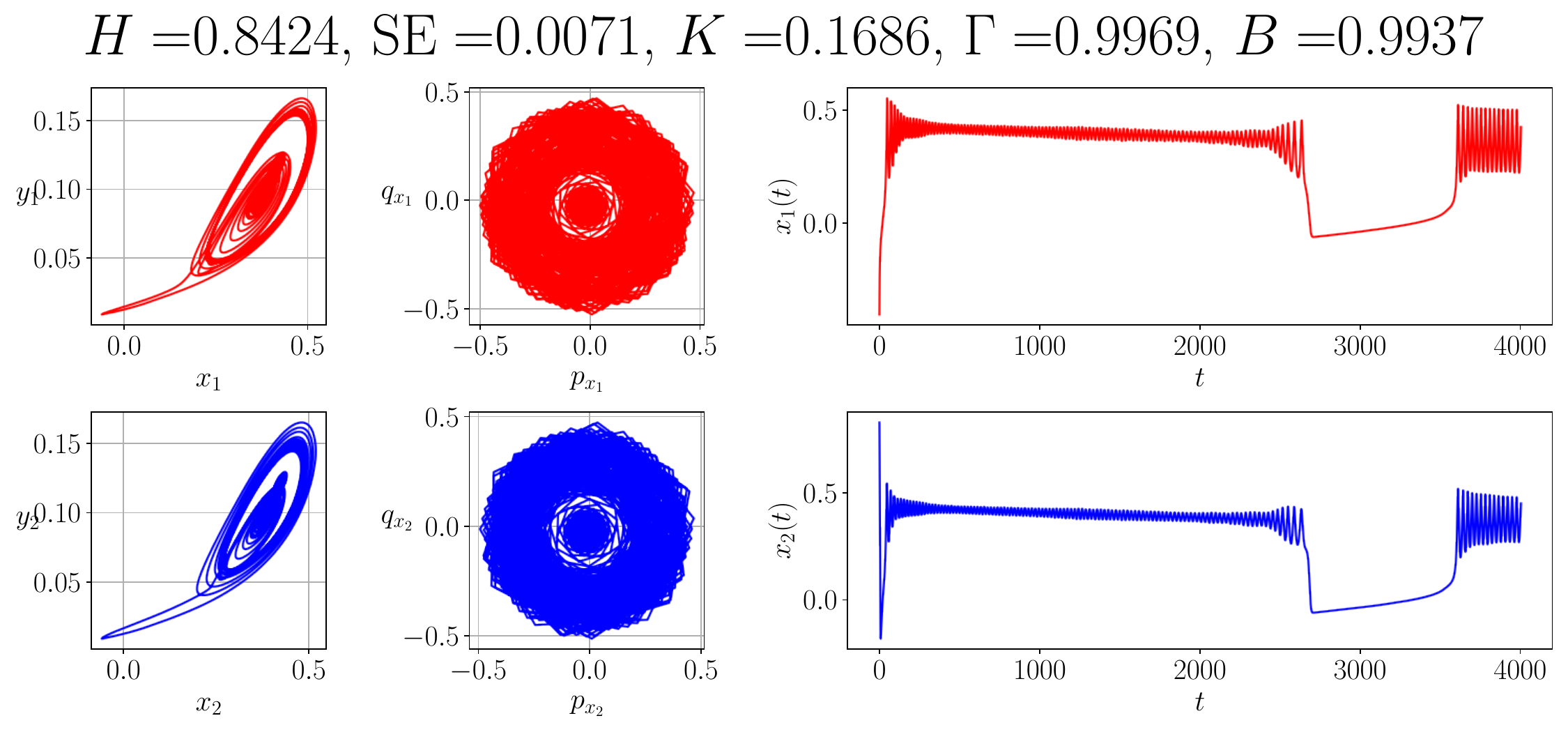} \\
(d) $\theta = 0.1$ & (e) $\theta = 2$ & (f) $\theta = 5$\\
\includegraphics[width=0.3\linewidth]{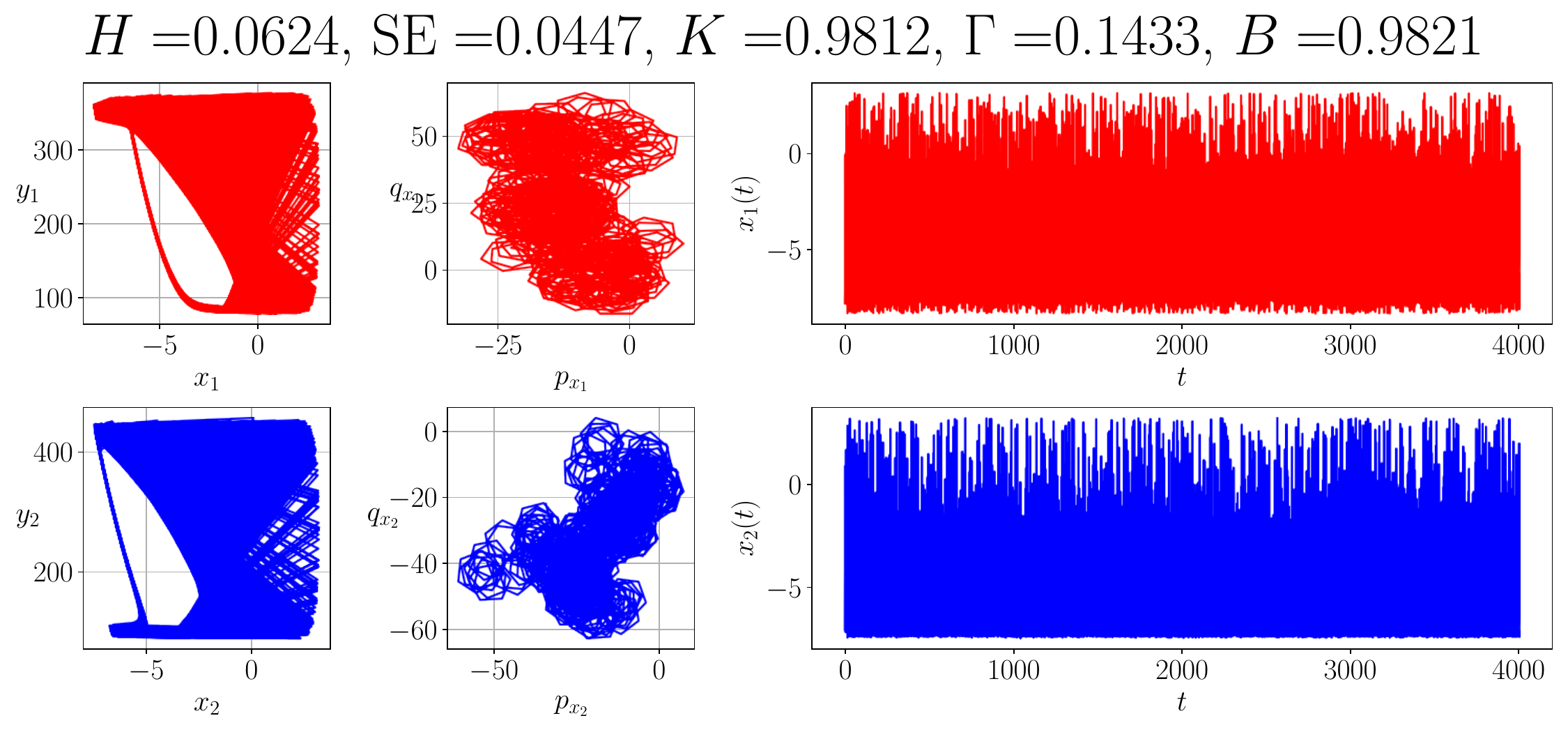} &  \includegraphics[width=0.3\linewidth]{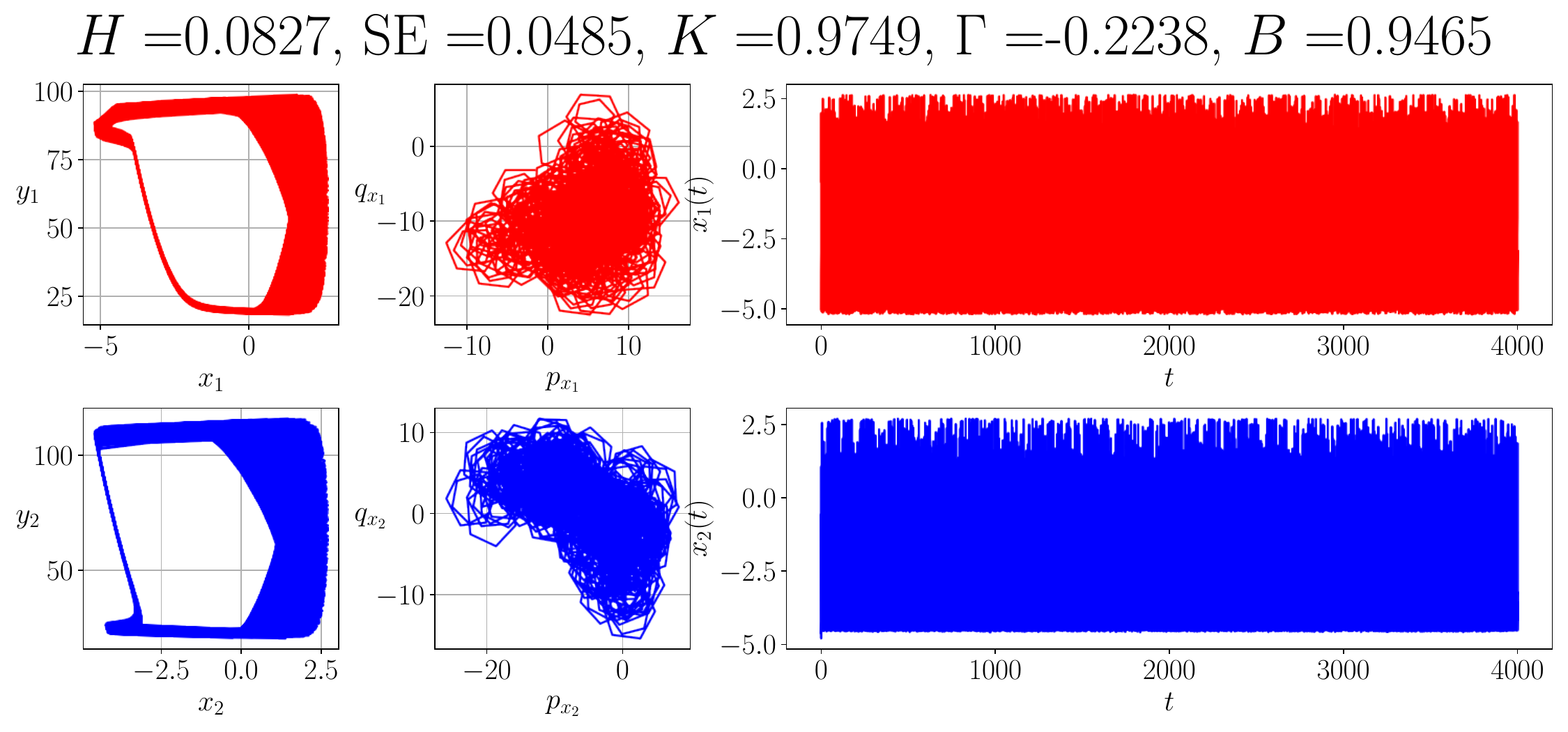} &  \includegraphics[width=0.3\linewidth]{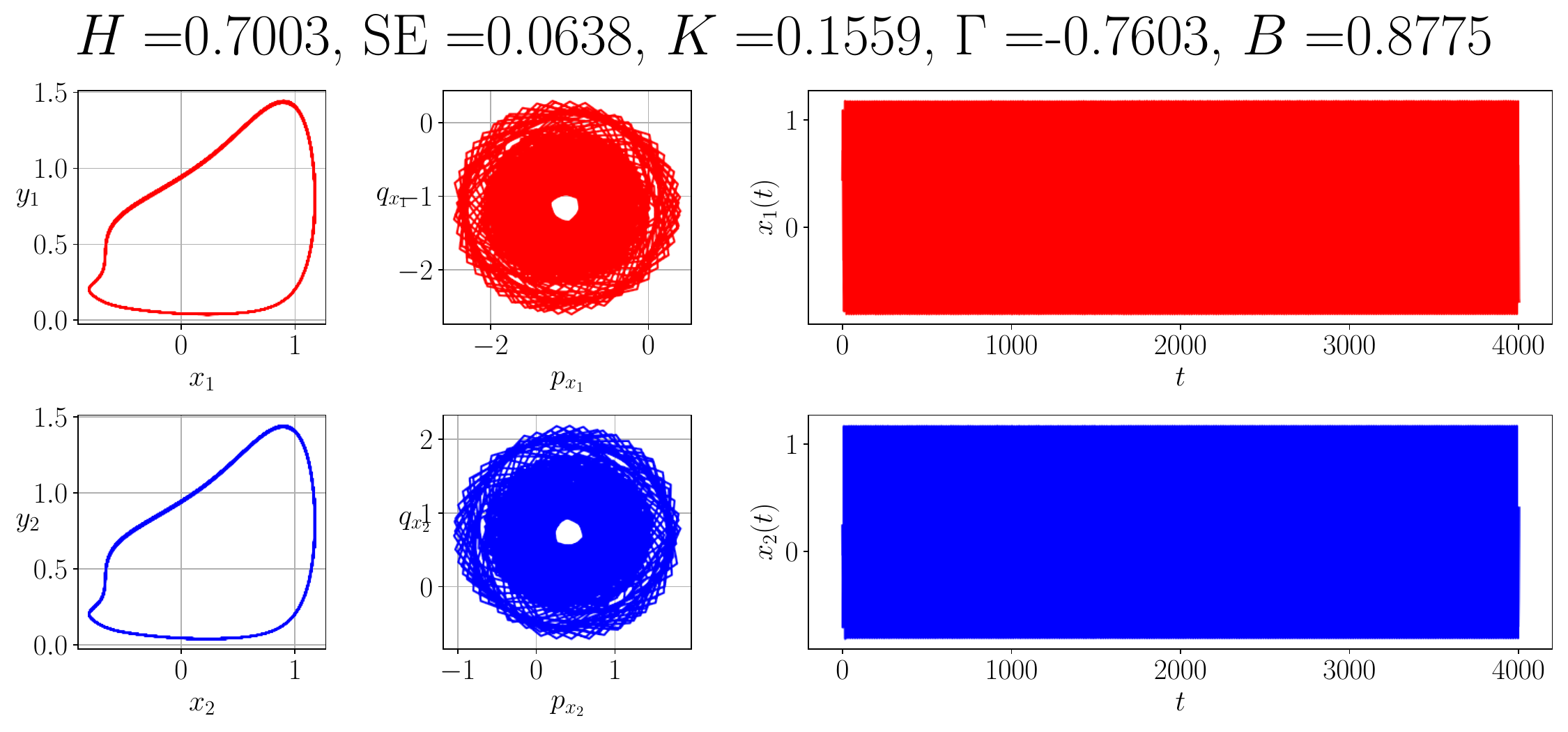} \\
(g) $\theta = -5$ & (h) $\theta = -2$ & (i) $\theta = -0.1$ \\
\includegraphics[width=0.3\linewidth]{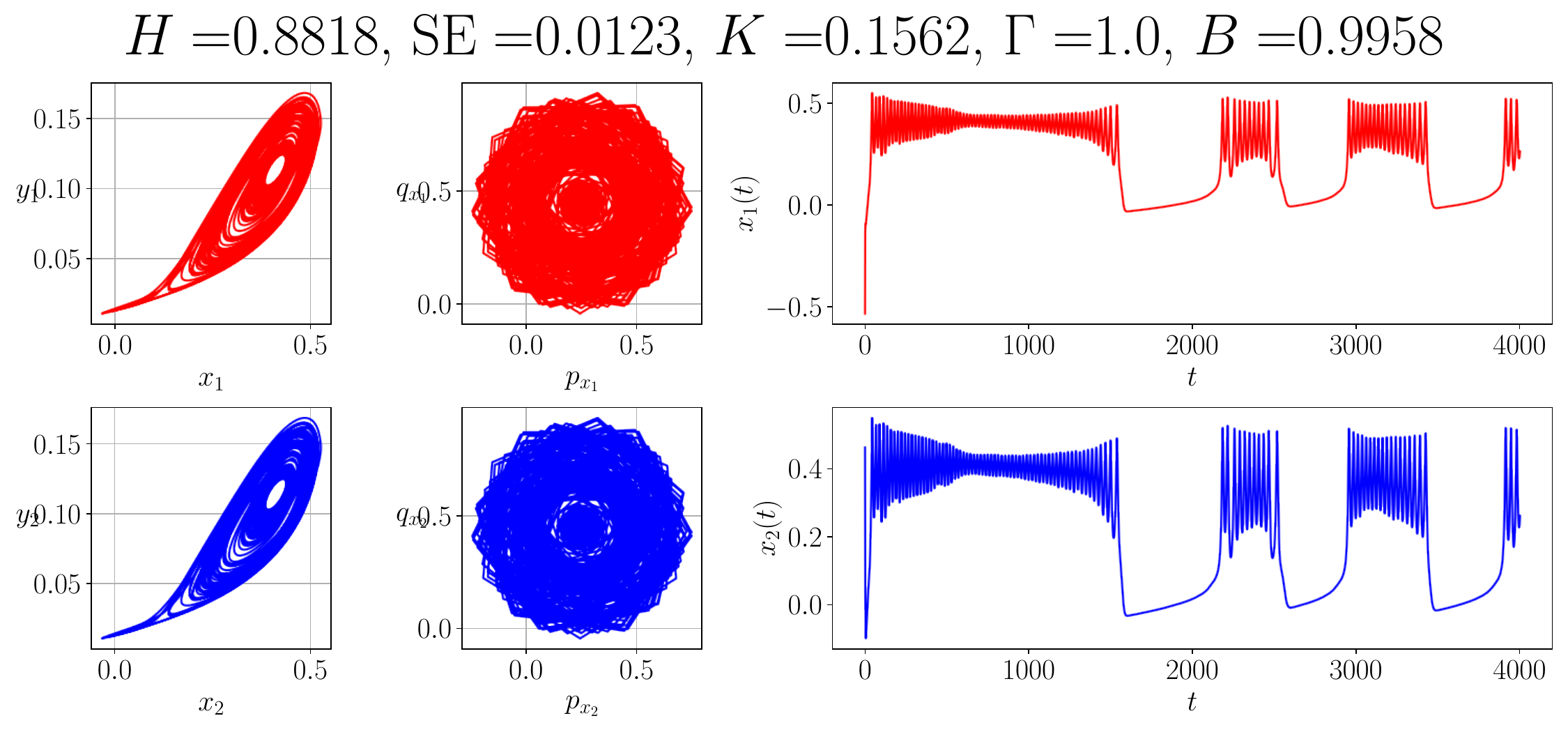} &  \includegraphics[width=0.3\linewidth]{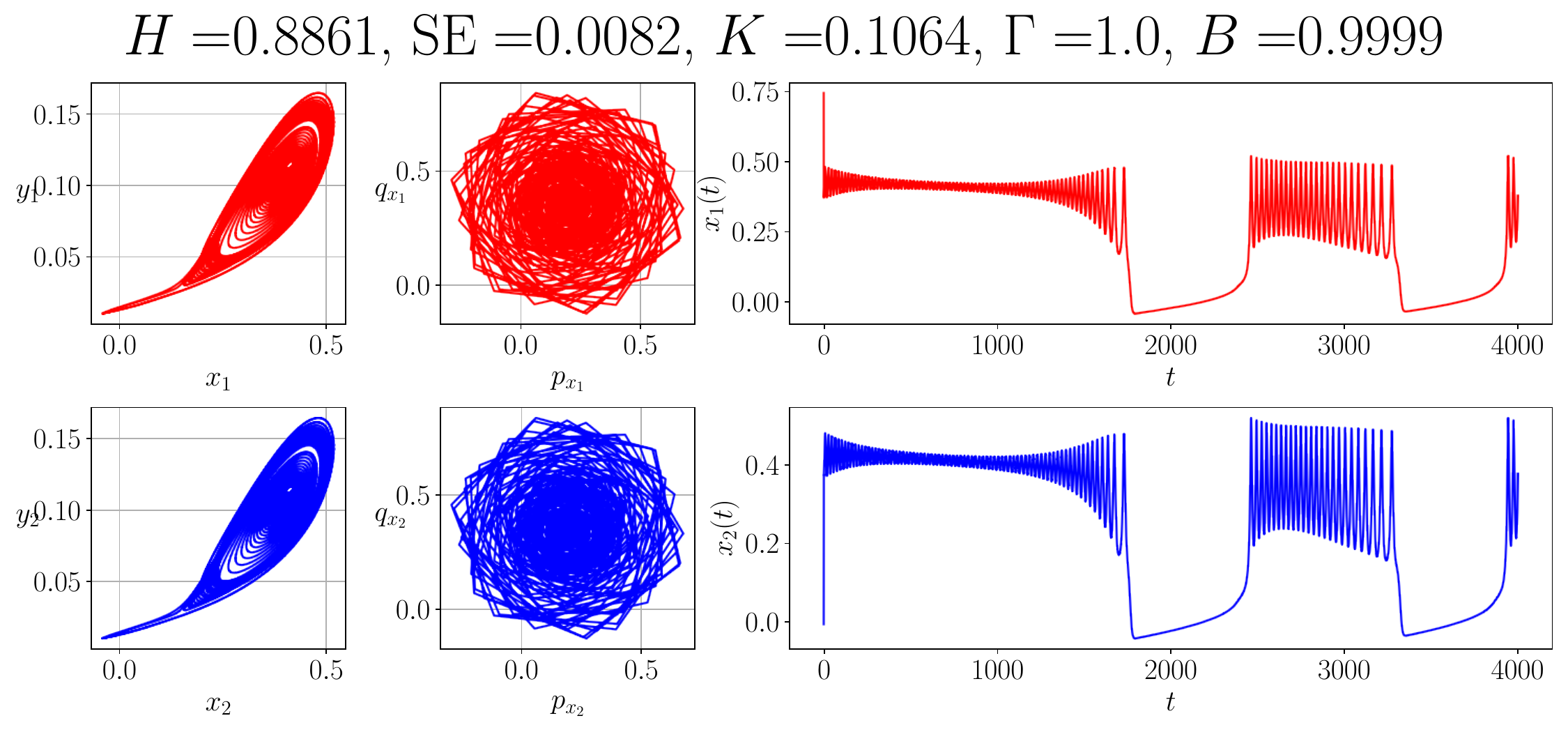} &  \includegraphics[width=0.3\linewidth]{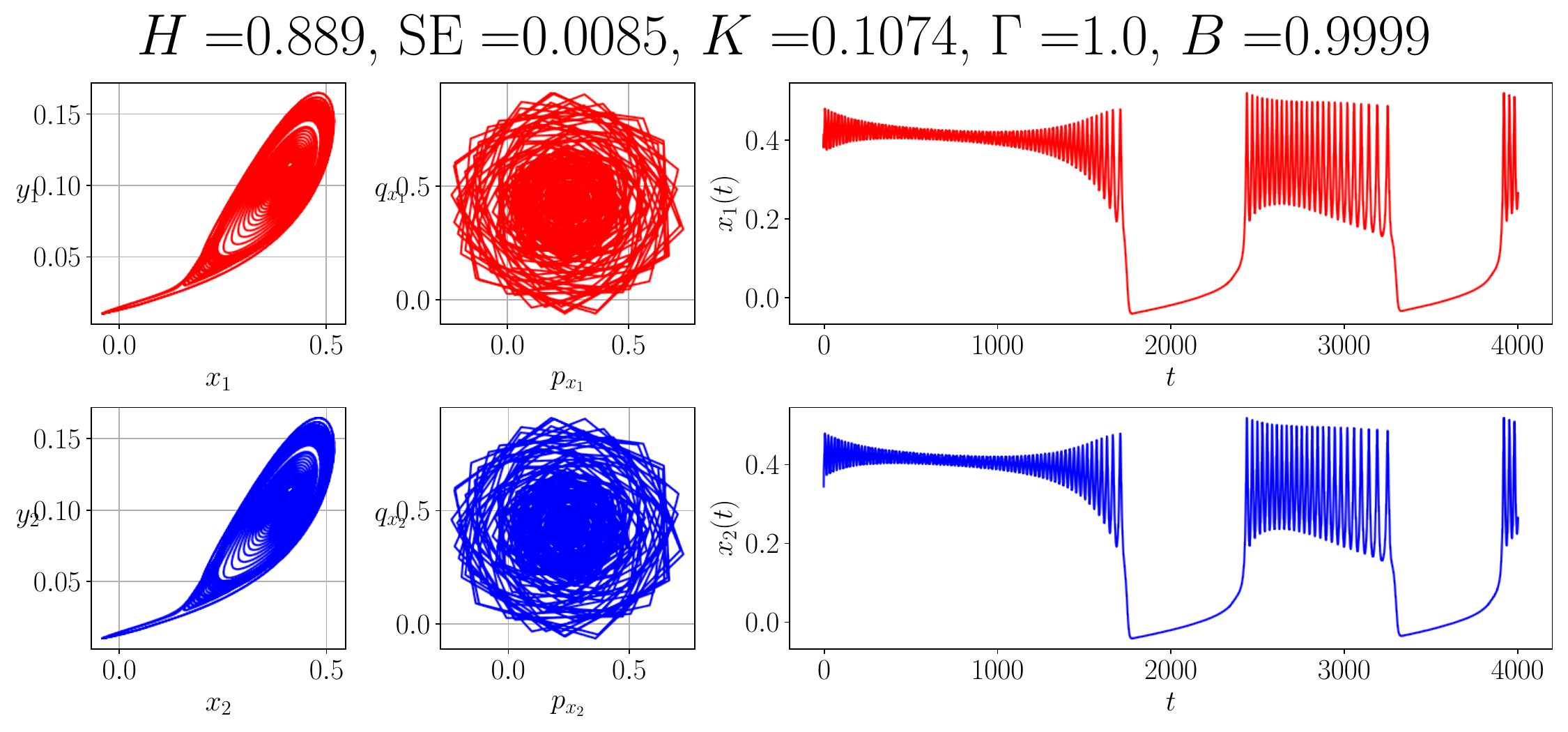} \\
(j) $\theta = 0.1$ & (k) $\theta = 2$ & (l) $\theta = 5$\\
\end{tabular}
\caption{Phase portraits, $p$ vs $q$ plots, and time series of~\eqref{eq:temperature} with varying $\theta \in [-5, 5]$ for two different temperature values: $T = 10^{\circ}$ for panels (a)--(f) and $T = 35^{\circ}$ for panels (g)--(l). Other parameters are set in table~\ref{tab:params}. The initial conditions $x_1(0)$ and $x_2(0)$ are sampled randomly from the continuous uniform distribution over the interval $[-1, 1]$. Other initial conditions are fixed as $y_1(0) = y_2(0)= 0.1$, $I_1(0) = 0.019$, and $ I_2(0)= 0.022$. For $T = 10^{\circ}$, the model exhibits anti-phase oscillations between the nodes for inhibitory coupling, which slowly aligns to show in-phase synchronized oscillations as $\theta$ slowly becomes positive. Throughout the coupling strength regime, both nodes exhibit bursting. For $T = 35^{\circ}$, the behavior is similar to that exhibited in Fig.~\ref{fig:Gap_pp}. The corresponding two-dimensional parameter sweep plots for different metrics are shown in Fig.~\ref{fig:Bif2D}.}
\label{fig:Temp_pp}
\end{figure*}

We then move on to the thermally sensitive gap junction model of two coupled dML neurons given by~\eqref{eq:temperature}. In this case we have two bifurcation parameters: the coupling strength $\theta$ which we vary in range $[-5, 5]$ and the temperature $T \in [0, 40]^\circ C$. We show twelve instances of varying $\theta$ and $T$ in Fig.~\ref{fig:Temp_pp}. The initial conditions $x_1(0)$ and $x_2(0)$ are sampled randomly from the continuous uniform distribution over the interval $[-1, 1]$. Other initial conditions are fixed as $y_1(0) = y_2(0)= 0.1$, $I_1(0) = 0.019$, and $ I_2(0)= 0.022$. The first six panels have $T = 10^\circ C < T_{\rm ref}$ and the last six panels have $T = 35 ^\circ C > T_{\rm ref}$. We report for $\theta = -5, -2, -0.1, 0.1, 2, 5$. When $T = 10^\circ C$, we do not observe any chaotic patterns. The signature fold/homoclinic bursting behavior is exhibited in both neurons along with a high persistence in the time series ($H \in (0.84, 0.9)$ approximately). The time series for both neurons are very regular as exhibited by the sample entropy not exceeding $0.0124$ and $K$ not exceeding $0.17$ approximately. This is clearly evident from the highly bounded $p$ vs $q$ plots. For a highly inhibitory $\theta \in [-5, -2]$, we see anti-phase oscillation in the nodes. Value of $\Gamma \approx -0.9$ confirms an anti-phase oscillatory behavior, though not fully synchronized. As $\theta$ becomes more positive the nodes become more asynchronous with $\Gamma$ reaching $0.7$ approximately for $\theta = 0.1$. As $\theta$ becomes stronger approaching $\theta = 0.5$, the nodes oscillate in phase and show total synchronization characterized by $\Gamma \approx 1$. The value of $B$ becomes very close to $1$ as well when $\Gamma \approx 1$ confirming a coherent phase locking. As soon as $T=35^\circ C > T_{\rm ref}$, the system behaves similarly to what we observed in Fig.~\ref{fig:Gap_pp}.

\begin{figure}[h]
\begin{tabular}{cc}
  \includegraphics[scale=0.18]{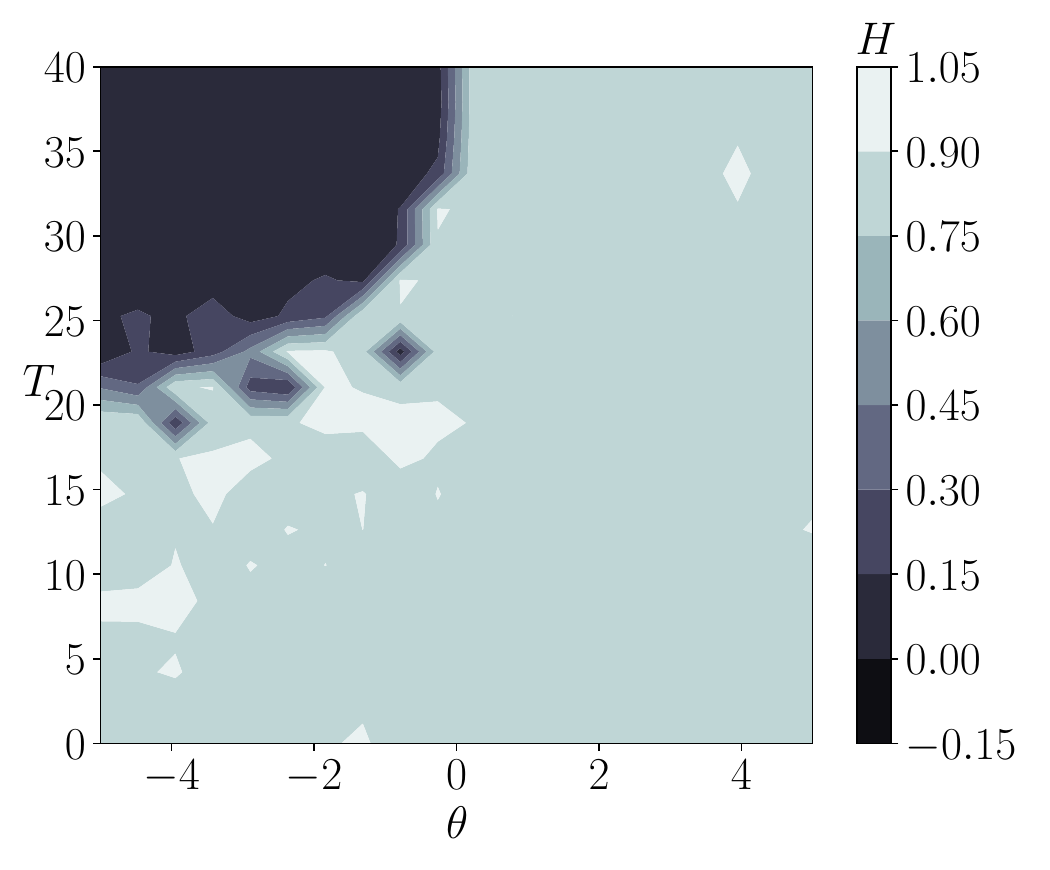} &  \includegraphics[scale=0.18]{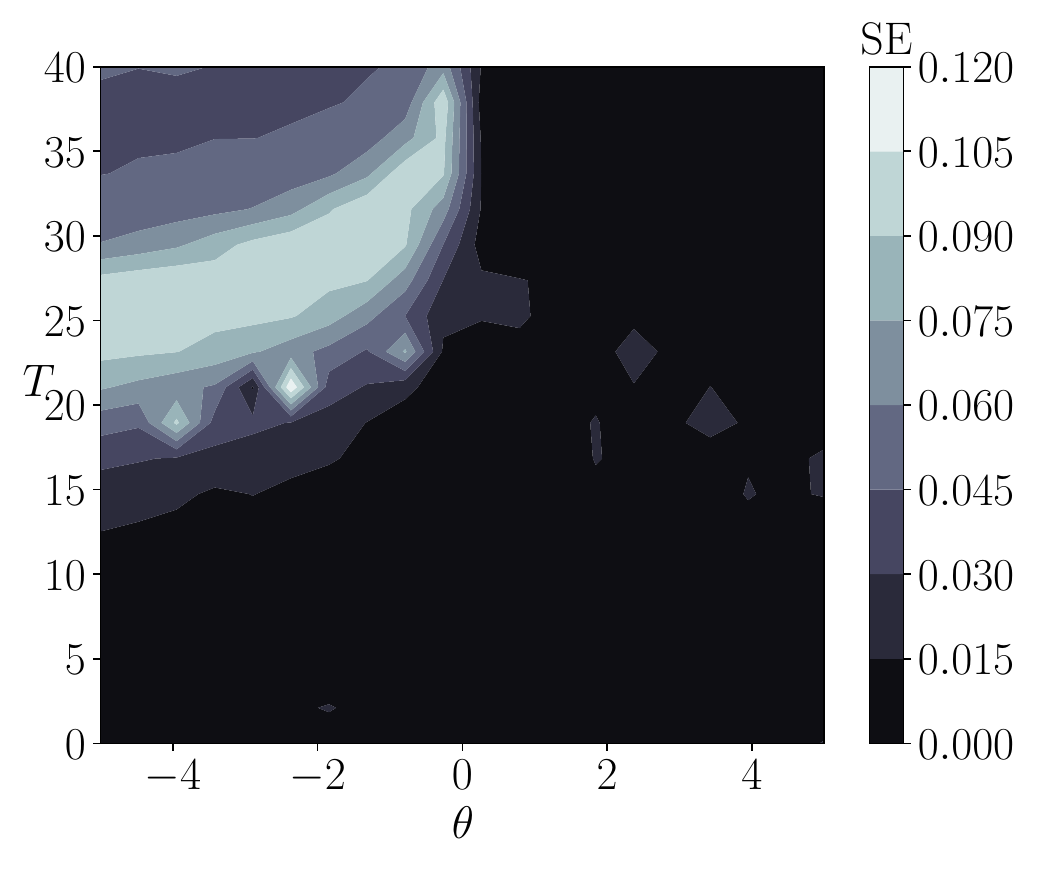}\\
  \includegraphics[scale=0.18]{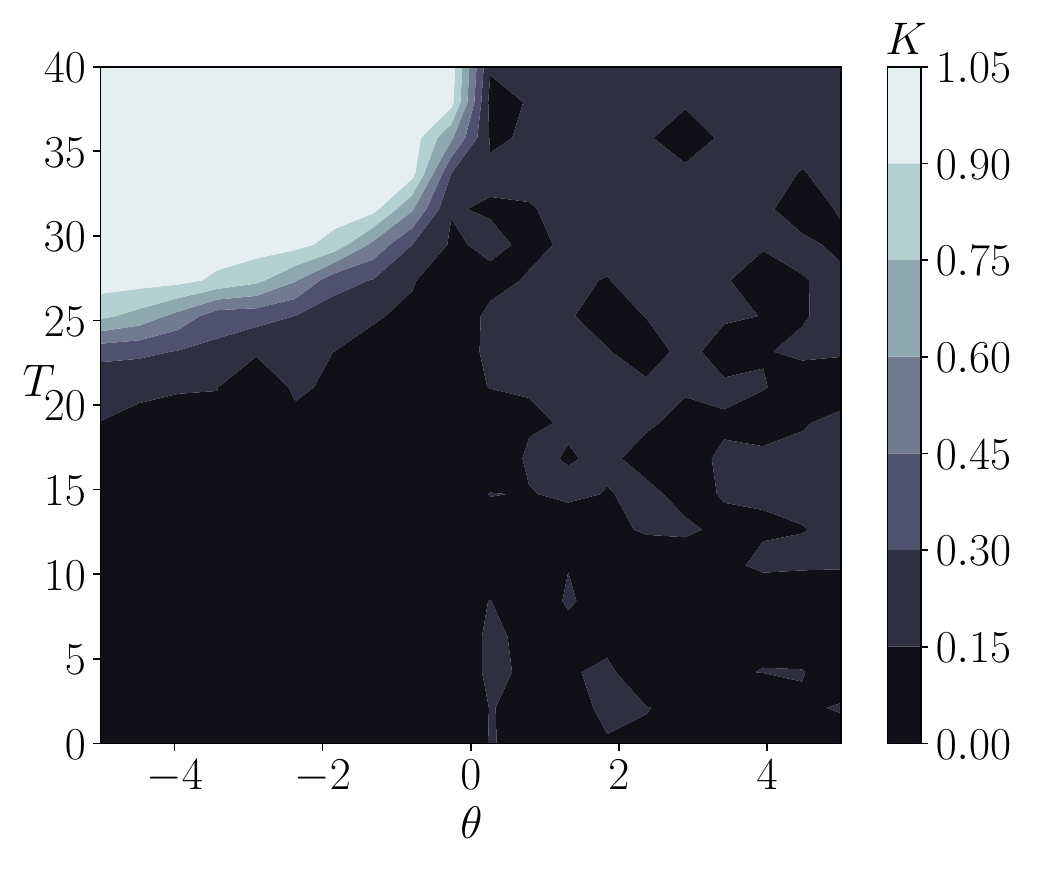} &  \includegraphics[scale=0.18]{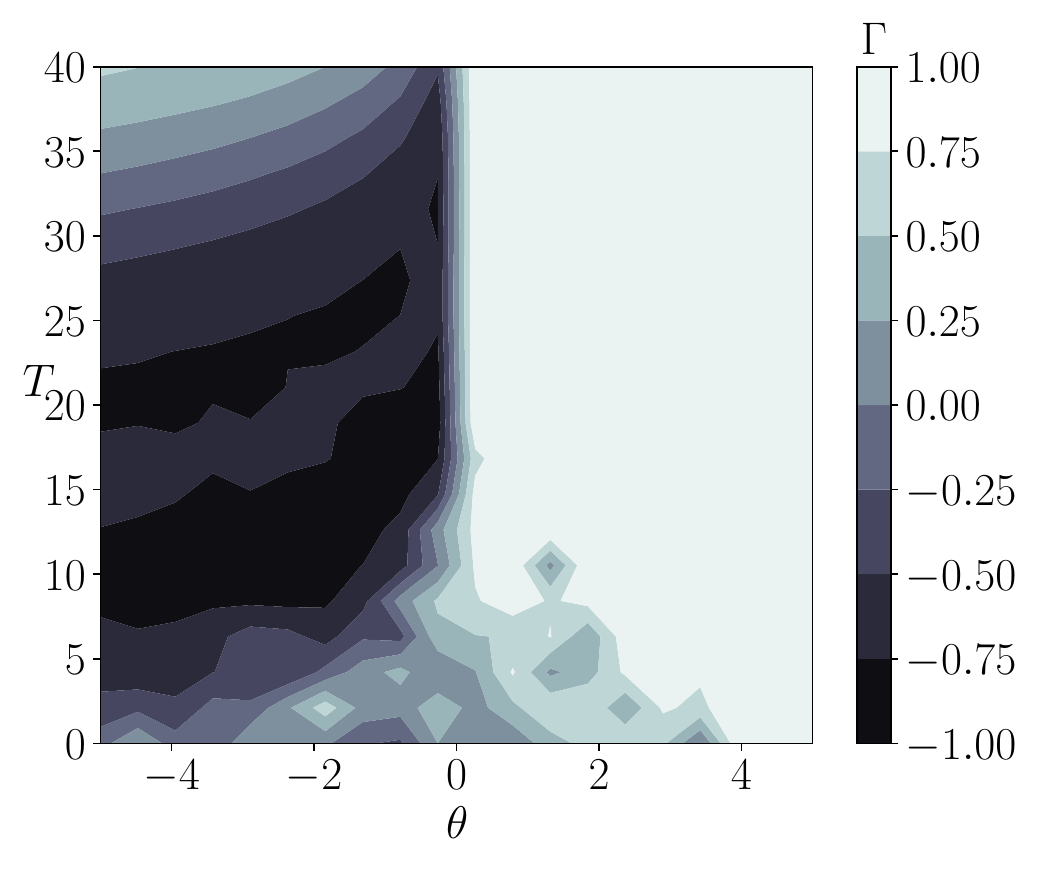}\\
  \includegraphics[scale=0.18]{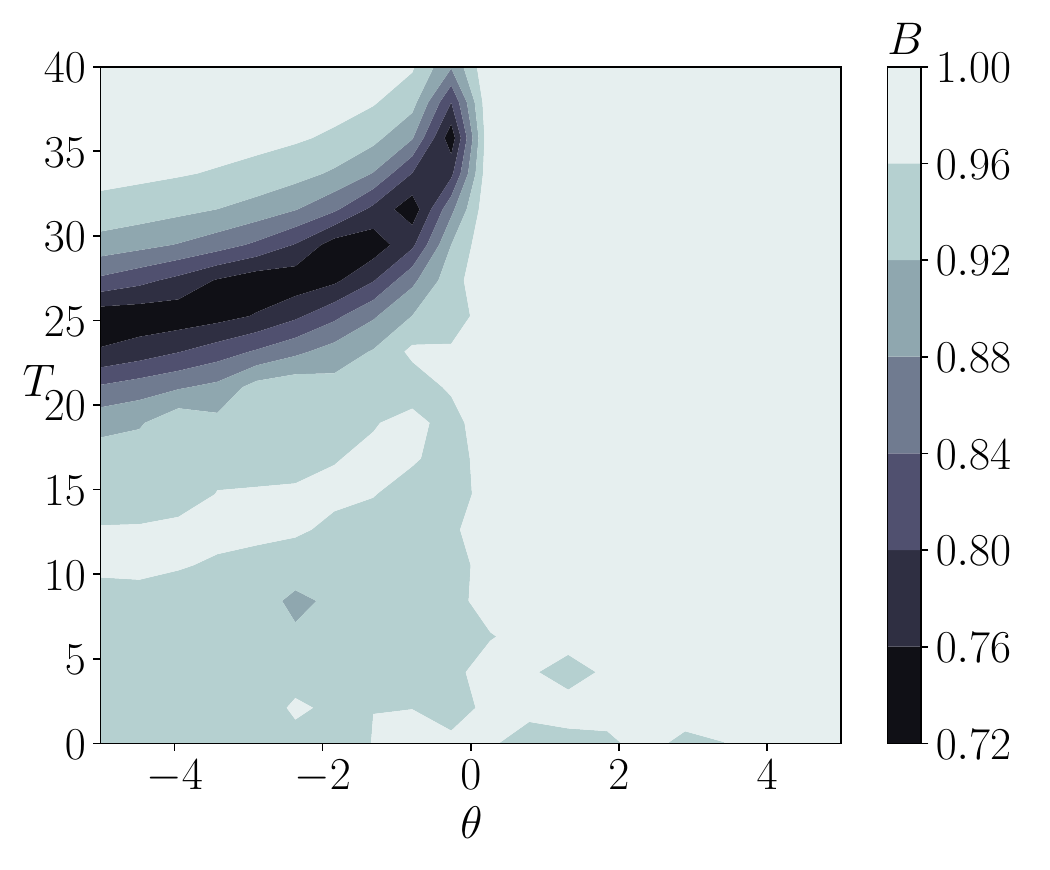}
\end{tabular}
\caption{Two-dimensional bifurcation plots of different metrics performing a parameter sweep on $\theta \in [-5, 5]$ and $T \in [0, 40]^\circ C$ for model~\eqref{eq:gapJunction}. The value of the $0-1$ test $K$ has been preprocessed for $\theta \approx -2.9$ and $T \approx 21.1 ^\circ$ to $0$ because it had a spurious negative value of $\approx -0.48$.}
\label{fig:Bif2D}
\end{figure}

Fig.~\ref{fig:Bif2D} is a collection of two-parameter bifurcation plots of the metrics, where we vary $\theta$ and $T$ along the axes. This is done on a $20 \times 20$ grid for a total of $400$ $(\theta, T)$ pairs. We see a clear bifurcation boundary for the Hurst exponent (panel (a)) when $\theta<0$ and $T > 20^\circ C = T_{\rm ref}$. In this region both time series behave in an anti-persistent manner exhibiting irregularity, i.e, $H \ll 0.5$. There might be some spurious values like $H = -0.004116$ which can be the result of the algorithm underperforming and can be safely ignored. Otherwise $H\approx 0.9 \gg 0.5$ which represents a persistent behavior. This is also attested by the ${\rm SE}$ values where we see a high complexity in the region $\theta <0$ and $T > T_{\rm ref}$ with ${\rm SE}$ reaching $\approx 0.118$. In the rest of the region the time series of the system is highly regular. When it comes to chaos, we see indeed in the region where the time-series is anti-persistent ($H <0.5$) and highly complex (${\rm SE}$ very high) $K$ is close to $1$ indicating chaos. There are some pixels where the $0-1$ test underperforms and gives erratic values like $K \approx -0.48$ at $(\theta, T) \approx (-2.9, 21.1^\circ C)$. We replace spurious values where $K$ is negative with $0$ and where $K>1$ with $1$. Note that these occurrences are very rare. The two nodes are either asynchronous (in the domain of chaos) or are oscillating anti-phase when $\theta <0$. As soon as $\theta >0$ the two nodes start oscillating in-phase with each other and in full synchrony $(\Gamma \approx 1)$. In the region of chaos and high complexity of the time series, we see $B$ is very low $\approx 0.733$, meaning the nodes are incoherent. As the nodes oscillate in anti-phase outside the chaotic regime, $B$ values increases slightly. Now for $\theta >0$ where the nodes are fully synchronized in phase, $B \approx 1$ indicating total coherence.

\begin{figure*}[h]
\begin{tabular}{ccc}
  \includegraphics[width=0.3\linewidth]{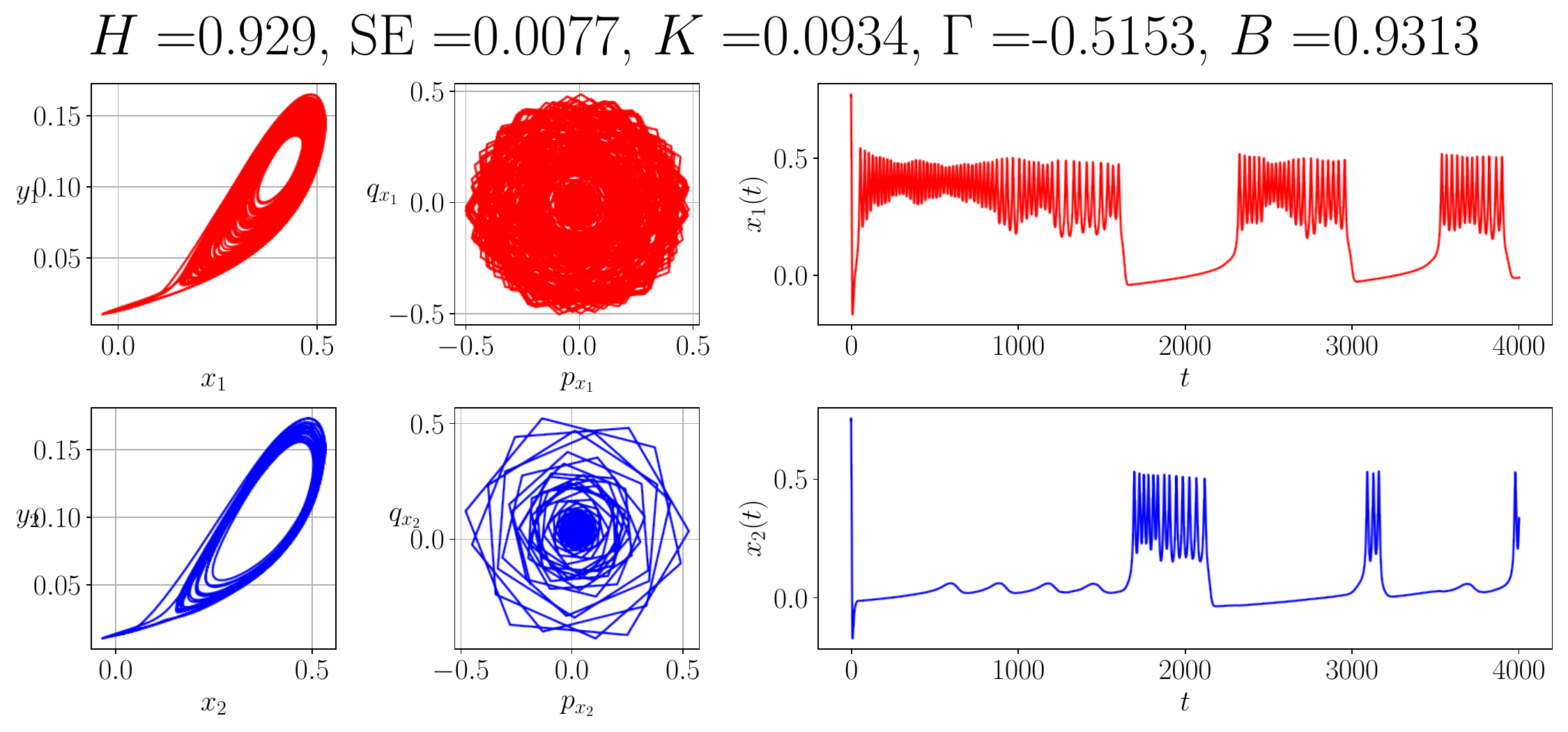} &  \includegraphics[width=0.3\linewidth]{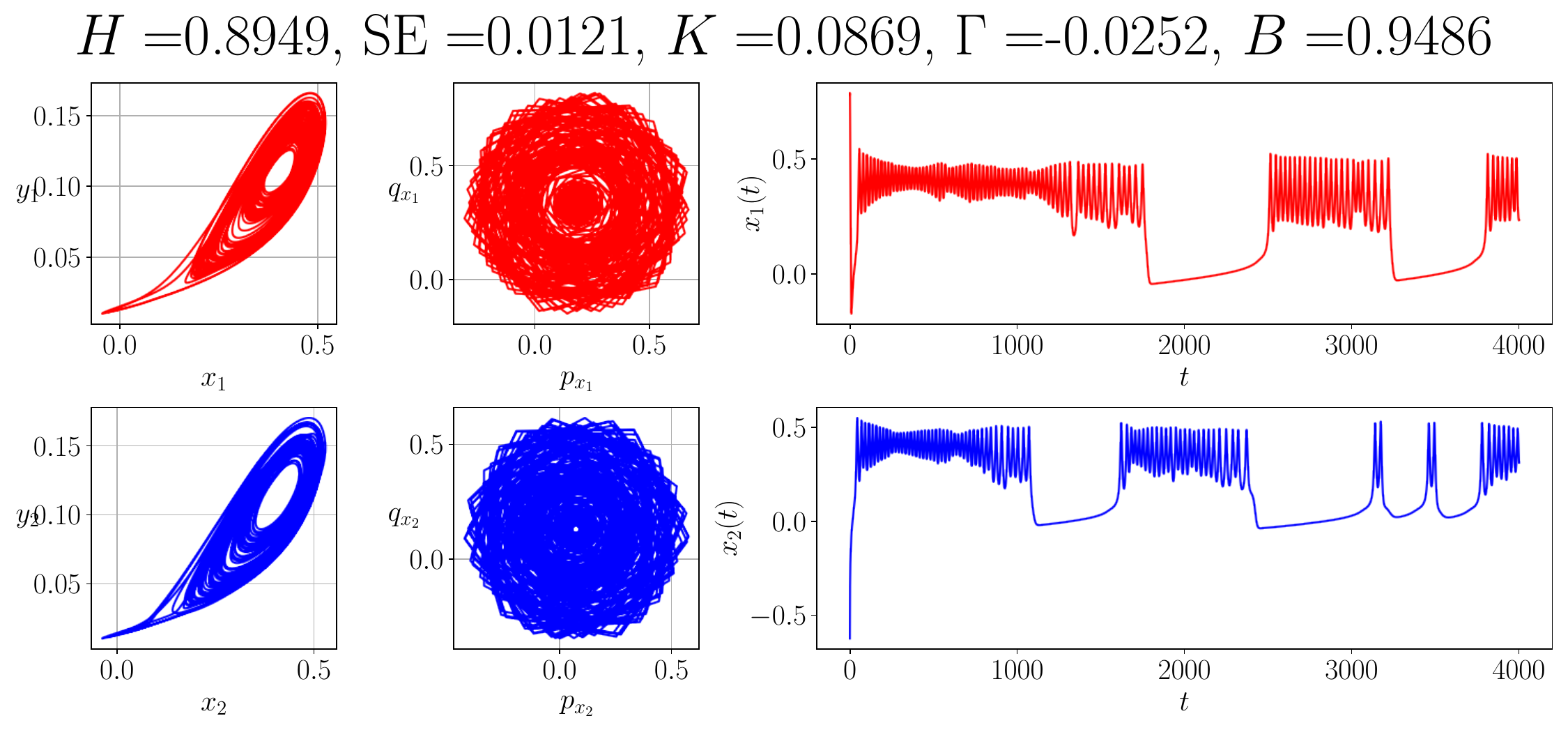} &  \includegraphics[width=0.3\linewidth]{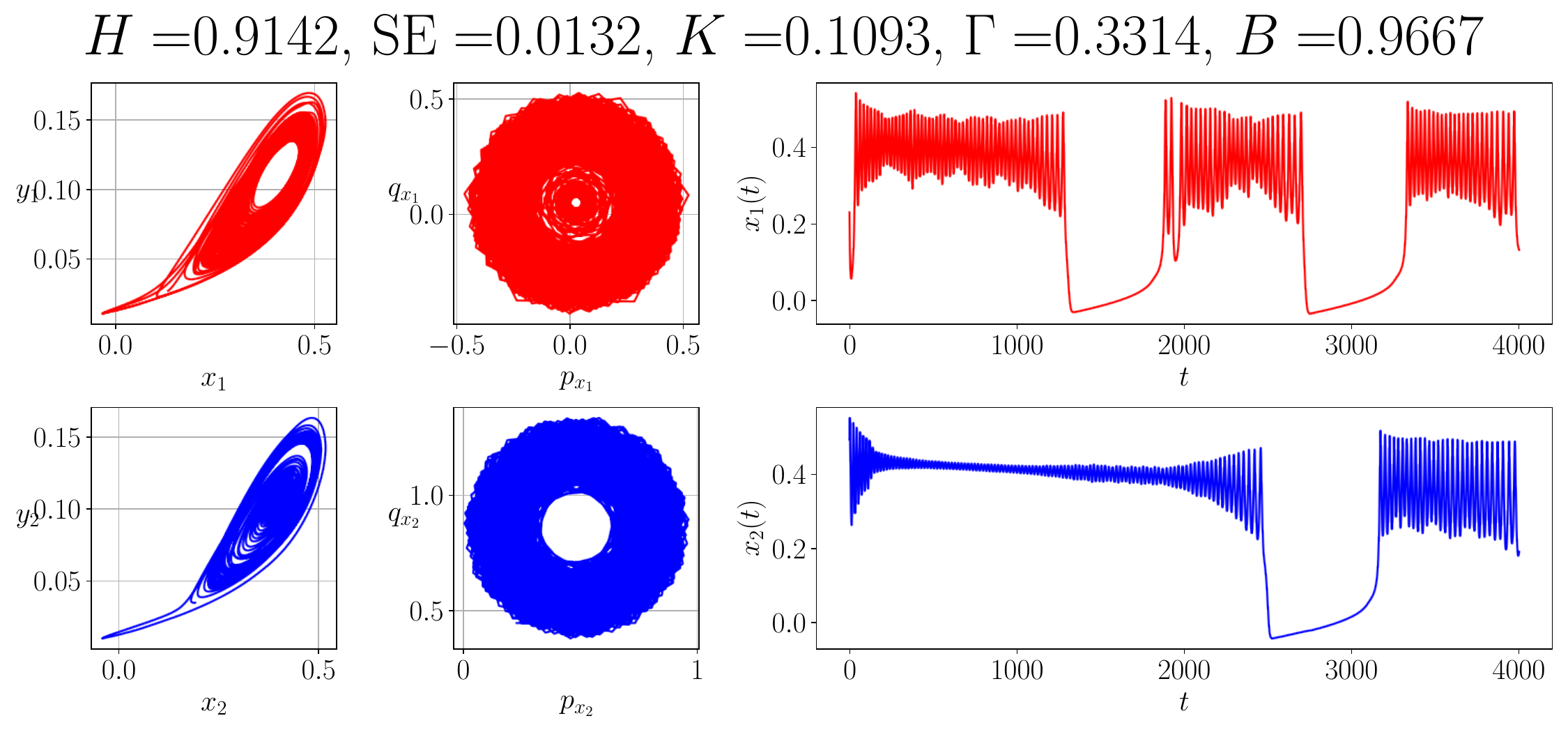} \\
(a) $\theta = -0.005$ & (b) $\theta = -0.001$ & (c) $\theta = 0.001$ \\
\includegraphics[width=0.3\linewidth]{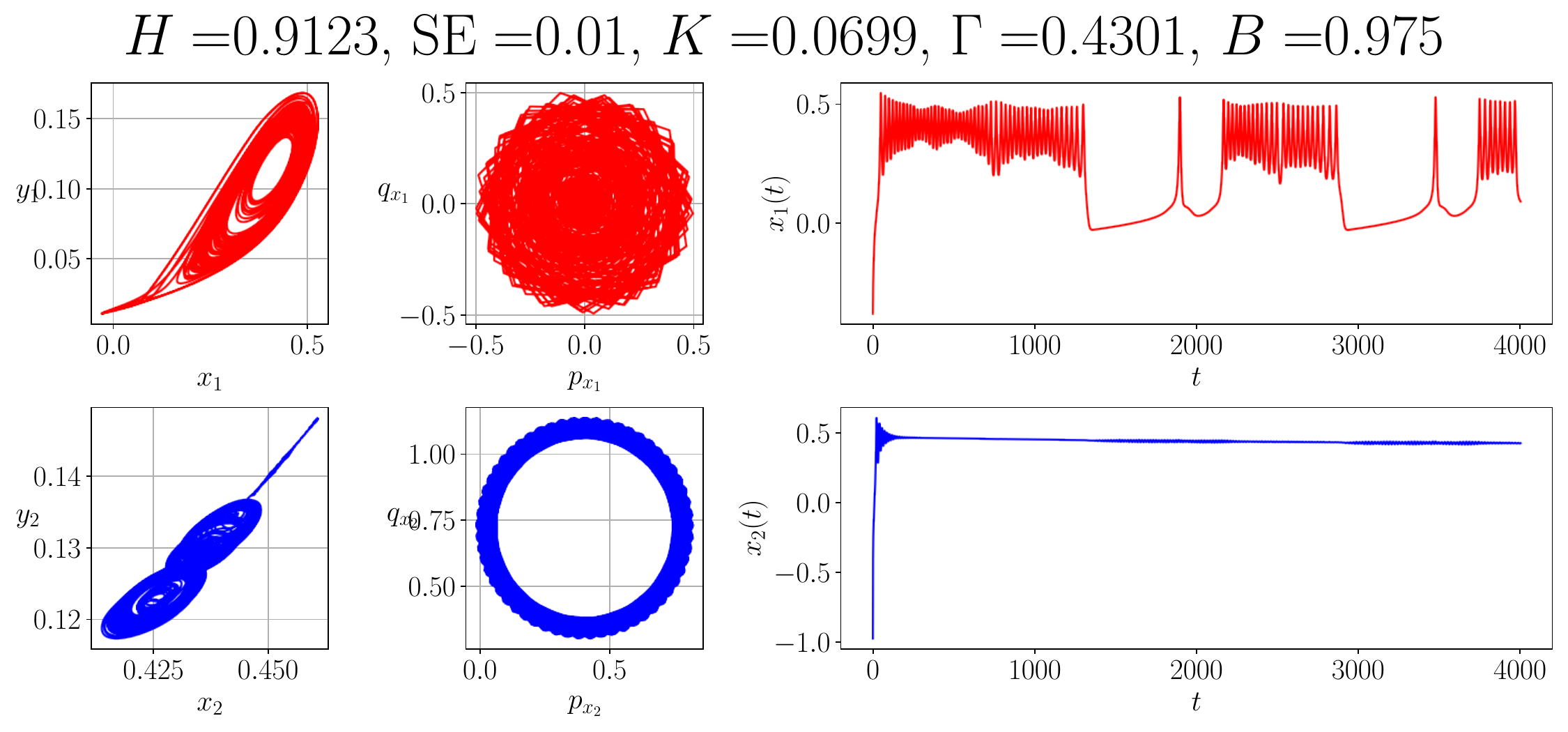} &  \includegraphics[width=0.3\linewidth]{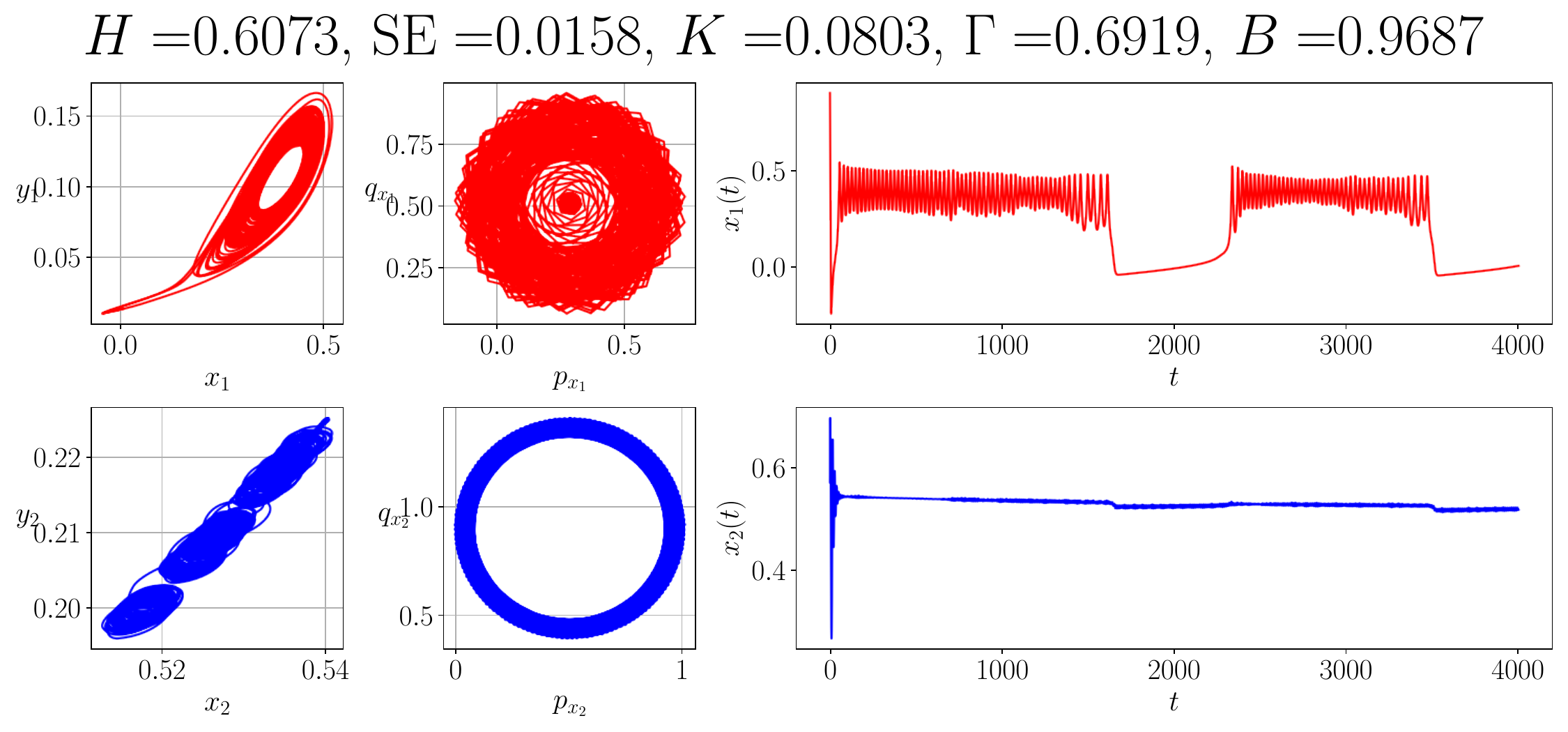} &  \includegraphics[width=0.3\linewidth]{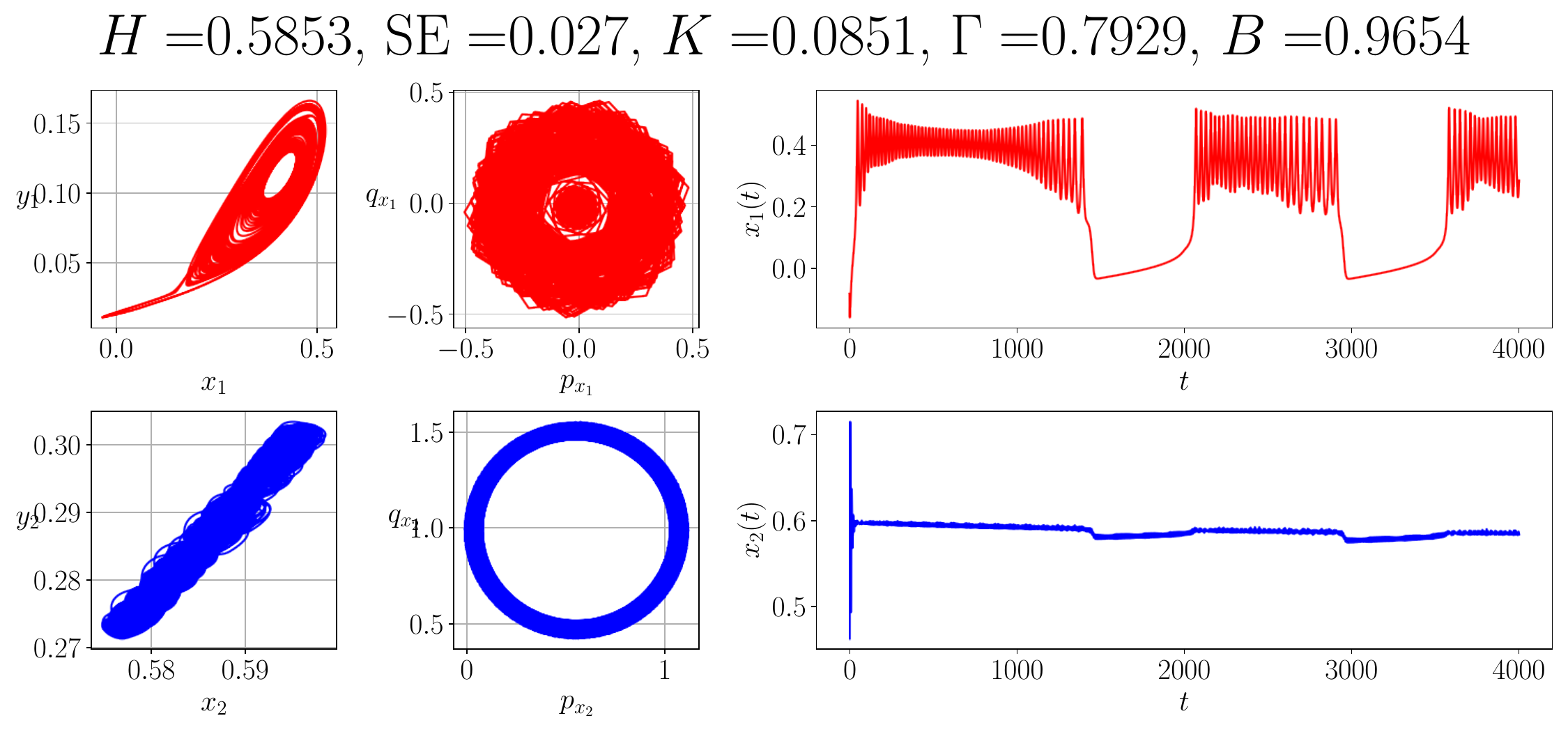} \\
(d) $\theta = 0.01$ & (e) $\theta = 0.05$ & (f) $\theta = 0.1$\\
\end{tabular}
\caption{Phase portraits, $p$ vs $q$ plots, and time series of the non-local chemical coupling~\eqref{eq:dML_chemical} with varying $\theta \in [-0.005, 0.1]$. Other parameters are fixed as in table~\ref{tab:params}, and $x_1(0)$ and $x_2(0)$ are sampled randomly from the continuous uniform distribution over the interval $[-1, 1]$. Other initial conditions are fixed as $y_1(0) = y_2(0)= 0.1$, $I_1(0) = 0.019$, and $ I_2(0)= 0.022$. The time series of both neurons remain highly persistent for inhibitory coupling and become less persistent for higher $\theta$. For higher coupling, the first node controls the dynamics in the second, with the second node highly regular. The nodes are usually unsynchronized. The corresponding parameter sweep plots are shown in Fig.~\ref{fig:BifChem}.}
\label{fig:chem_pp}
\end{figure*}

We now move on to non-local chemical coupling strategy~\eqref{eq:dML_chemical}. We show six instances of the coupling strength $\theta = -0.005, -0.001, 0.001, 0.01, 0.05, 0.1$, see Fig.~\ref{fig:chem_pp}. The initial conditions $x_1(0)$ and $x_2(0)$ are sampled randomly from the continuous uniform distribution over the interval $[-1, 1]$. Other initial conditions are fixed as $y_1(0) = y_2(0)= 0.1$, $I_1(0) = 0.019$, and $ I_2(0)= 0.022$. For inhibitory coupling strength $\theta <0$ and a weak excitatory coupling strength of order $10^{-2}$, the time series of both neurons remain highly persistent. As $\theta$ increases to $0.1$, the time series become less persistent but still remains $H>0.5$. The time series for both nodes exhibit self-similarity as denoted by the sample entropy values which fluctuate in the range $[0.0077, 0.027]$, which is very low to moderately low. The nodes act in a regular fashion as portrayed by the bounded $p$ vs $q$ plots and $K \approx 0.1$ for all cases of $\theta$. Both nodes exhibit irregular bursting until $\theta =0.001$. For higher values of $\theta$ we observe that the dynamics in first node controls the dynamics in second. When $\theta \ge 0.01$, the second node exhibits a highly controlled and bounded time series. As a whole, the nodes are never synchronized as can be reported from $\Gamma$ values, the highest being $0.7929$ and the lowest being $-0.0252$. The nodes do not oscillate in phase as values of $B$ are greater than $0.9$, but never more than $0.975$.

\begin{figure}[h]
    \centering
    \includegraphics[width=0.7\linewidth]{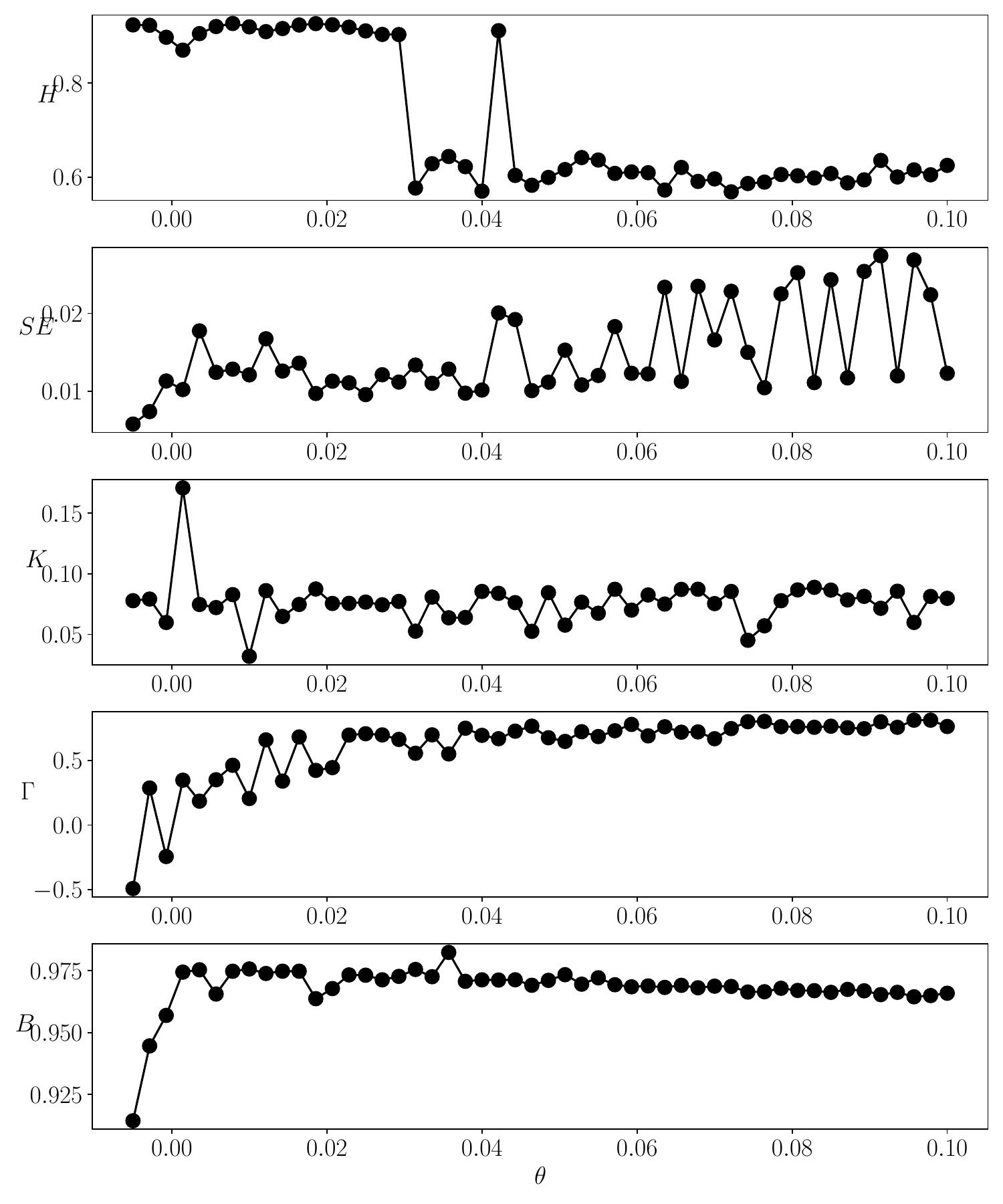}
    \caption{Bifurcation plots of different metrics performing a parameter sweep on $\theta \in [-0.005, 0.1]$ for model~\eqref{eq:dML_chemical}.}
    \label{fig:BifChem}
\end{figure}

The above characteristics are clearly portrayed in the parameter sweep plots in Fig.~\ref{fig:BifChem}. Both time series are always persistent with higher persistence for $\theta \le 0.03$ where $H \approx 0.9$ and moderate persistence for $\theta >0.03$ with $H \approx 0.6$. The sample entropy fluctuate between low to moderately low in the range $[0.0058, 0.0274]$ approximately. Chaos does not occur for our setting of chemical coupling as the system exhibits a moderately regular behavior ($K \in [0.032, 0.171]$ approx). Both nodes are asynchronized with $\Gamma$ approximately in the range $[-0.4933, 0.8126]$, and they are incoherent with $B$ in range $[0.91434, 0.98243]$. We see that $B$ asymptotes roughly at a value of $0.975$ as $\theta$ increases.

\begin{figure*}[h]
\begin{tabular}{ccc}
  \includegraphics[width=0.3\linewidth]{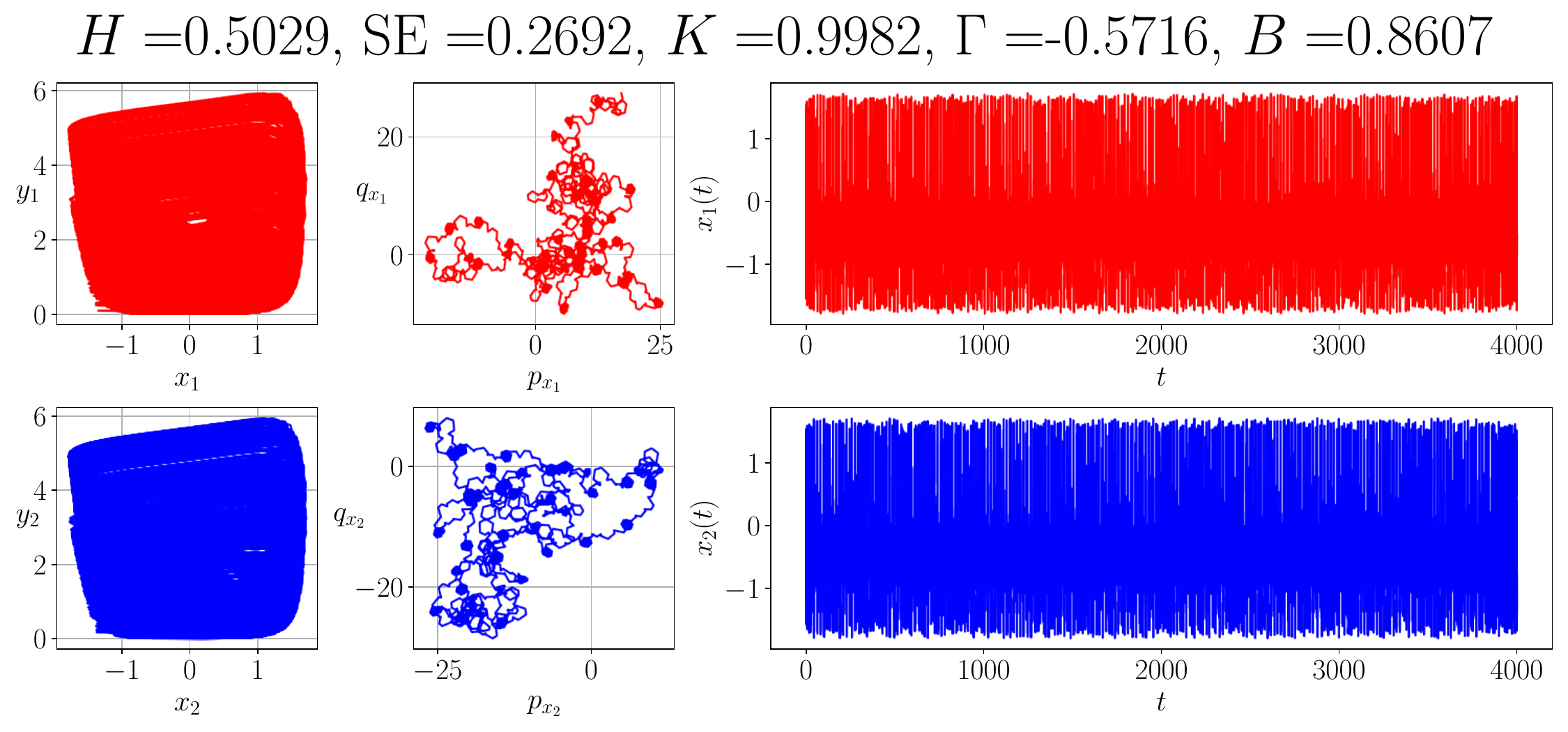} &  \includegraphics[width=0.3\linewidth]{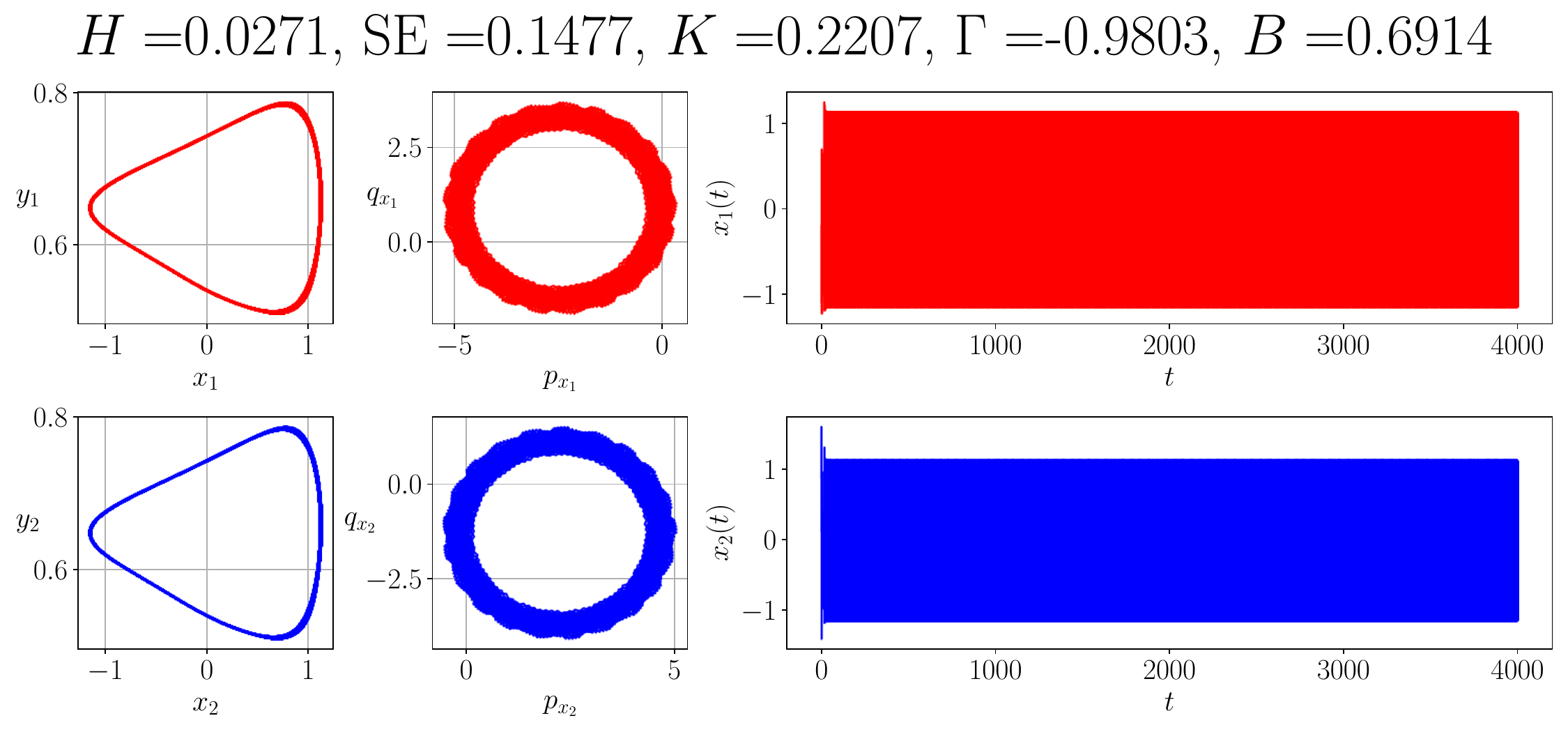} &  \includegraphics[width=0.3\linewidth]{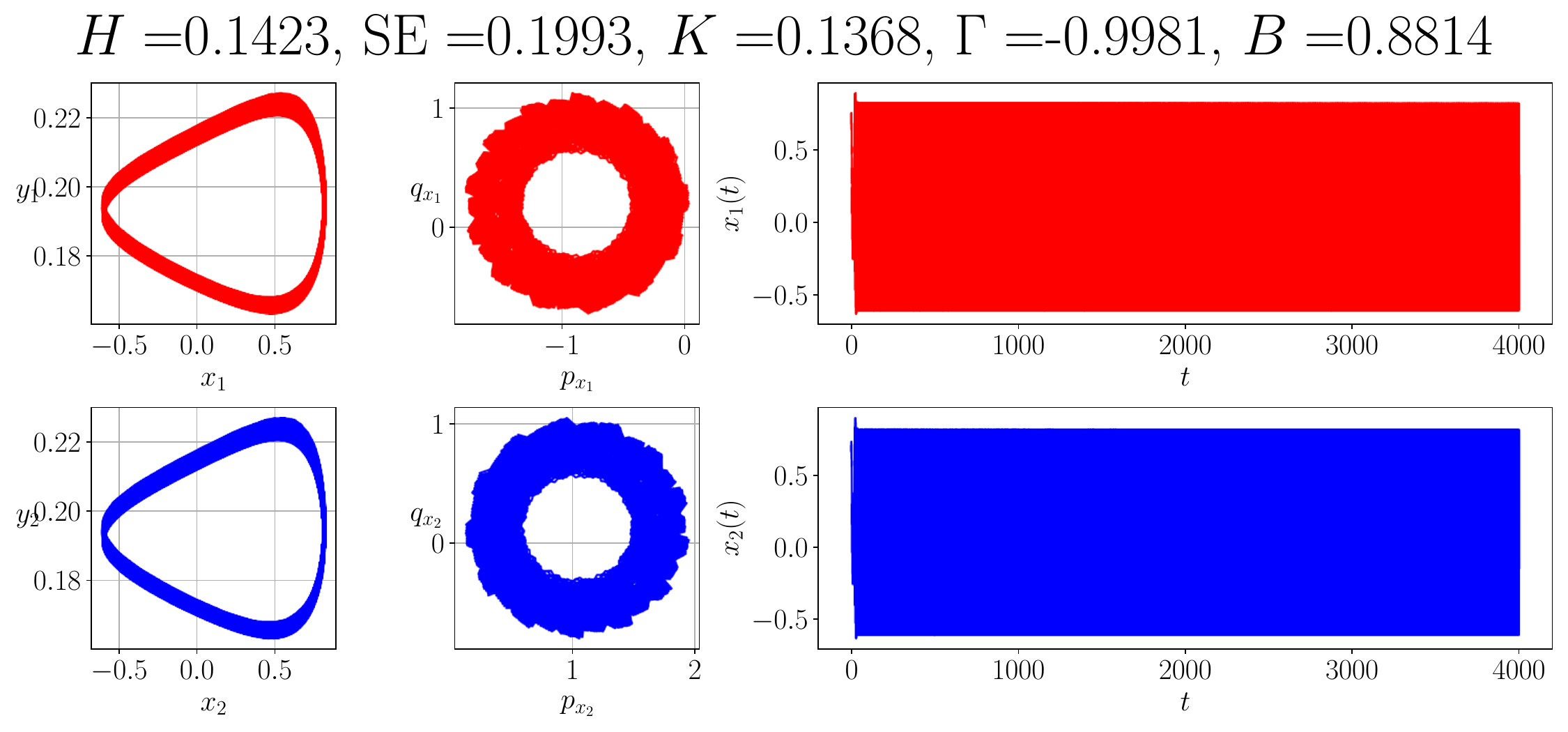} \\
(a) $\theta = -1$ & (b) $\theta = -0.5$ & (c) $\theta = -0.1$ \\
\includegraphics[width=0.3\linewidth]{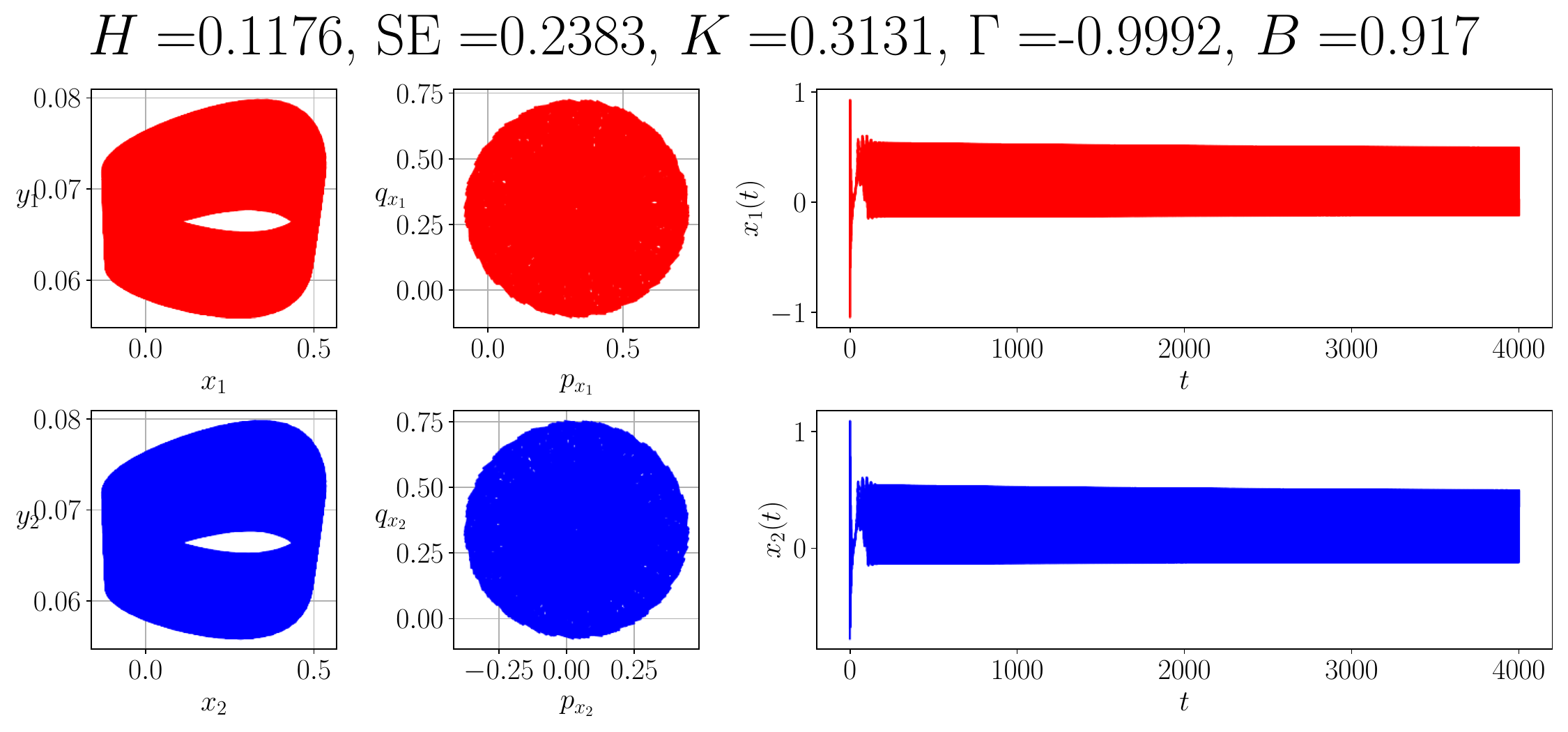} &  \includegraphics[width=0.3\linewidth]{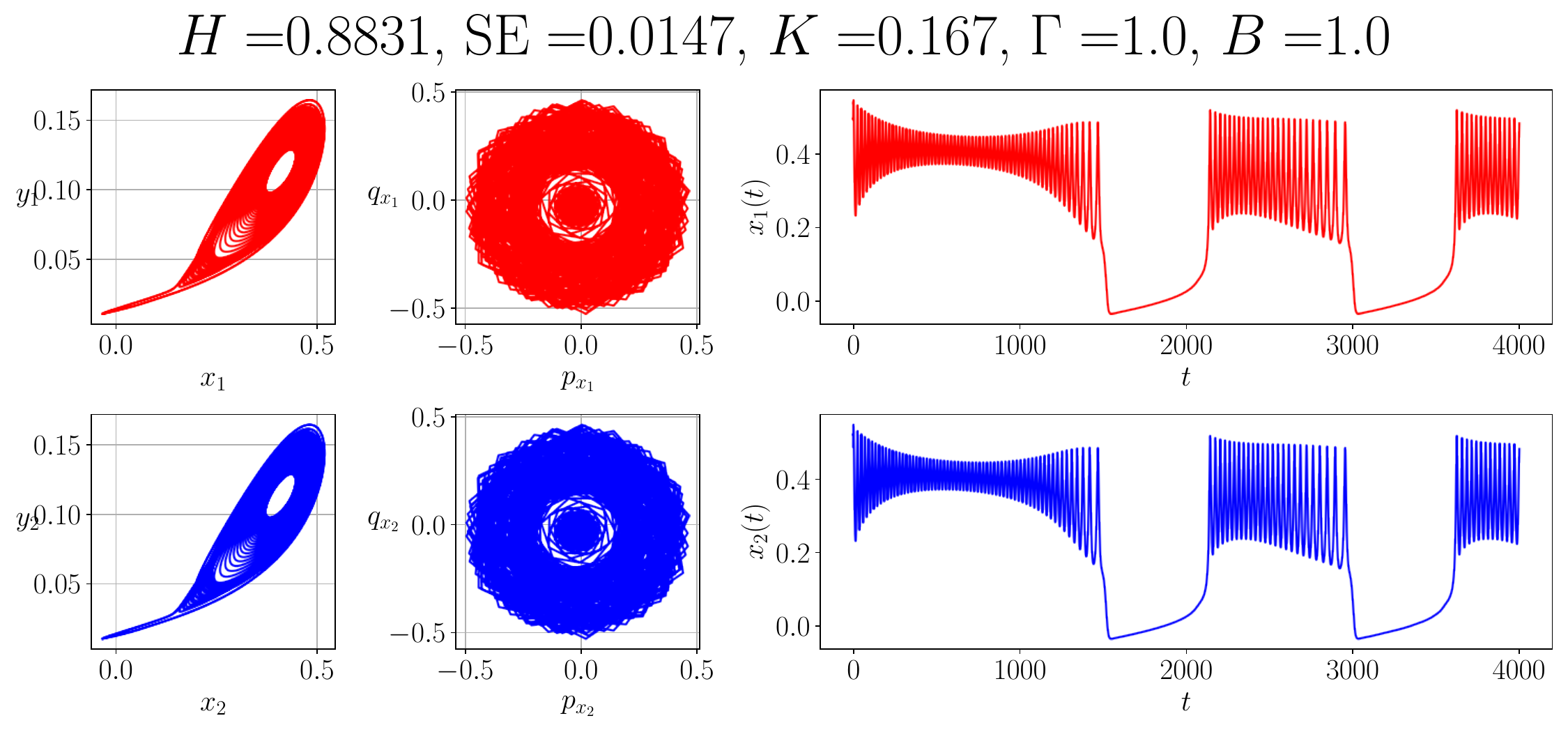} &  \includegraphics[width=0.3\linewidth]{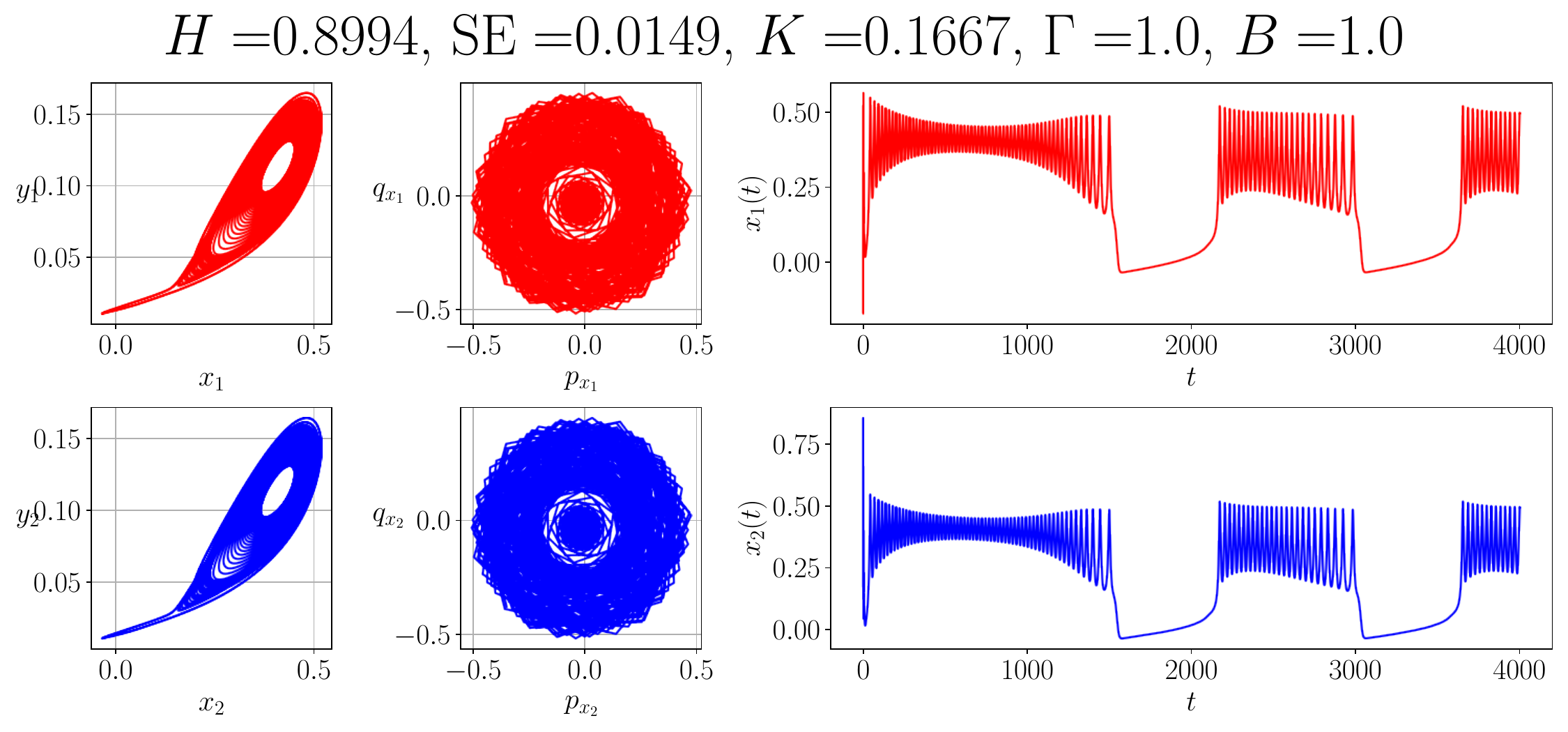} \\
(d) $\theta = 0.1$ & (e) $\theta = 0.5$ & (f) $\theta = 1$\\
\end{tabular}
\caption{Phase portraits, $p$ vs $q$ plots, and time series of the Josephson junction coupling~\eqref{eq:JJ} with varying $\theta \in [-1, 1]$. Other parameters are fixed as in table~\ref{tab:params}, and $x_1(0)$ and $x_2(0)$ are sampled randomly from the continuous uniform distribution over the interval $[-1, 1]$. Other initial conditions are fixed as $y_1(0) = y_2(0)= 0.1$, $I_1(0) = 0.018$, and $ I_2(0)= 0.022$. Also $\phi(0) = \mu(x_1(0) - x_2(0))$. For a large inhibitory coupling, both nodes are chaotic. The nodes generate more regular time series for weaker coupling close to $\theta \approx 0$. Nodes oscillate in anti-phase for inhibitory coupling and display fully synchronized bursting behavior when $\theta$ becomes excitatory. The corresponding parameter sweep plots are shown in Fig.~\ref{fig:BifJJ}.}
\label{fig:JJ_pp}
\end{figure*}

Next model at hand is the Josephson junction coupling given by~\eqref{eq:JJ}. The coupling strength is varied in the range $\theta \in [-1, 1]$. The initial conditions $x_1(0)$ and $x_2(0)$ are sampled randomly from the continuous uniform distribution over the interval $[-1, 1]$. Other initial conditions are fixed as $y_1(0) = y_2(0)= 0.1$, $I_1(0) = 0.018$, and $ I_2(0)= 0.022$. Also $\phi(0) = \mu(x_1(0) - x_2(0))$. First, we show six instances of phase portraits, $p$ vs $q$ plots and time series from both nodes with fixed $\theta$ values $\theta = -1, -0.5, -0.1, 0.1, 0.5, 1$, see Fig.~\ref{fig:JJ_pp}. At $\theta =-1$, we see a random Brownian motion in the time series of both nodes having $H \approx 0.5$. They are highly irregular which is represented by a high sample entropy value of ${\rm SE = 0.2692}$. Both nodes are chaotic as represented by the unbounded chaotic $p$ vs $q$ plots and $K \approx 1$. The nodes are asynchronized ($\Gamma = 0.5716$) and highly incoherent ($B = 0.8607$). As $\theta$ is increased to $\theta = -0.1$, we observe a more regular oscillation in the time series. The phase portraits of both nodes indicate a quasi-periodic orbit with the $p$ vs $q$ plots bounded. The time series become anti-persistent with $H$ values $0.0271$ for $\theta =-0.5$ and $0.1423$ for $\theta =-0.1$. The sample entropy values also decrease as expected to $0.1477$ for $\theta =-0.5$ and $0.1993$ for $\theta =-0.1$. Bounded quasiperiodic oscillations are supported by $K$ values of $0.2207$ and $0.1368$ as well. The nodes start oscillating in anti-phase reaching almost a full synchronization as $\Gamma \approx 1$ for $\theta =-0.5, -0.1$. As they are out of phase, we have $B$ exhibiting small values $0.6914$, $0.8814$. We see a similar behavior for $\theta = 0.1$, however the oscillation is bounded in a smaller domain compared to panels (b) and (c). As $\theta$ further increases to $\theta = 0.5, 1$, we see fold/homoclinic bursting in both nodes as was observed before for gap junction coupling, for example. Both nodes are again fully synchronized but in phase, exhibiting $\Gamma =1$ and $B = 1$.

\begin{figure}[h]
    \centering
    \includegraphics[width=0.7\linewidth]{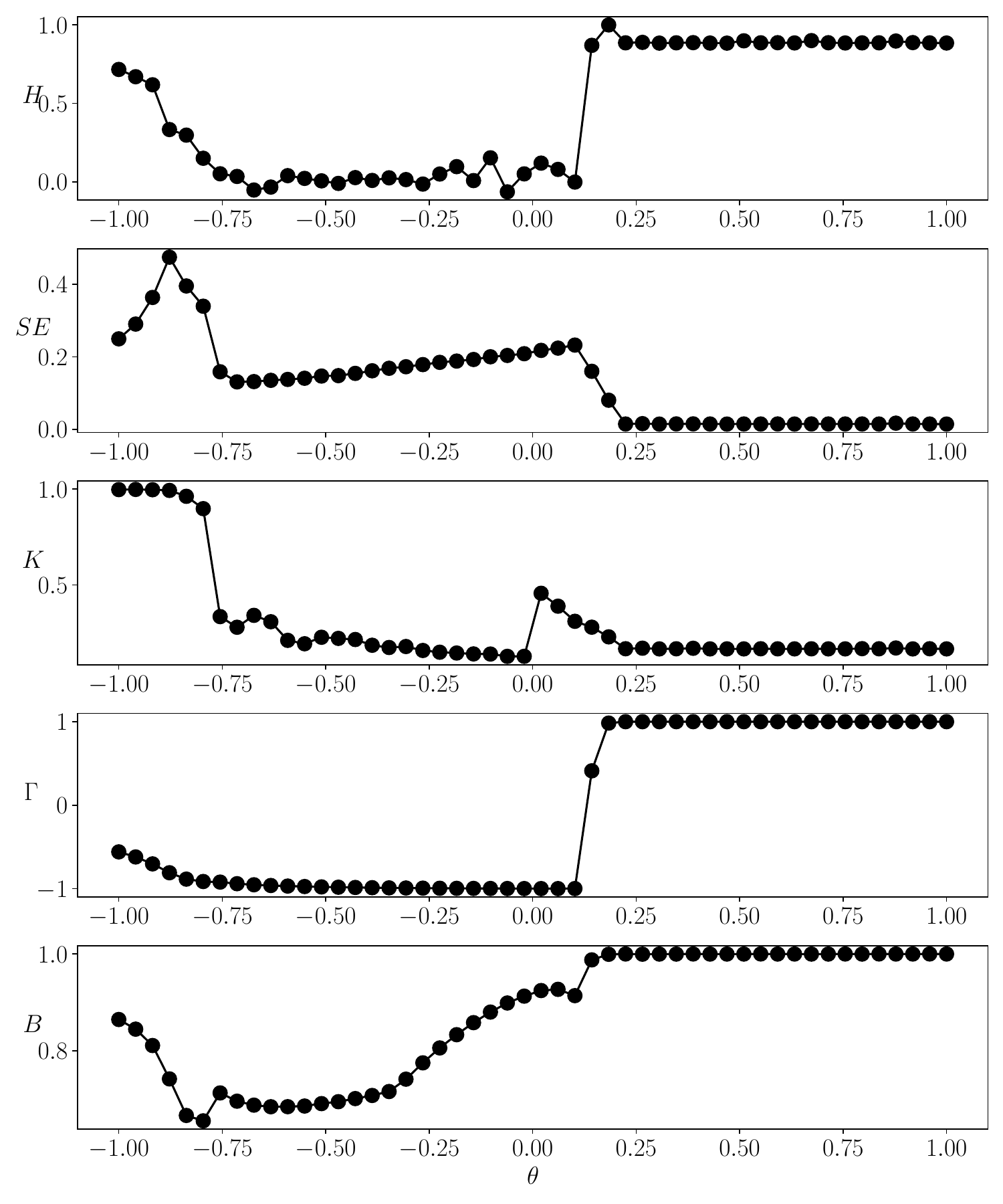}
    \caption{Bifurcation plots of different metrics performing a parameter sweep on $\theta \in [-1, 1]$ for model~\eqref{eq:JJ}.}
    \label{fig:BifJJ}
\end{figure}

Following this, we can look at the bigger picture from the plots presented in Fig.~\ref{fig:BifJJ}. Note that we observe some spurious values like $H \approx -0.053$ at $\theta \approx -0.673$ or $H \approx -0.00225$ at $\theta \approx 0.102$. These are due to the algorithm not performing well. However, the algorithm exhibits a correct behavior for most values of $\theta$ represented. Starting at $\theta =-1$, the time series of the coupled system exhibit a random browninan motion, and eventually values of $H$ drops down close to $0$ till $\theta \approx 0.1$. After that the system starts exhibiting regular repeated bursting the time series becomes more regular with $H$ values escalating to $0.88$ on average. The sample entropy remains quite high for the random Brownian motion in the time series and starts alleviating when the nodes starts a more regular oscillatory behavior in anti-phase. Sample entropy values remain in the range of $[0.23, 0.16]$ until $\theta \approx 0.1$. After that, it sharply falls down for the regime of bursting behavior, with ${\rm SE} \approx 0.01$. The $0-1$ test generates values of $K \approx 1$ close to the regime of irregular chaotic time series. However, it starts dropping to close to $0.2$, indicating the quasi-periodic behavior around where the time-series exhibits a more regular anti-phase oscillation. It then settles down to around $K \approx 0.166$ eventually. Values of $\Gamma$ represent a small regime of asynchrony in the regime where both nodes behave chaotically. However, both nodes quickly synchronize but they oscillate anti-phase, meaning $\Gamma \approx -1$. This occurs until $\theta \approx 0.1$. On further increase in $\theta$, the two nodes quickly start synchronizing but now in phase, meaning $\Gamma$ becomes $\approx 1$. The Kuramoto order parameter $B$ is very less around $[0.655, 0.9138]$ starting from where the nodes are chaotic, to synchronized anti-phase. After $\theta >0.1$ where the nodes starts fully synchronizing in phase, $B$ becomes $1$ indicating phase-locked behavior. 

\begin{figure*}[h]
\begin{tabular}{ccc}
  \includegraphics[width=0.3\linewidth]{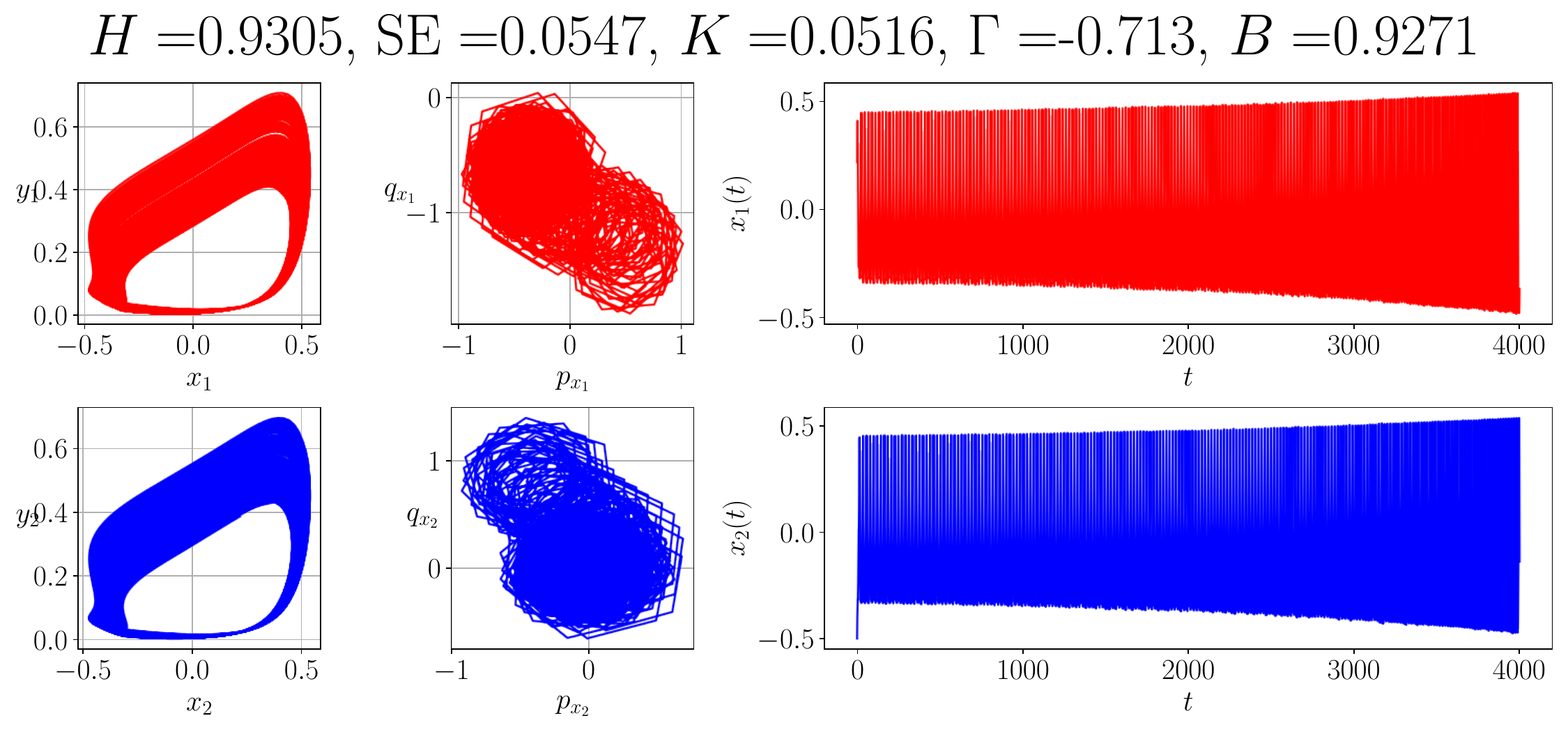} &  \includegraphics[width=0.3\linewidth]{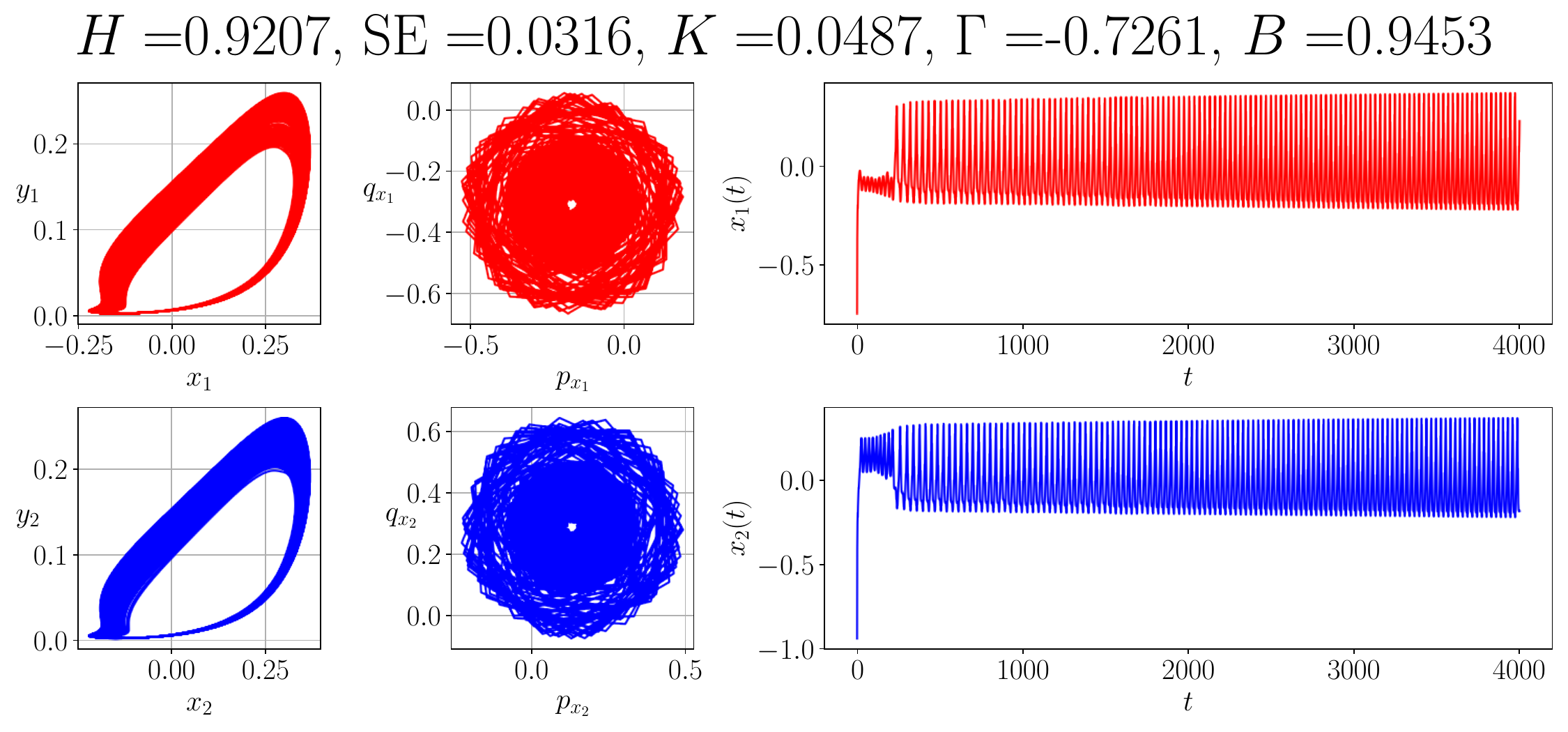} &  \includegraphics[width=0.3\linewidth]{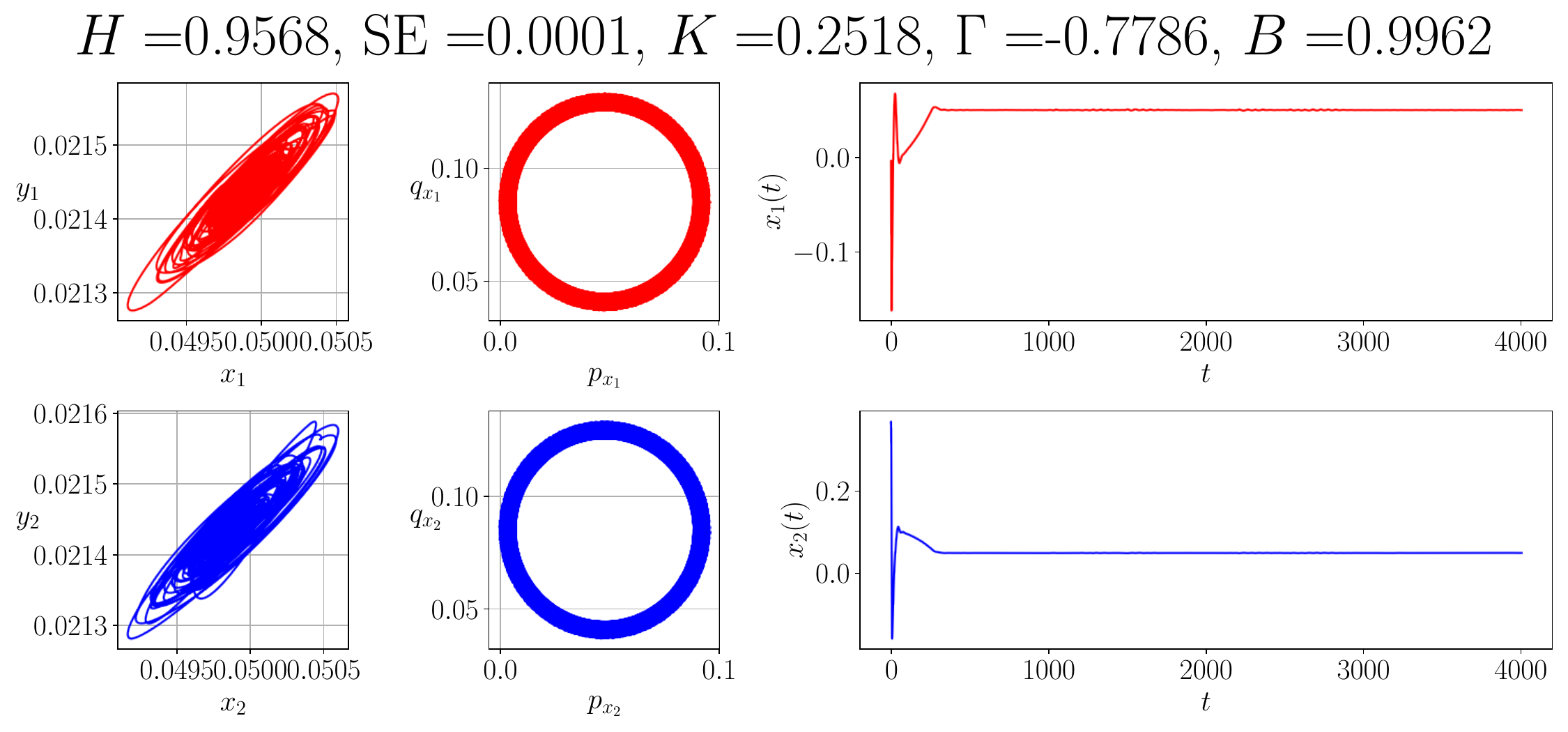}\\
(a) $\theta = -0.02$ & (b) $\theta = -0.01$ & (c) $\theta = -0.005$\\
\includegraphics[width=0.3\linewidth]{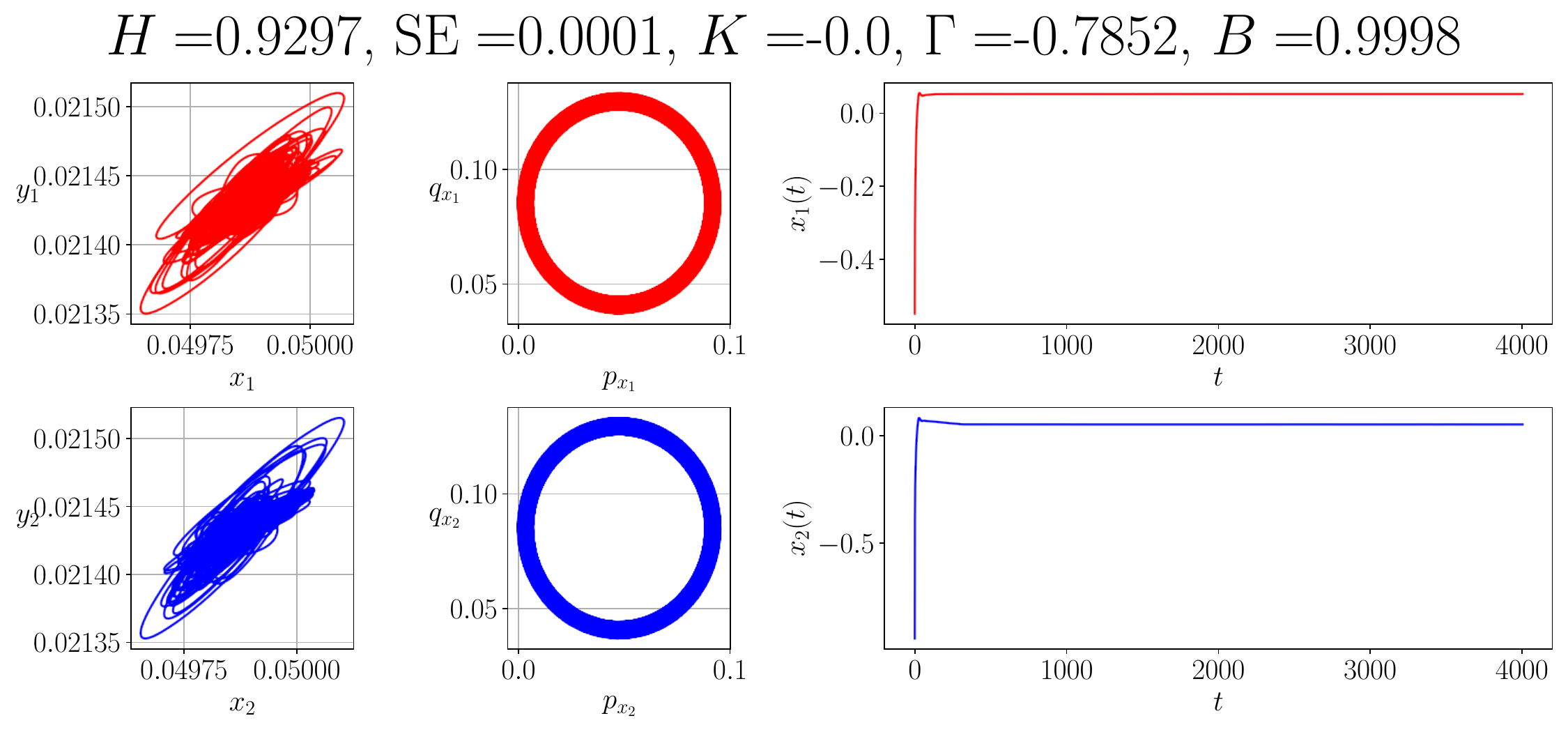} &  \includegraphics[width=0.3\linewidth]{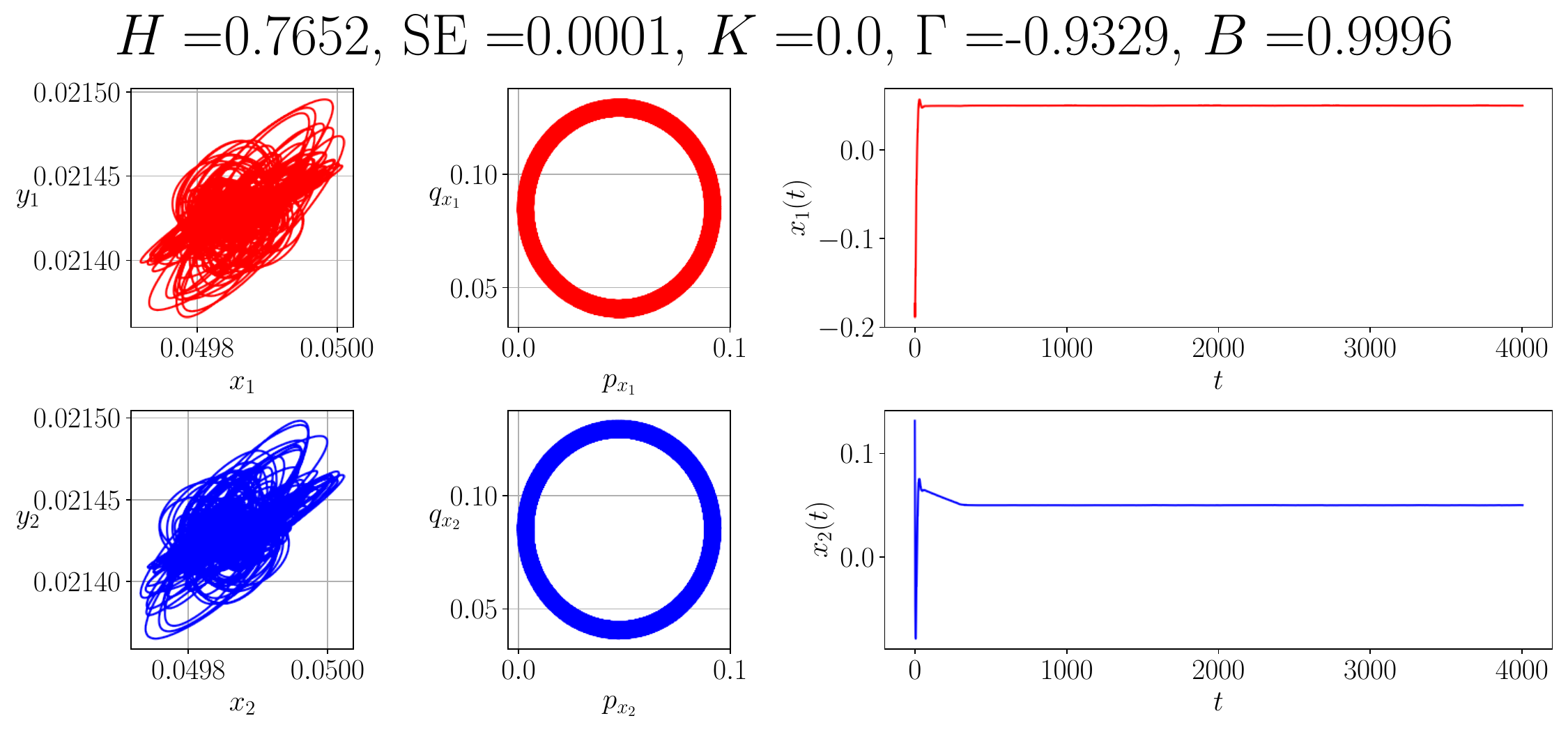} &  \includegraphics[width=0.3\linewidth]{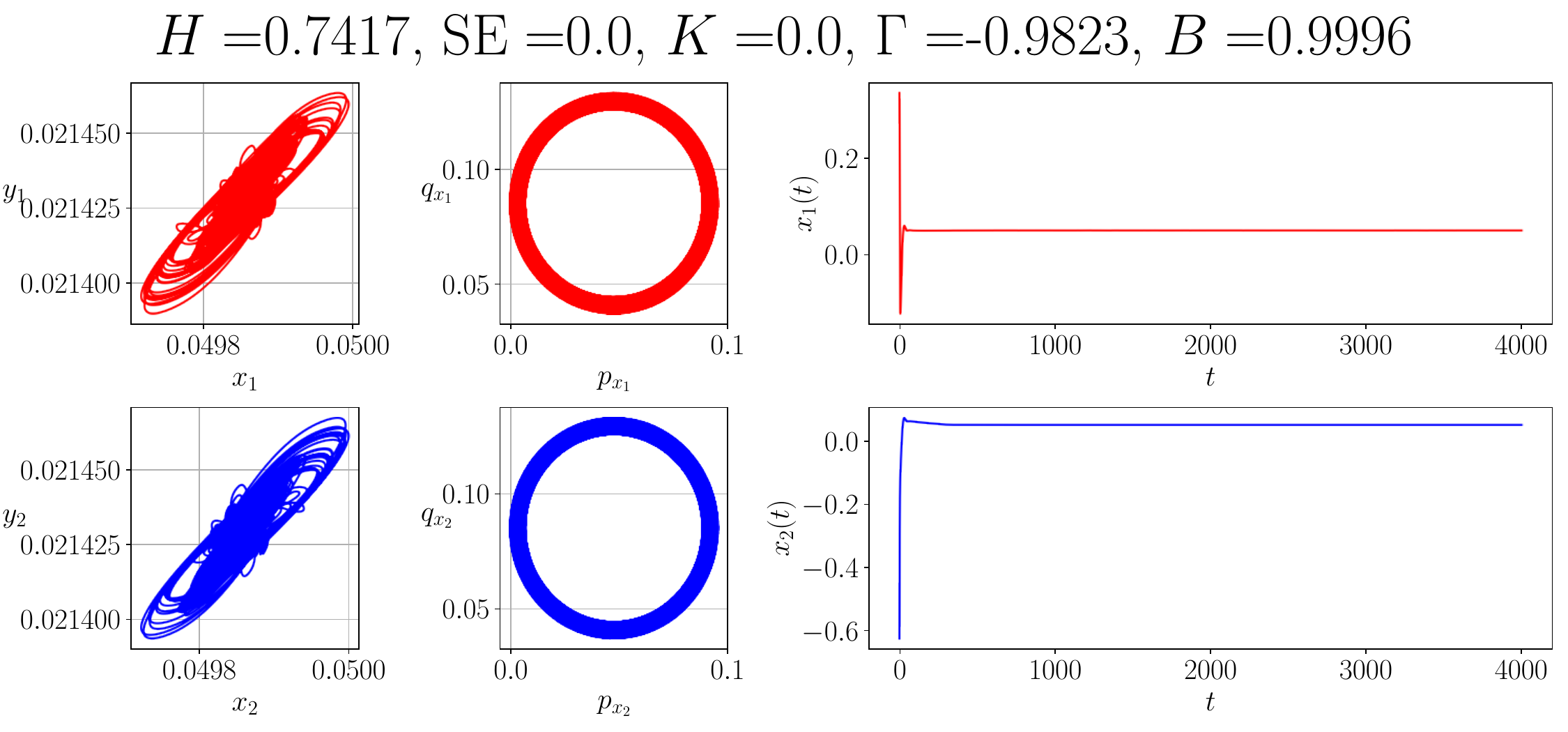}\\
(d) $\theta = 0.002$ & (e) $\theta = 0.005$ & (f) $\theta = 0.01$\\
\end{tabular}
\caption{Phase portraits, $p$ vs $q$ plots, and time series of the memristive coupling~\eqref{eq:mem_coupling} with varying $\theta \in [-0.02, 0.01]$. Other parameters are fixed as in table~\ref{tab:params}, and $x_1(0)$ and $x_2(0)$ are sampled randomly from the continuous uniform distribution over the interval $[-1, 1]$. Other initial conditions are fixed as $y_1(0) = y_2(0)= 0.1$, $I_1(0) = 0.018$, and $ I_2(0)= 0.022$. Also $\phi(0) = \theta(x_1(0) - x_2(0))$. Both time series are persistent throughout. For excitatory coupling, the system exhibits a decay oscillation in both nodes, where the trajectories tend to fall into a symmetric equilibrium point $(x^*, y^*, x^*, y^*)$. Also the nodes are oscillating in anti-phase with full anti-phase synchronization for a high excitatory coupling. The corresponding parameter sweep plots are shown in Fig.~\ref{fig:BifMem}.}
\label{fig:mem_pp}
\end{figure*}

The last coupling type that we consider for a two-node network is the memristive coupling, given by~\eqref{eq:mem_coupling}. The coupling strength is varied in the range $\theta \in [-0.02, 0.01]$, see Fig.~\ref{fig:mem_pp}. The initial conditions $x_1(0)$ and $x_2(0)$ are sampled randomly from the continuous uniform distribution over the interval $[-1, 1]$. Other initial conditions are fixed as $y_1(0) = y_2(0)= 0.1$, $I_1(0) = 0.018$, and $ I_2(0)= 0.022$. Also $\phi(0) = \theta(x_1(0) - x_2(0))$.  Like the last few models, we first show six instances of phase portraits, $p$ vs $q$ plots, and time series for both nodes of the coupled system. We have fixed $\theta = -0.02, -0.01, -0.005, 0.002, 0.005, 0.01$. At $\theta<-0.005$, we see a persistent behavior in the time series of both nodes having high Hurst exponent values $H >0.9$. Both time series are regular with ${\rm SE} = 0.0547$ for $\theta =-0.02$ and ${\rm SE} = 0.0316$ for $\theta =-0.01$. Both nodes show moderately bounded $p$ vs $q$ plots with $K \approx 0.05$. These are very small values of $K$ close to $0$. The two nodes have a tendency to oscillate out of phase, thus $\Gamma = -0.713, -0.7261$. Values of Kuramoto order parameter $B$ lies in a moderately smaller range of $B = 0.9271, 0.9453$ because of the anti-phase oscillatory tendency. For $\theta \ge -0.005$, the system exhibits a decay oscillation in both nodes, where the trajectories tend to fall into a symmetric equilibrium point $(x^*, y^*, x^*, y^*)$. The Hurst exponent indicates a highly persistent time series with $H >0.7$ throughout and the time-series is highly self similar with ${\rm SE} \approx 0$. Note that the $p$ vs $q$ plots indicate highly bounded trajectories with $K \approx 0$. We observe a spurious value of $K \approx 0.2518$ at $\theta = -0.005$, which we believe is the result of very weak coupling close to $\theta= 0$. The anti-phase dynamics among the nodes is maintained, with more synchronization observed for $\theta =0.005, 0.01$ (with $\Gamma = -0.9329, -0.9823$ respectively). Also note that $B \approx 1$ meaning the nodes are phase locked however in anti-phase synchrony. 

\begin{figure}[h]
    \centering
    \includegraphics[width=0.7\linewidth]{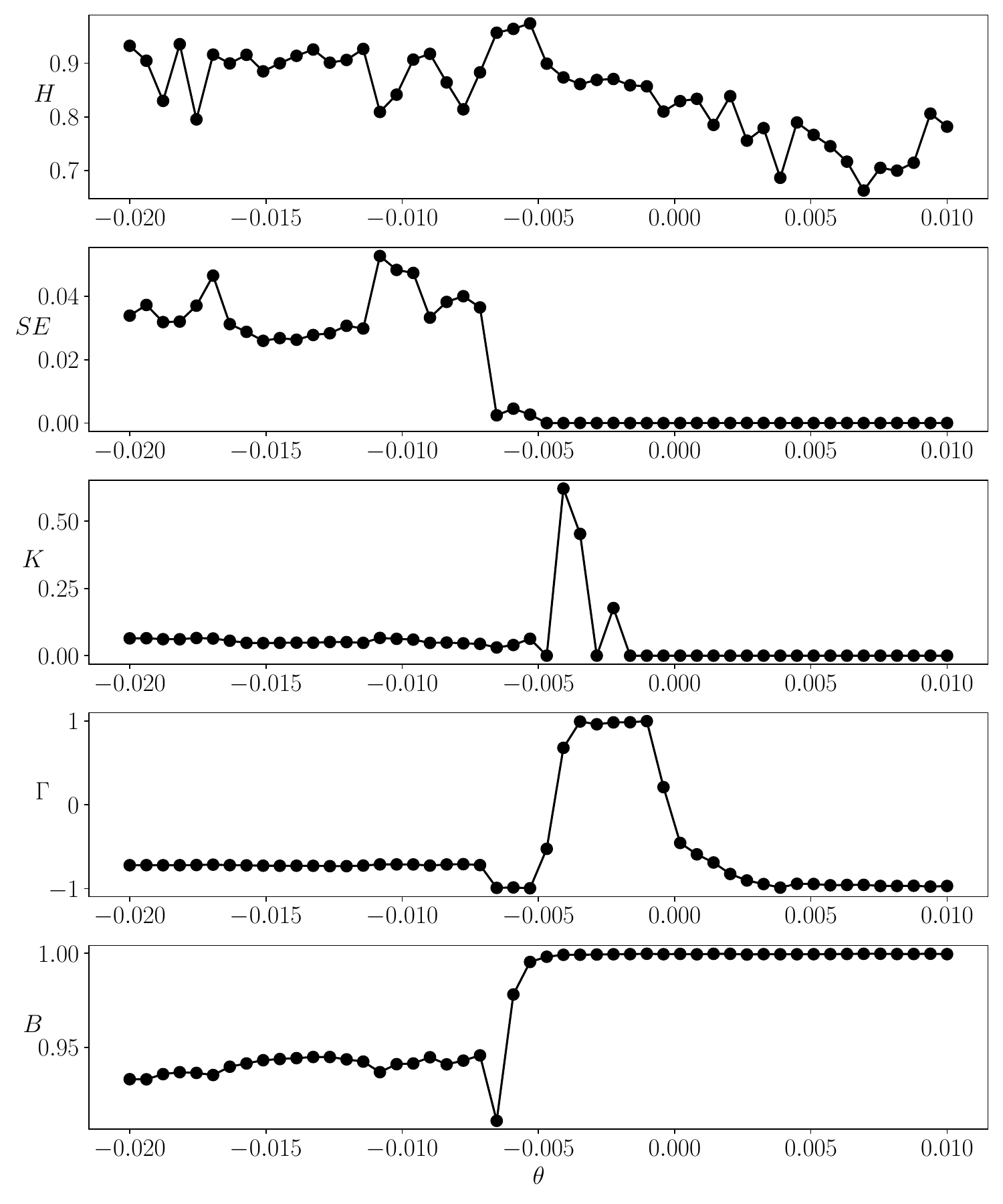}
    \caption{Bifurcation plots of different metrics performing a parameter sweep on $\theta \in [-0.02, 0.01]$ for model~\eqref{eq:mem_coupling}. The spurious negative values in $K$ ($\theta \approx -0.005$) due to the algorithm underperforming have been replaced by $0$.}
    \label{fig:BifMem}
\end{figure}

When we do a parameter sweep of these metrics on $\theta$ (Fig.~\ref{fig:BifMem}), we see a persistence in the time series ($0.66245 \le H \le 0.9745$) throughout the range of the coupling strength. The time series are highly regular and self similar. As $-0.02\le \theta <-0.005$, the sample entropy value fluctuates in the range $[0.02, 0.053]$ approximately. As soon as $\theta>0.005$, the sample entropy becomes $0$ pertaining to decay oscillation to an equilibrium point in the system. In terms of the $0-1$ test, it remains in the order of $10^{-2}$ indicating a regular time series from both nodes. As $\theta$ approaches $=-0.005$, the algorithm for the $0-1$ test starts under-performing spewing some spurious values, for example $K \approx -0.5566$ ($\theta \approx -0.0047$), $K \approx -0.677$ ($\theta =-0.00286$), $K \approx -0.2265$ ($\theta = -0.00163$), and so on. There for five to six such occurrences. So we post-process these values to $K=0$ and then plot our result. As $\theta$ increases to $\theta = 0.01$, the $0-1$ test generates $K = 0$, indicating a highly regular dynamics. Values of $\Gamma$ show an anti-phase tendency ($\Gamma \approx 0.73$) between the two nodes until $\theta \approx -0.005$. As $\theta$ increases beyond that, we see a regime of very weak coupling till $\theta =0$ where both nodes are in phase ($\Gamma \in [0.95, 0.996]$ approximately), after which the nodes are again in anti-phase ($\Gamma \in [-1, -0.9]$). The Kuramoto order parameter for the anti-phase tendency, where $\theta <-0.005$ has values $B \in [0.91, 0.98]$ approximately. Beyond that, $B \approx 1$.

\begin{figure*}[h]
\begin{tabular}{ccc}
  \includegraphics[width=0.3\linewidth]{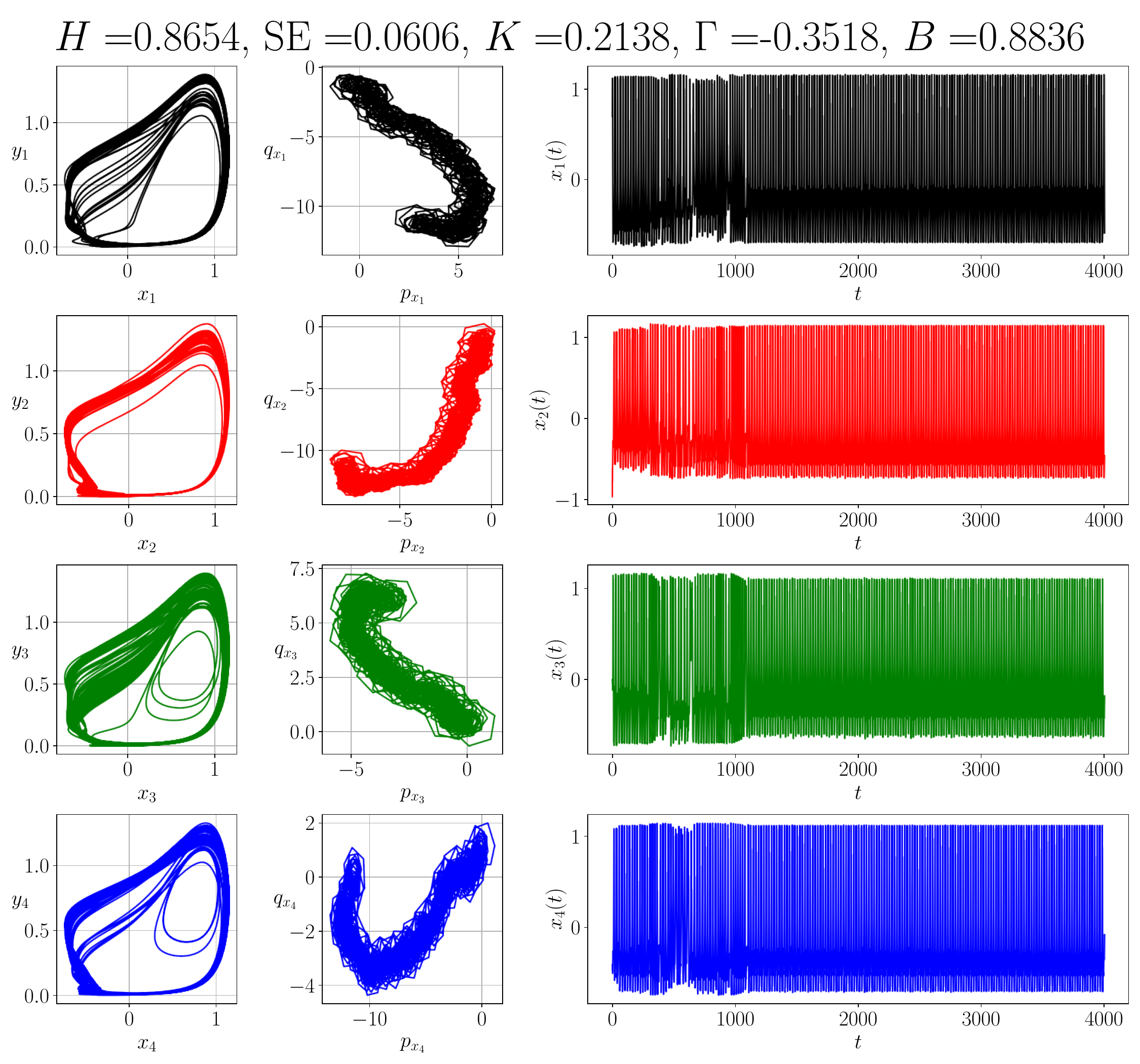} &  \includegraphics[width=0.3\linewidth]{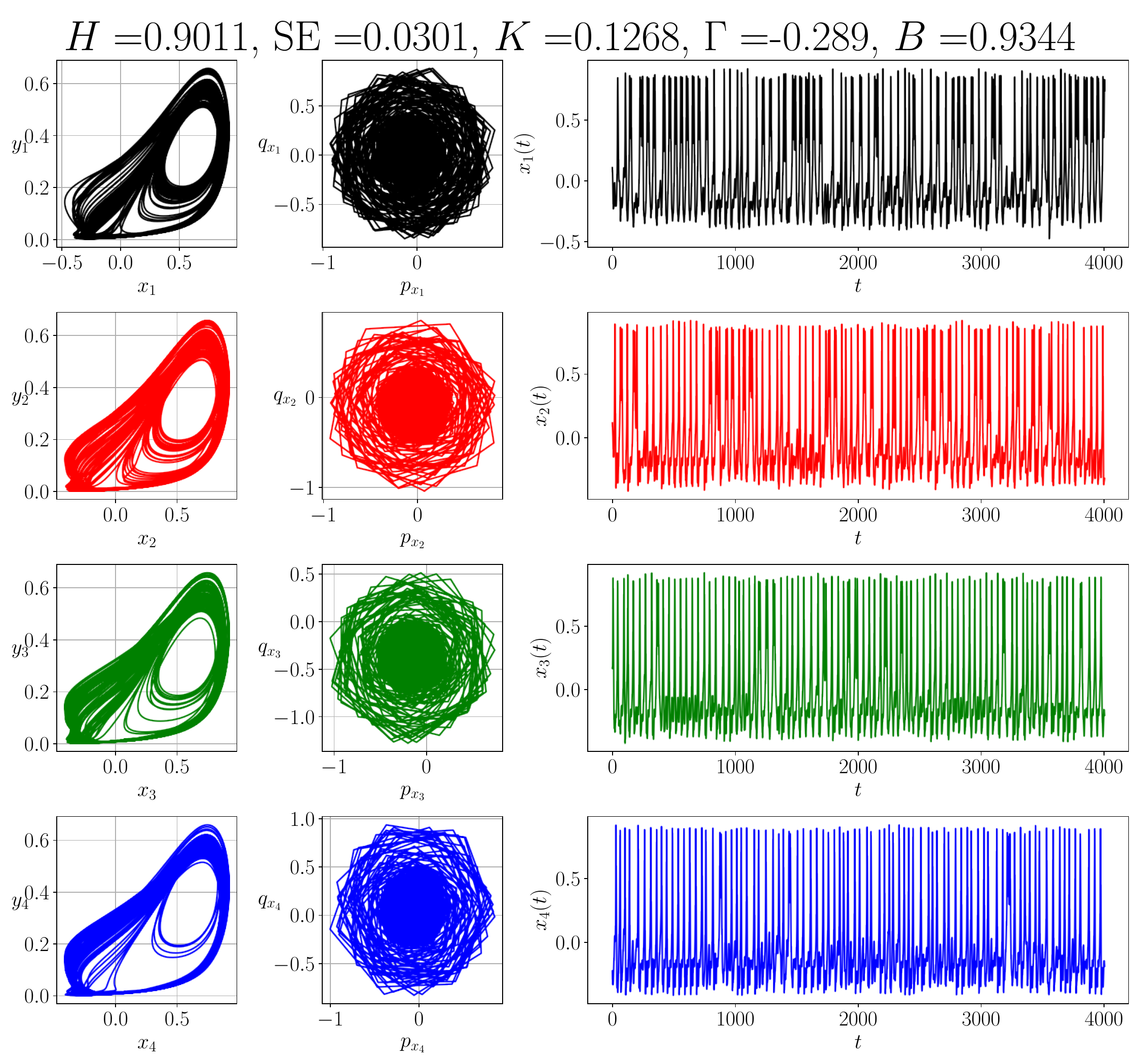} &  \includegraphics[width=0.3\linewidth]{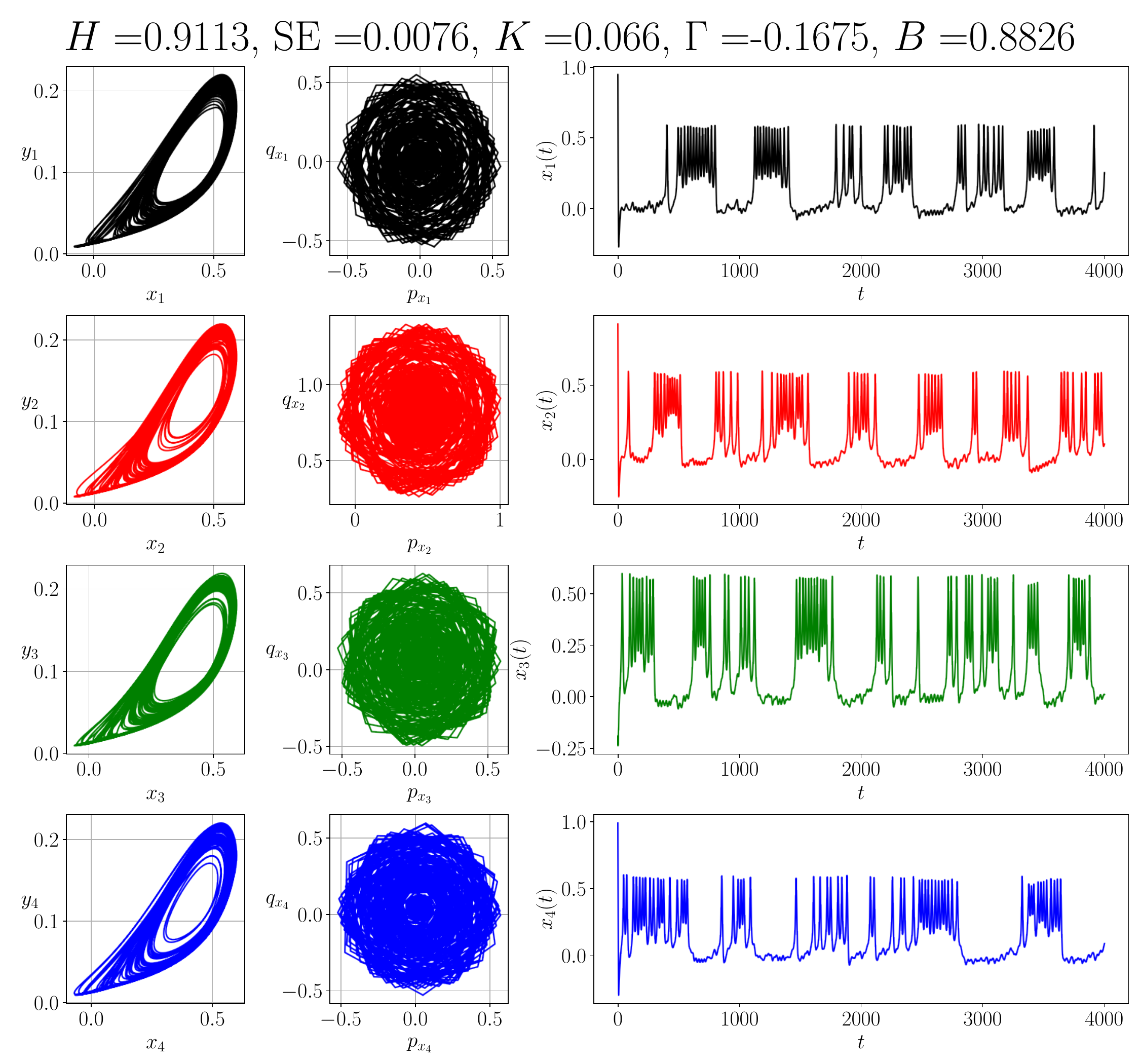}\\
(a) $\theta = -0.1$ & (b) $\theta = -0.05$ & (c) $\theta = -0.01$\\
\includegraphics[width=0.3\linewidth]{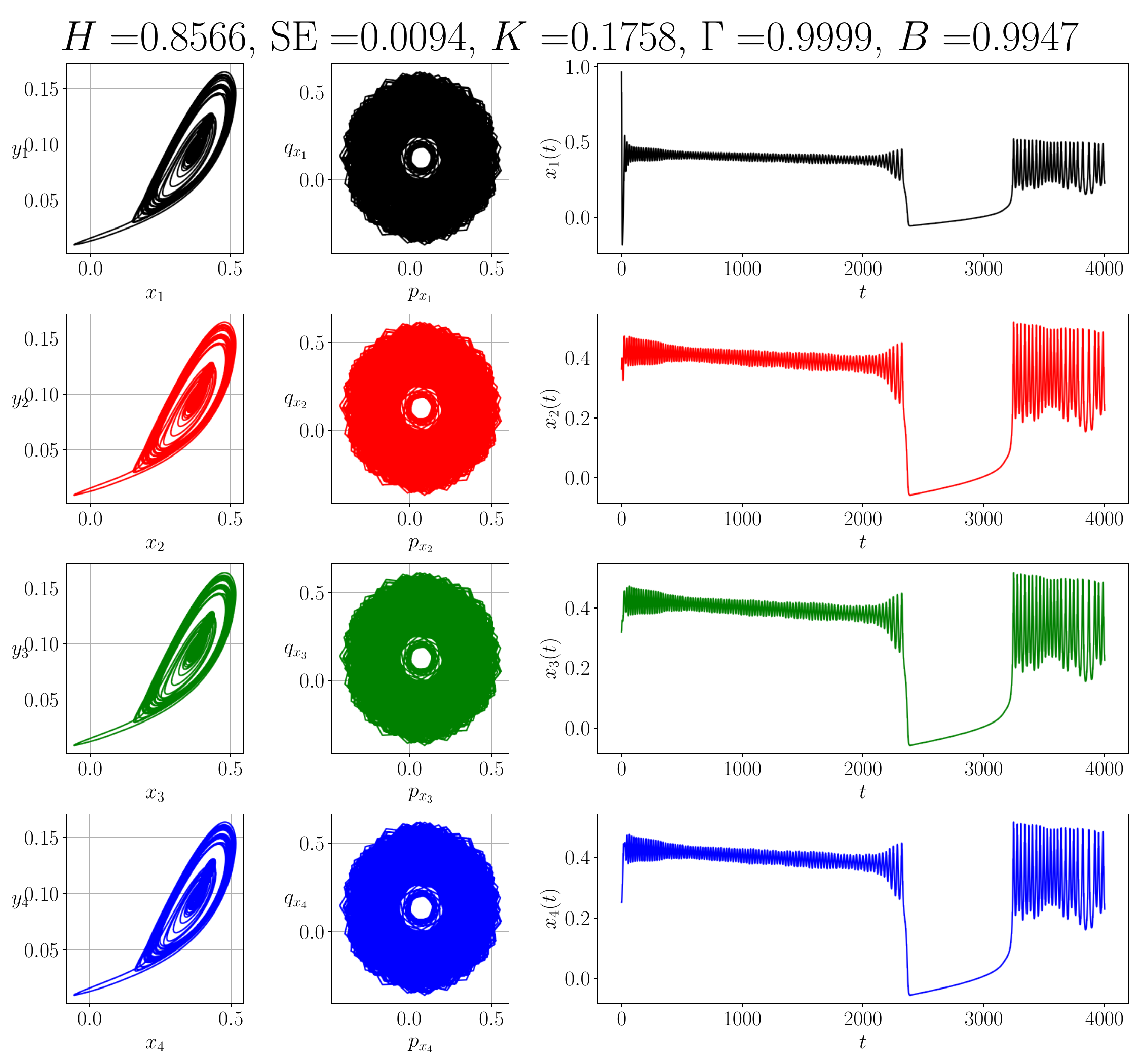} &  \includegraphics[width=0.3\linewidth]{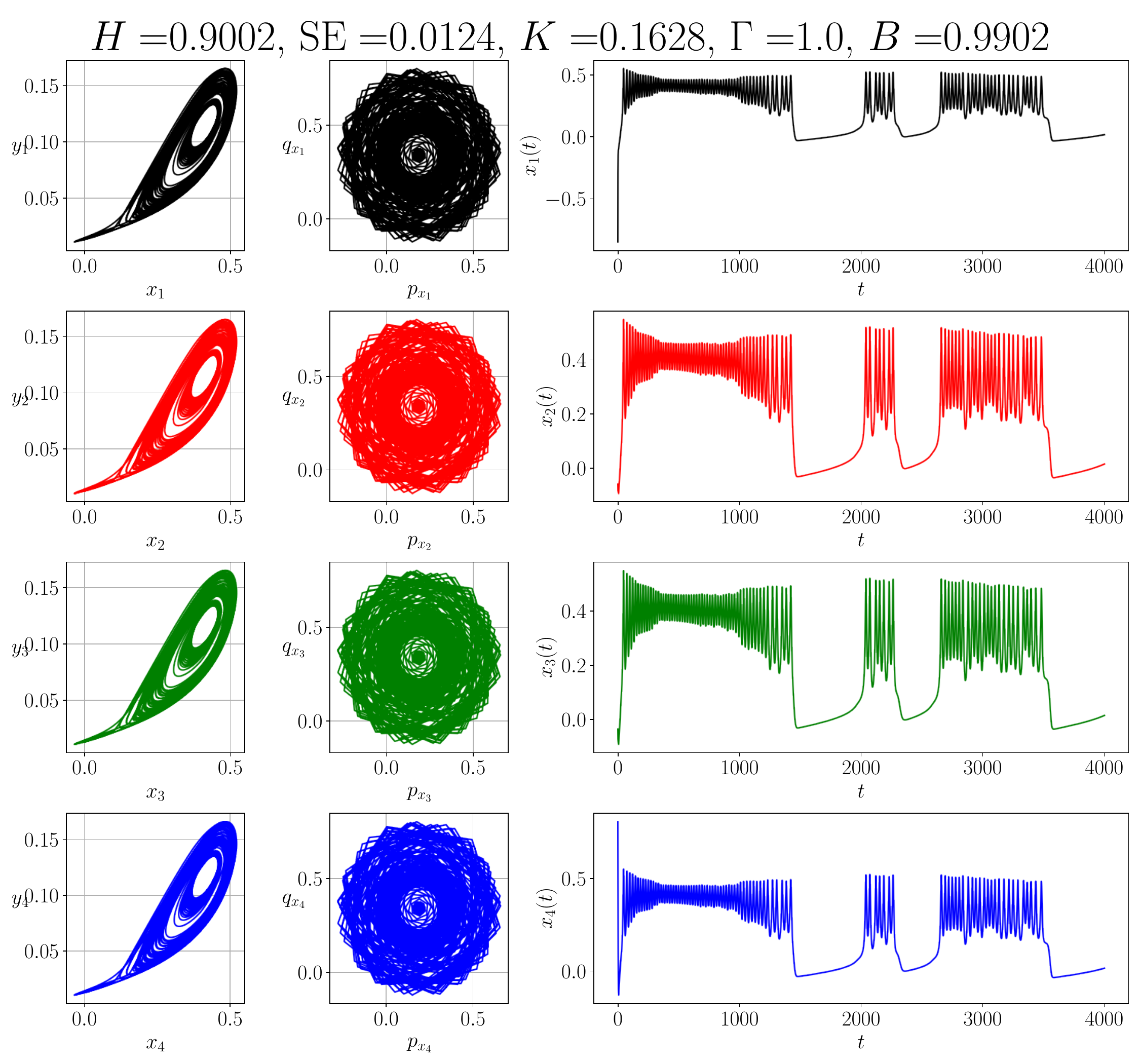} &  \includegraphics[width=0.3\linewidth]{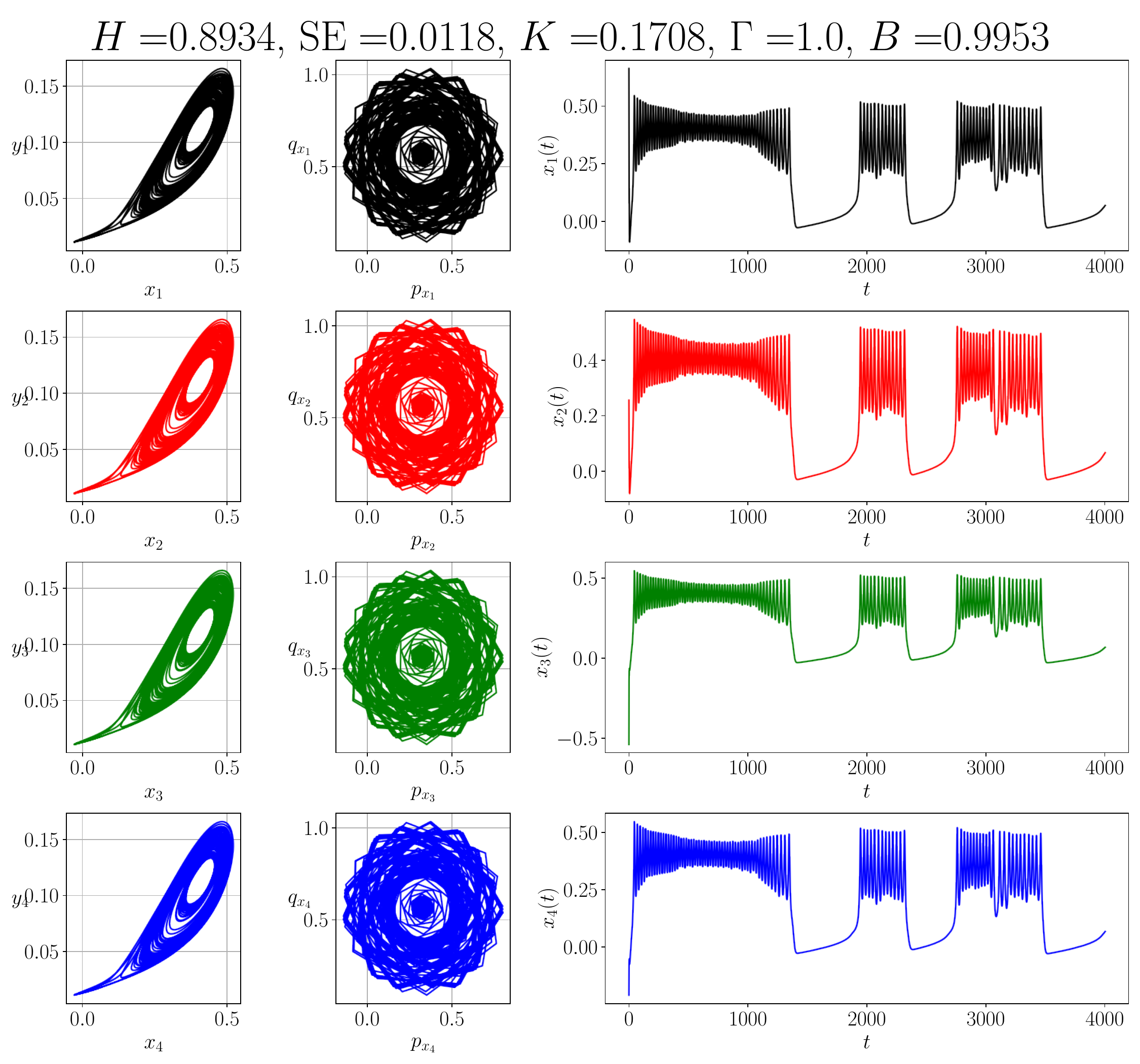}\\
(d) $\theta = 0.01$ & (e) $\theta = 0.05$ & (f) $\theta = 0.1$\\
\end{tabular}
\caption{Phase portraits, $p$ vs $q$ plots, and time series of the higher-order gap junction coupling in the smallest ring-star network~\eqref{eq:higher_order} with varying $\theta \in [-0.1, 0.1]$. Other parameters are fixed as in table~\ref{tab:params}, and $x_n(0), \quad n = 1, \ldots 4,$ are sampled randomly from the continuous uniform distribution over the interval $[-1, 1]$. Other initial conditions are fixed as $y_1(0) = y_2(0)= y_3(0) = y_4(0) = 0.1$, $I_1(0) = 0.018$, $I_2(0) = 0.019$, $I_3(0) = 0.02$, and $ I_4(0)= 0.022$. Both time series are persistent throughout. Higher-order excitatory coupling promotes synchronization. A stronger higher-order coupling leads to bursting patterns across every node. The corresponding parameter sweep plots are shown in Fig.~\ref{fig:BifHO}.}
\label{fig:ho_pp}
\end{figure*}

Next, we consider the small network consisting of four dML neurons, coupled via high-order simplicial complexes~\eqref{eq:higher_order}. The initial conditions $x_n(0), \quad n = 1, \ldots 4,$ are sampled randomly from the continuous uniform distribution over the interval $[-1, 1]$. Other initial conditions are fixed as $y_1(0) = y_2(0)= y_3(0) = y_4(0) = 0.1$, $I_1(0) = 0.018$, $I_2(0) = 0.019$, $I_3(0) = 0.02$, and $ I_4(0)= 0.022$. A reminder that we are back to implementing gap junction coupling in this model. The difference is that we now have four nodes, and there is an extra level of complexity implemented by two-simplex gap junction couplings beyond the pairwise ones. We show six instances of $\theta$ values, $\theta = -0.1, -0.05, -0.01, 0.01, 0.05, 0.1$. Without breaking down the properties exhibited by every metric as we have been doing so far, we want to point out that higher-order excitatory coupling promotes synchronization in this scenario~\citep{ZhLu23}. On increasing the strength of the higher-order coupling, the network demonstrates bursting patterns across every node, see Fig.~\ref{fig:ho_pp}.

\begin{figure}[h]
    \centering
    \includegraphics[width=0.7\linewidth]{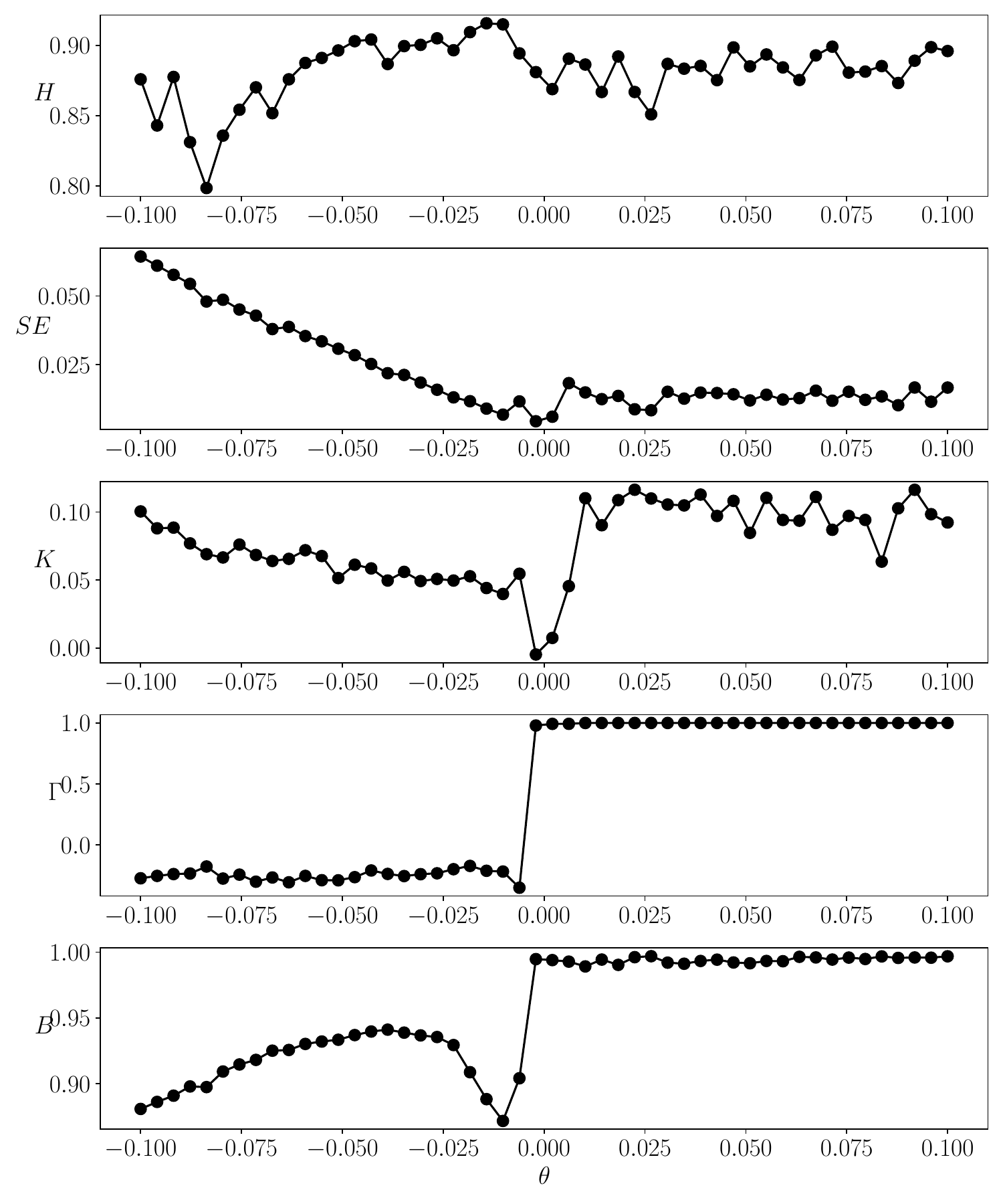}
    \caption{Bifurcation plots of different metrics performing a parameter sweep on $\theta \in [-0.1, 0.1]$ for model~\eqref{eq:higher_order}.}
    \label{fig:BifHO}
\end{figure}

The parameter sweep plots indicate a consistent persistence among the time series of every node in the network. The sample entropy lowers down to a moderately high value as the nodes start exhibiting bursting behavior. Chaos does not exist in this regime, and the nodes synchronize in phase as soon as the higher-order coupling becomes excitatory, see Fig.~\ref{fig:BifHO}.

\begin{figure*}[h]
\begin{tabular}{ccc}
  \includegraphics[width=0.3\linewidth]{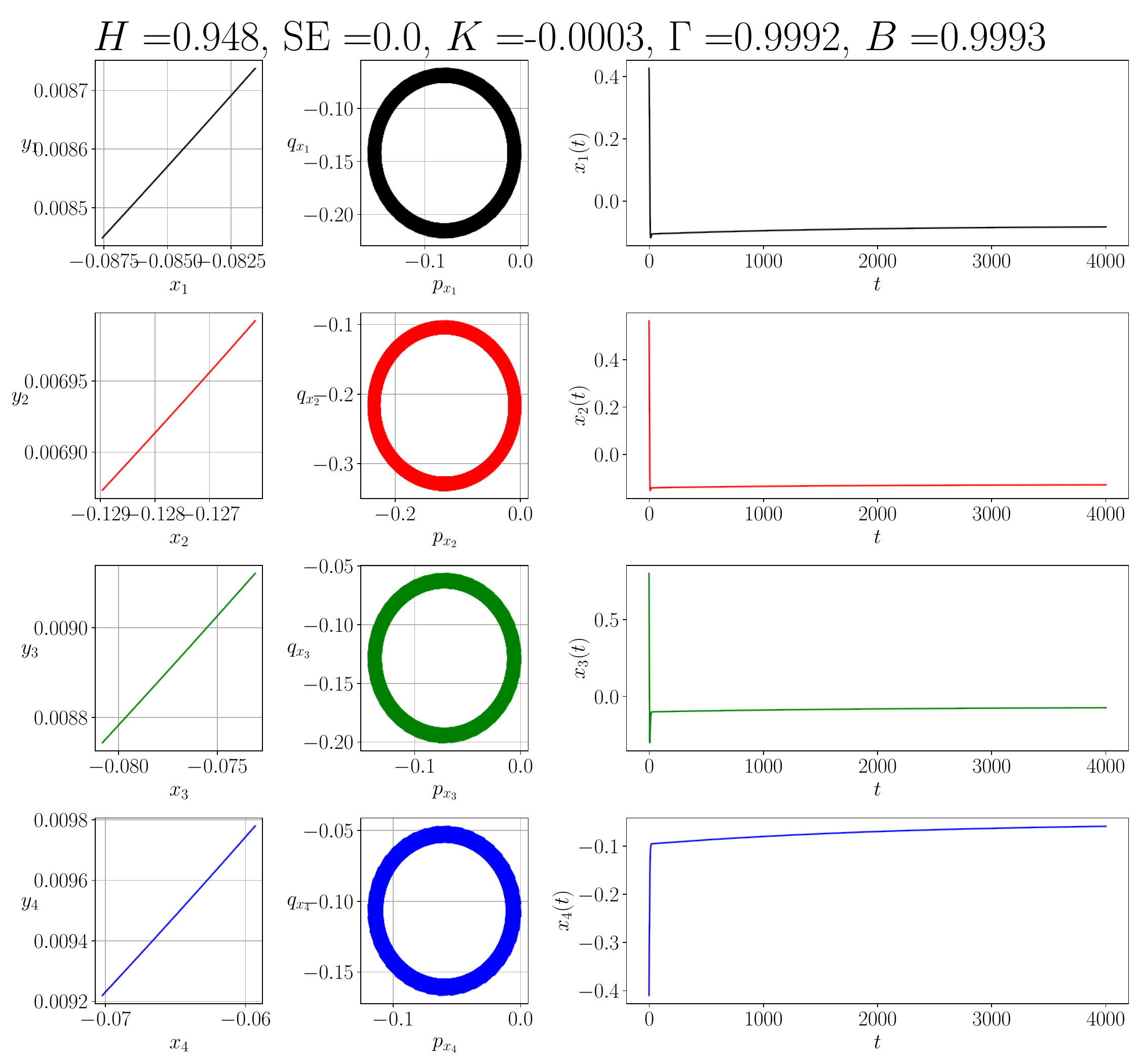} &  \includegraphics[width=0.3\linewidth]{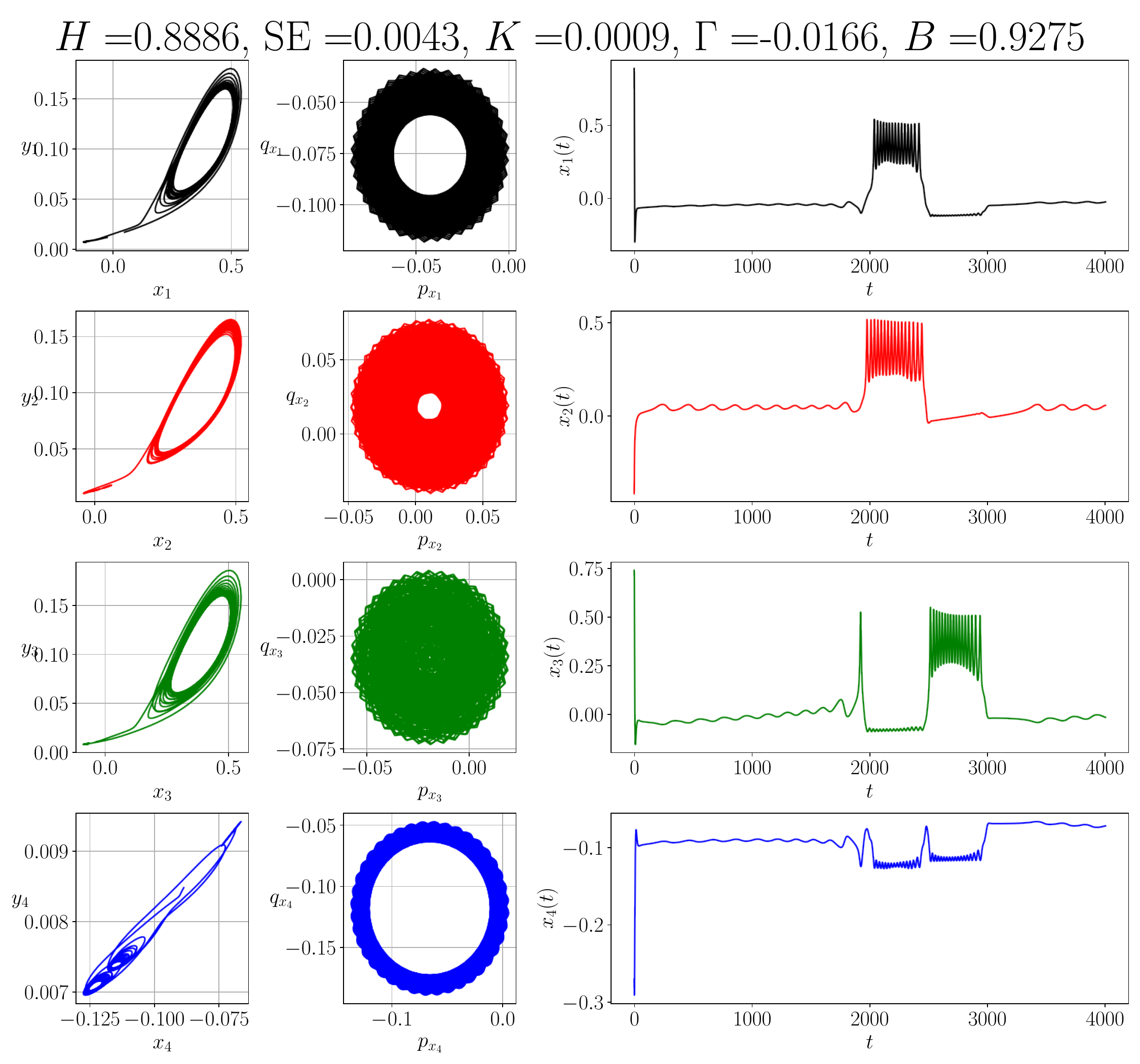} &  \includegraphics[width=0.3\linewidth]{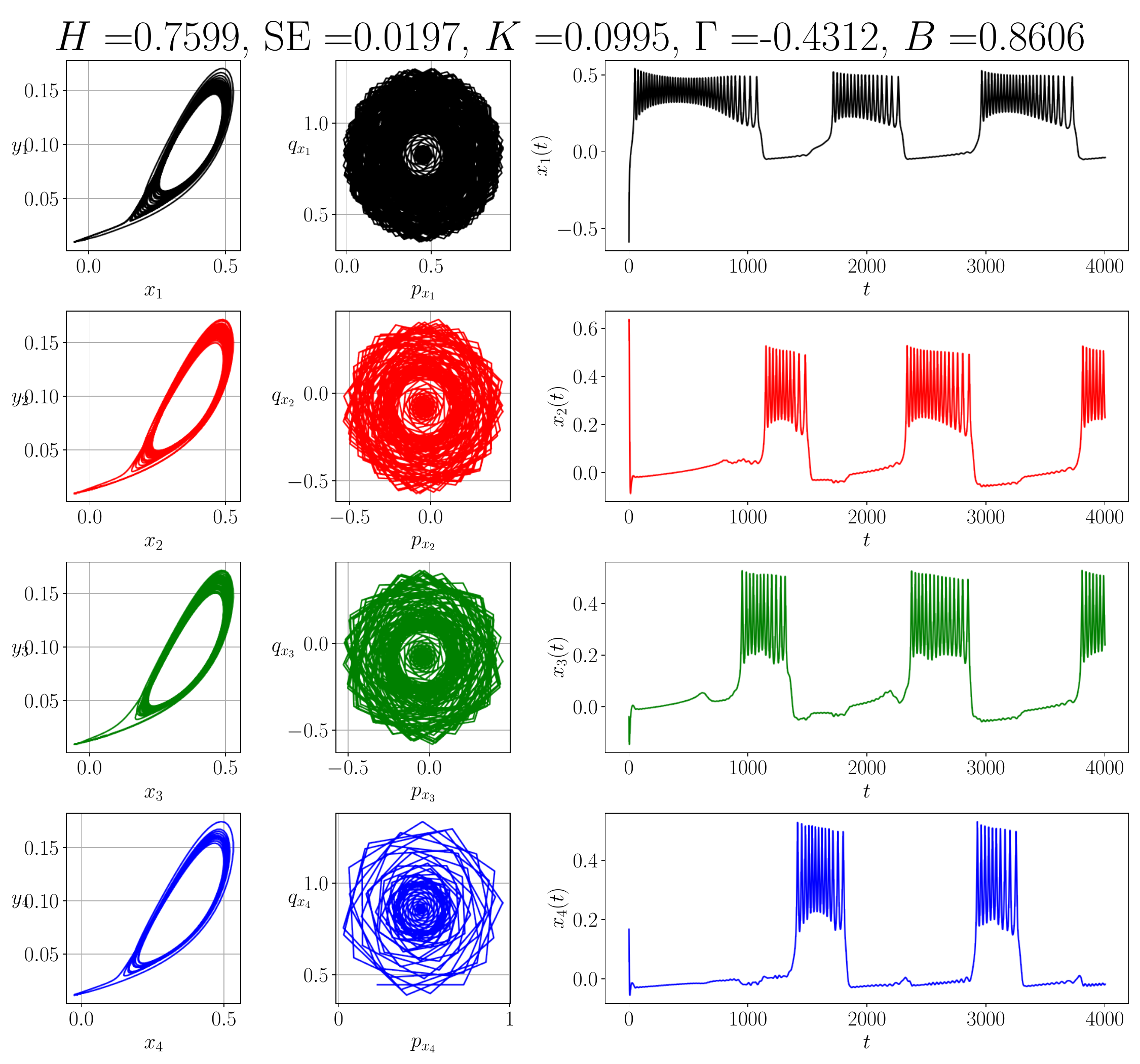}\\
(a) $\theta = -0.01$ & (b) $\theta = -0.005$ & (c) $\theta = -0.001$\\
\includegraphics[width=0.3\linewidth]{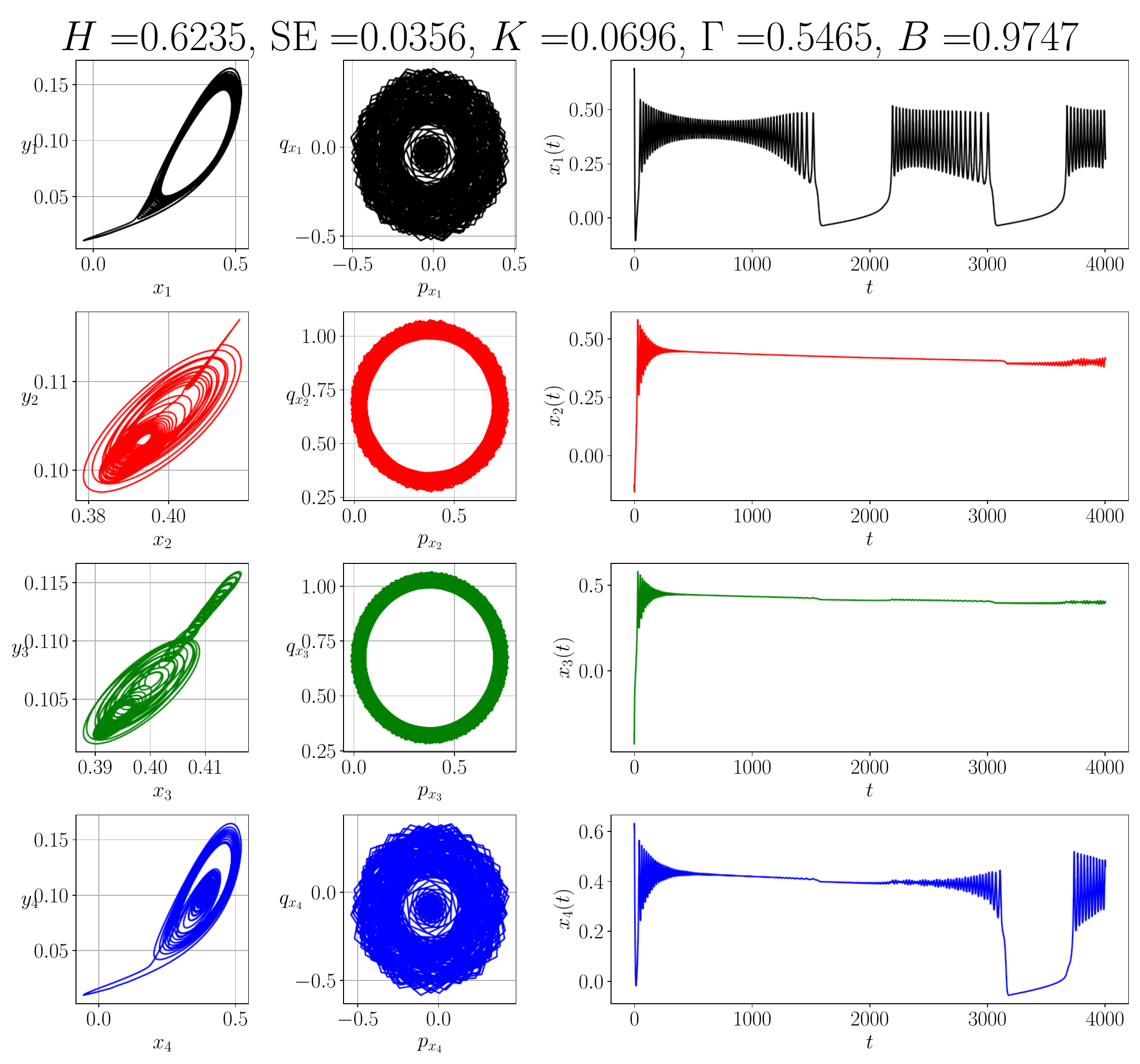} &  \includegraphics[width=0.3\linewidth]{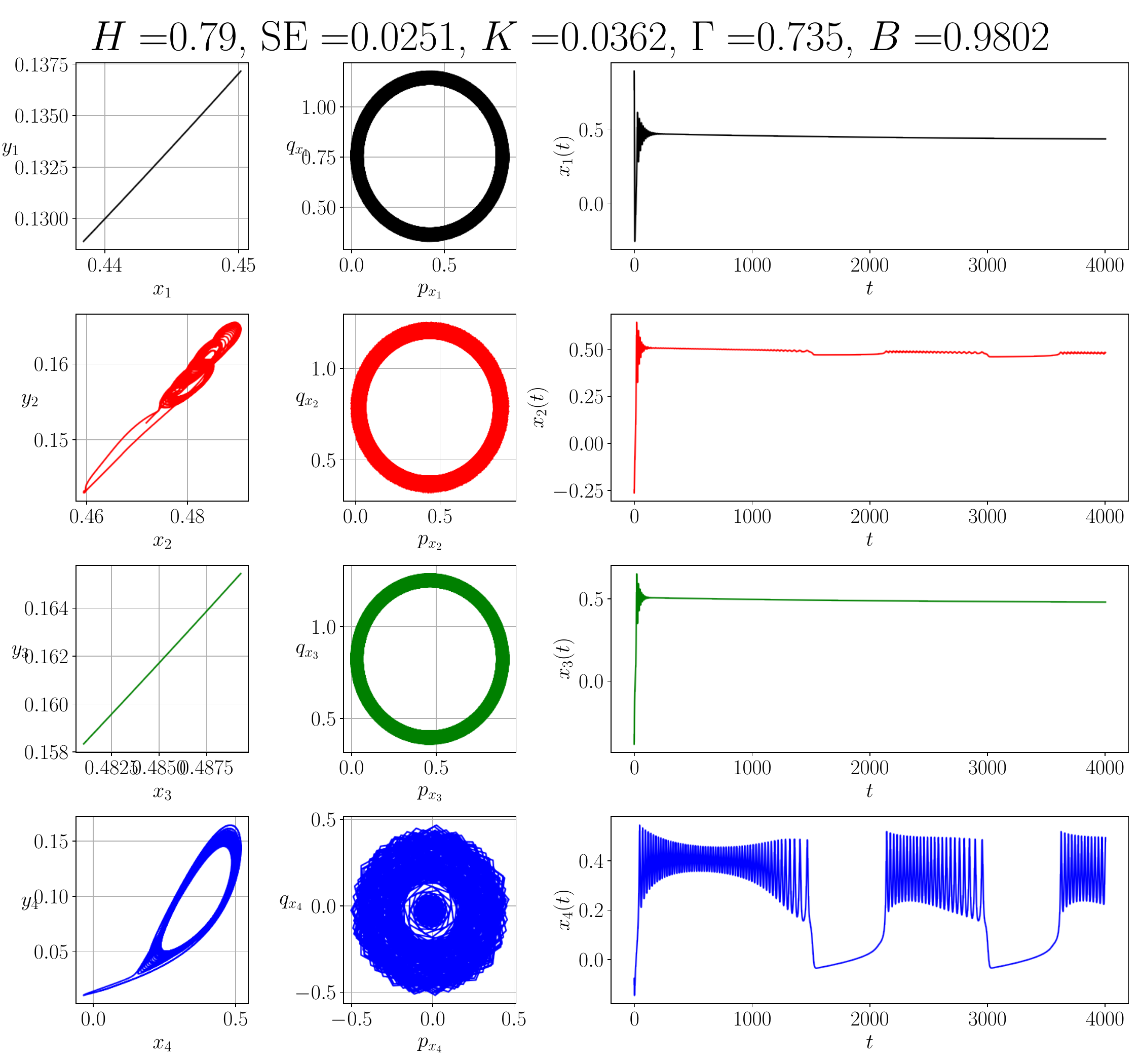} &  \includegraphics[width=0.3\linewidth]{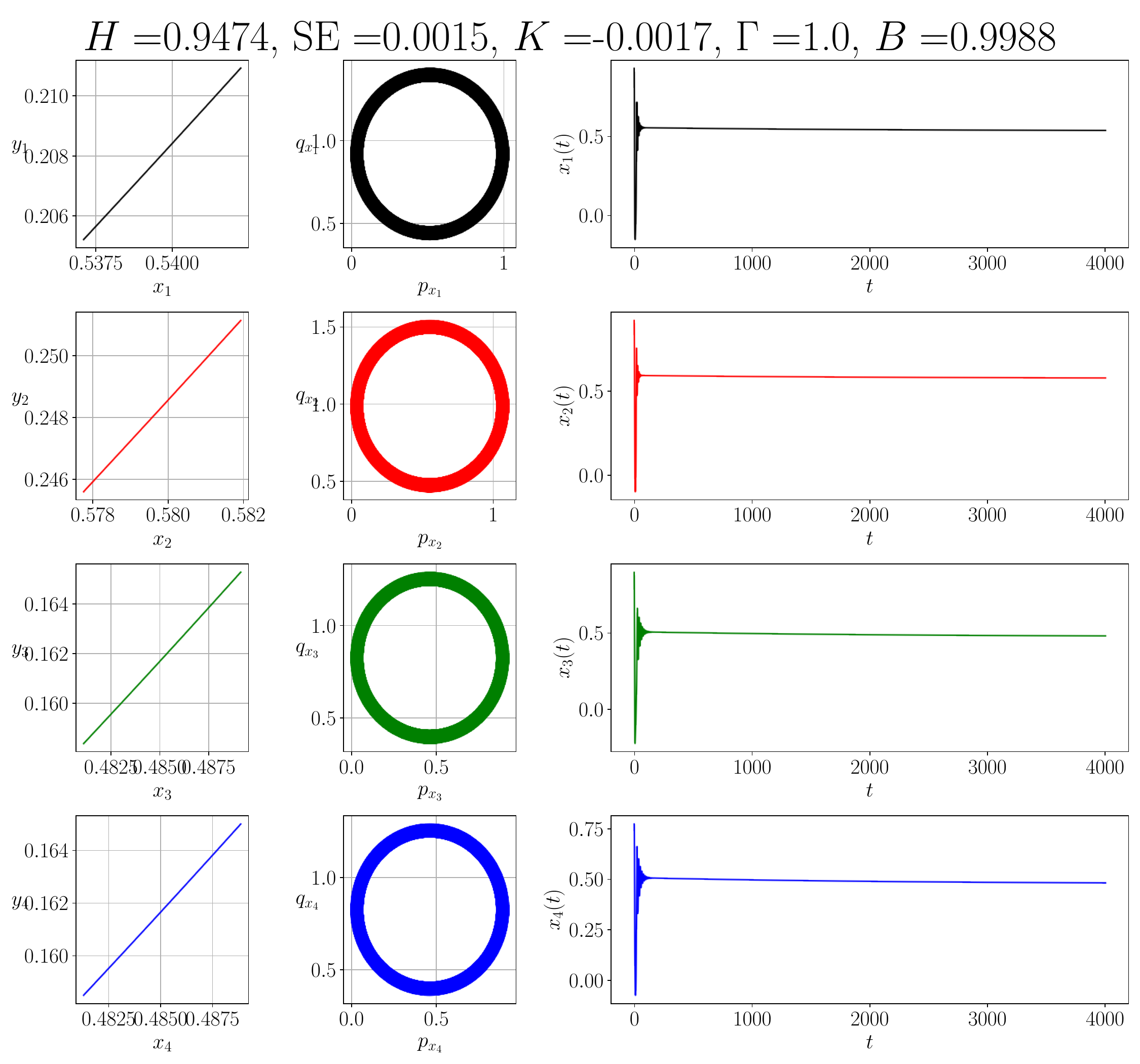}\\
(d) $\theta = 0.001$ & (e) $\theta = 0.005$ & (f) $\theta = 0.01$\\
\end{tabular}
\caption{Phase portraits, $p$ vs $q$ plots, and time series of four-node random network of chemical couplings with autapses~\eqref{eq:random} with varying $\theta \in [-0.01, 0.01]$. Other parameters are fixed as in table~\ref{tab:params}, and $x_n(0), \quad n = 1, \ldots 4,$ are sampled randomly from the continuous uniform distribution over the interval $[-1, 1]$. Other initial conditions are fixed as $y_1(0) = y_2(0)= y_3(0) = y_4(0) = 0.1$, and $I_n(0), \quad n = 1, \ldots 4,$ are sampled randomly from the continuous uniform distribution over the interval $[0.019, 0.022]$. Regular decaying dynamics among all are observed for $|\theta|=0.01$. Nodes exhibit a mix of bursting and decay oscillation for $|\theta|<0.01$. They are synchronized and phase-locked when $|\theta| \approx 0.01$. The corresponding parameter sweep plots are shown in Fig.~\ref{fig:BifRandom}.}
\label{fig:random_pp}
\end{figure*}

Finally, we consider a four-node small network of random pairwise couplings (no higher-order coupling in this case), where the strategy is modeled via non-local chemical synapse~\eqref{eq:random}. A random network is simulated by generating a random $4\times 4$ adjacency matrix of binary values $0$ and $1$ using \texttt{numpy}'s \texttt{random.randint()} function. Note that in this model we allow autapses, thus the diagonal of the adjacency matrix can have either $0$ (no self-loop) or $1$ (with self loop). Also, chemical coupling is unidirectional, and thus the matrix is asymmetric. The initial conditions $x_n(0), \quad n = 1, \ldots 4,$ are sampled randomly from the continuous uniform distribution over the interval $[-1, 1]$. Other initial conditions are fixed as $y_1(0) = y_2(0)= y_3(0) = y_4(0) = 0.1$, and $I_n(0), \quad n = 1, \ldots 4,$ are sampled randomly from the continuous uniform distribution over the interval $[0.019, 0.022]$. Note that for the computation of $\Gamma$, we remove the first $500$ data points from all four time series. We show six instances of how every node in the network considers $\theta = -0.01, -0.005, -0.001, 0.001, 0.005, 0.01$, see Fig.~\ref{fig:random_pp}. For every value of $\theta$, we have randomized the adjacency matrix at the start of the simulation, which remains unchanged throughout the run of a simulation. For example the adjacency matrix of the network represented in Fig.~\ref{fig:random_schem} is $A = \begin{bmatrix}
    0 & 1 & 1 & 0 \\
    0 & 1& 1& 1\\
    1 & 0 & 1& 1\\
    1& 0 &0 &0
\end{bmatrix}$. Note that $A$ clearly conveys that self-loops exist for nodes $2$ and $3$ as $A_{2, 2} = A_{3, 3} = 1$. One thing worth mentioning is that for the implementation of $0-1$ test, we have just utilized the regression method. We notice a very regular dynamics among all nodes for $|\theta|=0.01$, that is, every node displays decay. As $|\theta|<0.01$, we see some inconsistencies among the dynamics of each node, pertaining to weak coupling strength (both inhibitory and excitatory). A mix of bursting and decay oscillation is seen among the nodes when $|\theta|<0.01$. Nodes are synchronized and phase-locked when $|\theta| \approx 0.01$.

\begin{figure}[h]
    \centering
    \includegraphics[width=0.7\linewidth]{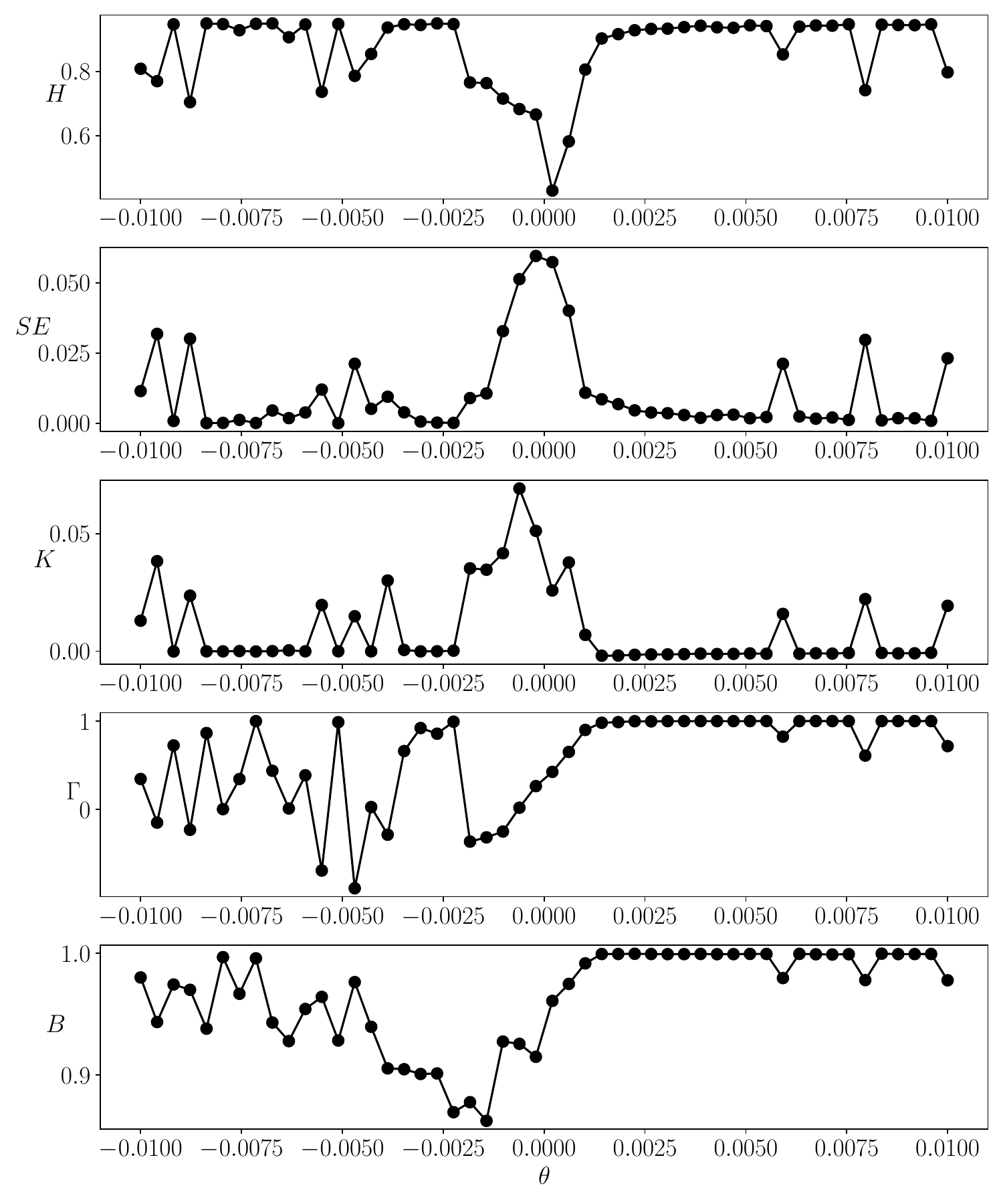}
    \caption{Bifurcation plots of different metrics performing a parameter sweep on $\theta \in [-0.01, 0.01]$ for model~\eqref{eq:higher_order}.}
    \label{fig:BifRandom}
\end{figure}

From Fig.~\ref{fig:BifRandom}, we can see from the $H$ and ${\rm SE}$ values that for $\theta$ values away from a very weak coupling $\theta \not\in [-0.0025, 0.002]$ approximately, the time series from the node are persistent with high regularity. As $\theta \approx 0$, the persistence drops and the sample entropy sees a hike, pertaining to a mix of decay and bursting behavior among dML nodes. We observe a hike in $K$ value close to $\theta \approx 0$ as well, due to this reason, which otherwise remains close to zero. The nodes are synchronized and phase-locked when $|\theta| \approx 0.01$.

\section{A note on Granger causality}
\label{sec:granger}

To statistically test for whether the activity of one neuron drives the activity in another in a coupled system, one can implement the {\em Granger causality} test~\citep{Gr69}. This test allows for identifying causal relationships between neurons based on temporal lags between their activities~\citep{CrTu15, DeSa11}. Seth~\citep{Se05}, and Seth and Edleman~\citep{SeEd07} applied Granger causality to realize causal relationships between neurons in simulated neural systems. Our motivation to test Granger causality on our network models stems from these lines of research. A thorough review of applying Granger causality in neural time series data is provided by Seth in Scholarpedia~\citep{Se07}.

In this manuscript we consider the Josephson junction coupling~\eqref{eq:JJ}, and the memristive coupling~\eqref{eq:mem_coupling} as test cases for applying Granger causality. We test whether the voltage signal in the first node ($x_1$) ``Granger-causes'' the voltage signal in the second node ($x_2$). In order to numerically implement the causality test, we use \texttt{Python}'s \texttt{statsmodels} package~\citep{SePe10} that provides a function called \texttt{grangercausalitytests()} for this purpose. More specifically, this is present within \texttt{statsmodels.tsa.stattools}. The function tests for Granger {\em non} causality for two time series, where the second time series is considered as the independent signal and the first time series is considered as the dependent signal. So we put the time series from the first node as the second column, and the time series from the second node as the first column as an input to the \texttt{grangercausalitytests()} function. The maximum lag we consider is five. For~\eqref{eq:JJ} we consider coupling strength $\theta = -0.01$, and for~\eqref{eq:mem_coupling} we consider $\theta = 1$. All other parameters are kept as is in table~\ref{tab:params}. We will observe that for both models, beyond the number of lag $\ge 2$, we get $p$-value $p \approx 0 \ll 0.05$, where $0.05$ is assumed to be the significance level. This means we can reject the null hypothesis that the time series $x_1$ {\em does not} cause time series $x_2$, which translates to the fact that time series $x_1$ Granger-causes time series $x_2$. This is intuitively what we expected from our simulated model: the activity in the first neuron drives the activity in the second as incorporated by the coupling.

% %===============================================================================
\section{Conclusion}
\label{sec:conclusion}
We aimed to popularize Schaeffer and Cain's slow-fast denatured Morris--Lecar model of neuron dynamics. In that regard, we first discussed its single-cell behavior and then put forward a collection of coupled small network models of dML neurons. We have collected some widely discussed coupling strategies from the literature and have shown how dML neurons form a connected ensemble and drive the system dynamics. The coupling strategies covered are gap junction, thermally sensitive gap junction, chemical, Josephson junction, memristive, high-order gap junction, and random chemical coupling. In this manuscript, we have shown how the time series generated from these simulated models can be further utilized to realize certain properties (persistence, complexity, chaos, synchronization, and phase coherence) of the small networks. To achieve this, we have implemented various tools and algorithms from the nonlinear science literature, specifically the Hurst exponent, sample entropy, $0-1$ test, Pearson's cross-correlation coefficient, and Kuramoto's order parameter. The coupling strength is considered the primary bifurcation parameter, and we have studied how varying this strength induces various routes to chaos, bursting, and decay oscillations.

In some strategies, we have seen that coupling induces chaos in its inhibitory regime and also when the temperature is higher than the reference temperature (particularly in the case of the thermally sensitive coupling). In most cases, excitatory coupling and its more positive values drive the coupled system into exhibiting fold/homoclinic bursting, aligned with the fact that the dML neurons synchronize simultaneously. In some cases, excitatory coupling drives the system to decay oscillation, falling into a symmetric equilibrium point (memristive coupling). When the inhibitory coupling is very weak, a chaotic system transitions to a quasi-periodic system before exhibiting bursting as coupling becomes excitatory. The $0-1$ test acts as the perfect tool to distinguish between chaotic and regular behavior. On top of that, the Hurst exponent and the sample entropy add an extra level of characterization to diagnose persistence and self-similarity in time series, respectively. Numerically, we are able to detect a high correlation between anti-persistence, irregularity, and chaotic behavior utilizing these metrics, and we believe that this package of metrics in tandem will be of high value to the neuroscience community to analyze neuron time series data, for example, electroencephalography data. Researchers are also encouraged to corroborate real world data with our simulated model dynamics and possibly report modified parameter values that fit the data more closely.

We have also performed hypothesis testing to investigate whether the time series in one neuron drives the dynamics of the other in a two-coupled system. In order to do so we look into the two-coupled Josephson junction and memristive coupling. The hypothesis testing is done via the technique of Granger causality. It is observed that the $p$-value for our coupled neuron setting is much less than the assumed significance level, indicating causality.

All our coupled models have a static coupling configuration over time. This is a limitation and can be further improved where the coupling strength is itself a dynamical variable and adapts over time. This can be done via Hebbian interactions between the neurons. Bronski {\em et al.}~\citep{BrHe17} introduced a network of Kuramoto oscillators coupled via Hebbian interactions, which are dynamic coupling strategies. Inspired by that, we can do a similar thing for our coupled dML neurons, whose model equations will be given by 
\begin{equation}
\label{eq:Hebbian}
  \begin{aligned}
    \dot{x}_i &= f(x_i, y_i, I_i) \\
    &+ \sum_{j \in B(i)}\sigma_{i, j} \sin\bigg(\bigg(\tan^{-1}\bigg(\frac{y_j}{x_j}\bigg) - \tan^{-1}\bigg(\frac{y_i}{x_i}\bigg)\bigg)\bigg),\\
    \dot{y}_i &= g(x_i, y_i, I_i), \\
    \dot{I}_i & = h(x_i, y_i, I_i), \\
    \dot{\sigma}_{i,j} &= \mu \cos\bigg(\bigg(\tan^{-1}\bigg(\frac{y_i}{x_i}\bigg) - \tan^{-1}\bigg(\frac{y_j}{x_j}\bigg)\bigg) \bigg) - \beta \sigma_{i, j}.
\end{aligned}  
\end{equation}
This is an interesting model to investigate in the future. Furthermore, we have only considered standard order differential equations in our models. It can be further modified to fractional-order systems which incorporate {\em memory effects} in the dML neurons. A fractional-order coupled system of dML neurons (two-dimensional version) was considered recently by Ghosh {\em et al.}~\citep{GhFa25b}. It would be interesting to implement this suite of time series analysis on the fractional-order coupled systems of dML neurons. Additionally, we have considered small-network models that act as bridges between a single-cell neuron behavior and a complex multi-cell neuron network. Small network models are more tractable in terms of analysis and give us an in-depth knowledge of the intricate details of neuron dynamics. However, the real nervous system is extremely complex and consists of billions of neurons. In order to simulate the behavior of these kinds of large scale models, we need to leverage parallel processing techniques with a large number of neuron nodes, where each nodes have its own independent dynamics and they are coupled following particular network topologies, particularly because It is computationally demanding to simulate larger networks. Similar to \texttt{Nolds} written in \texttt{Python}, there is another sophisticated package called \texttt{ComplexityMeasures.jl}~\citep{DaHa25} (written in \texttt{Julia}) which provides thousands of different kinds of entropy and complexity measures for analyzing time series data from complex dynamical systems.

There is still a lot to achieve in terms of the system dynamics of the coupled dML neurons introduced here. We believe this manuscript will be a starting point for researchers in this domain to pick these models up and analyze their indicate dynamics further. There is also a need to write more optimized software for large-scale simulations and their time series analysis. Our coupled system is homogeneous in the sense that the neurons are identical in terms of parameter values. A step further would be to consider a heterogeneous network which can be realized both on edge level and node level, and also on both space and time. Also, the network can be extended to multiple layers. Furthermore, one can also consider a reaction-diffusion system of these models. These neuronal dynamics produce time series, and it is obvious to analyze these time series using the combination of the metrics discussed here. Finally, it would be interesting to test our approach for map-based neuron models~\citep{IbCa11}.

\section*{Data and code availability}

All data files and \texttt{Python} scripts are openly available for download and implementation from our GitHub repository: \url{https://github.com/indrag49/TS-SlowFast-dML}.

\section*{Acknowledgement}
The authors thank Anjana S. Nair for converting \texttt{01chaostest.jl} from \texttt{Julia} to \texttt{Python} as part of her Master's thesis. 

% \appendix

% %===============================================================================
% \section{Appendix 1}
% \label{app:app1}

% %...............................................................................

\bibliographystyle{plainnat}
\bibliography{main}

\begin{thebibliography}{94}
\providecommand{\natexlab}[1]{#1}
\providecommand{\url}[1]{\texttt{#1}}
\expandafter\ifx\csname urlstyle\endcsname\relax
  \providecommand{\doi}[1]{doi: #1}\else
  \providecommand{\doi}{doi: \begingroup \urlstyle{rm}\Url}\fi

\bibitem[amJ()]{amJulia}
amitg7/01chaostest.jl: 0-1 test for chaos - distinguish between regular and chaotic dynamics in deterministic dynamical systems.
\newblock \url{https://github.com/amitg7/01ChaosTest.jl}.

\bibitem[Anesiadis and Provata(2022)]{AnPr22}
K~Anesiadis and A~Provata.
\newblock Synchronization in multiplex {Leaky Integrate-and-Fire} networks with nonlocal interactions.
\newblock \emph{Front. Netw. Physiol.}, 2:\penalty0 910862, 2022.

\bibitem[Ardhanareeswaran et~al.(2025)Ardhanareeswaran, Sudharsan, Senthilvelan, and Ghosh]{ArSu25}
R.S. Ardhanareeswaran, S.~Sudharsan, M.~Senthilvelan, and D.~Ghosh.
\newblock Intermittent cluster synchronization in a unidirectional ring of bursting neurons.
\newblock \emph{Phys. Rev. E}, 111\penalty0 (1):\penalty0 014215, 2025.

\bibitem[Belykh et~al.(2005)Belykh, De~Lange, and Hasler]{BeDe05}
I.~Belykh, E.~De~Lange, and M.~Hasler.
\newblock Synchronization of bursting neurons: What matters in the network topology.
\newblock \emph{Phys. Rev. Lett.}, 94\penalty0 (18):\penalty0 188101, 2005.

\bibitem[Berge(1984)]{Be84}
C.~Berge.
\newblock \emph{Hypergraphs: combinatorics of finite sets}, volume~45.
\newblock Elsevier, 1984.

\bibitem[Bick et~al.(2020)Bick, Goodfellow, Laing, and Martens]{BiGo20}
C.~Bick, M.~Goodfellow, C.R. Laing, and E.A. Martens.
\newblock Understanding the dynamics of biological and neural oscillator networks through exact mean-field reductions: a review.
\newblock \emph{J. Math. Neurosci.}, 10\penalty0 (1):\penalty0 9, 2020.

\bibitem[Bick et~al.(2023)Bick, Gross, Harrington, and Schaub]{BiGr23}
C.~Bick, E.~Gross, H.A. Harrington, and M.T. Schaub.
\newblock What are higher-order networks?
\newblock \emph{SIAM Rev.}, 65\penalty0 (3):\penalty0 686--731, 2023.

\bibitem[Boya et~al.(2024)Boya, Muni, Echenaus{\'\i}a-Monroy, and Kengne]{BoBe24}
B.F.B.A. Boya, S.S. Muni, J.L. Echenaus{\'\i}a-Monroy, and J.~Kengne.
\newblock Chaos, synchronization, and emergent behaviors in memristive {Hopfield} networks: bi-neuron and regular topology analysis.
\newblock \emph{Eur. Phys. J. Spec. Top.}, pages 1--15, 2024.

\bibitem[Bronski et~al.(2017)Bronski, He, Li, Liu, Sponseller, and Wolbert]{BrHe17}
J.C. Bronski, Y.~He, X.e Li, Y.~Liu, D.R. Sponseller, and S.~Wolbert.
\newblock The stability of fixed points for a {Kuramoto} model with {Hebbian} interactions.
\newblock \emph{Chaos}, 27\penalty0 (5), 2017.

\bibitem[Bukh et~al.(2017)Bukh, Rybalova, Semenova, Strelkova, and Anishchenko]{BuRy17}
A.~Bukh, E.~Rybalova, N.~Semenova, G.I. Strelkova, and V.~Anishchenko.
\newblock New type of chimera and mutual synchronization of spatiotemporal structures in two coupled ensembles of nonlocally interacting chaotic maps.
\newblock \emph{Chaos}, 27\penalty0 (11), 2017.

\bibitem[Castanedo-Guerra et~al.(2016)Castanedo-Guerra, Steur, and Nijmeijer]{CaSt16}
I.T. Castanedo-Guerra, E.~Steur, and H.~Nijmeijer.
\newblock Synchronization of coupled {Hindmarsh--Rose} neurons: effects of an exogenous parameter.
\newblock \emph{IFAC-PapersOnLine}, 49\penalty0 (14):\penalty0 84--89, 2016.

\bibitem[Chialvo(1995)]{Ch95}
D.R. Chialvo.
\newblock Generic excitable dynamics on a two-dimensional map.
\newblock \emph{Chaos Soliton Fract.}, 5\penalty0 (3-4):\penalty0 461--479, 1995.

\bibitem[Chua(1971)]{Ch71}
L.~Chua.
\newblock Memristor-the missing circuit element.
\newblock \emph{IEEE Trans. Circuit Theory}, 18\penalty0 (5):\penalty0 507--519, 1971.

\bibitem[Craddock et~al.(2015)Craddock, Tungaraza, and Milham]{CrTu15}
R.C. Craddock, R.L. Tungaraza, and M.P. Milham.
\newblock Connectomics and new approaches for analyzing human brain functional connectivity.
\newblock \emph{Gigascience}, 4\penalty0 (1):\penalty0 s13742--015, 2015.

\bibitem[Datseris and Haaga(2025)]{DaHa25}
G.~Datseris and K.A. Haaga.
\newblock {ComplexityMeasures. jl}: Scalable software to unify and accelerate entropy and complexity timeseries analysis.
\newblock \emph{PloS One}, 20\penalty0 (6):\penalty0 e0324431, 2025.

\bibitem[Deshpande et~al.(2011)Deshpande, Santhanam, and Hu]{DeSa11}
G.~Deshpande, P.~Santhanam, and X.~Hu.
\newblock Instantaneous and causal connectivity in resting state brain networks derived from functional {MRI} data.
\newblock \emph{Neuroimage}, 54\penalty0 (2):\penalty0 1043--1052, 2011.

\bibitem[Dormand and Prince(1980)]{DoPr80}
John~R Dormand and Peter~J Prince.
\newblock A family of embedded runge-kutta formulae.
\newblock \emph{J. Comput. Appl. Math.}, 6\penalty0 (1):\penalty0 19--26, 1980.

\bibitem[Fatoyinbo et~al.(2022)Fatoyinbo, Muni, Ghosh, Sarumi, and Abidemi]{FaMu22}
H.O. Fatoyinbo, S.S. Muni, I.~Ghosh, I.O. Sarumi, and A.~Abidemi.
\newblock Numerical bifurcation analysis of improved denatured {M}orris--{L}ecar neuron model.
\newblock In \emph{2022 International Conference on decision aid sciences and applications (DASA)}, pages 55--60. IEEE, 2022.

\bibitem[Feudel et~al.(2000)Feudel, Neiman, Pei, Wojtenek, Braun, Huber, and Moss]{FeNe00}
U.~Feudel, A.~Neiman, X.~Pei, W.~Wojtenek, H.~Braun, M.~Huber, and F.~Moss.
\newblock Homoclinic bifurcation in a {Hodgkin--Huxley} model of thermally sensitive neurons.
\newblock \emph{Chaos}, 10\penalty0 (1):\penalty0 231--239, 2000.

\bibitem[FitzHugh(1961)]{Fi61}
R.~FitzHugh.
\newblock Impulses and physiological states in theoretical models of nerve membrane.
\newblock \emph{Biophys. J.}, 1\penalty0 (6):\penalty0 445--466, 1961.

\bibitem[Follmann et~al.(2024)Follmann, Jaswal, Jacob, de~Oliveira, Herbert, Macau, and Rosa]{FoJa24}
R.~Follmann, T.~Jaswal, G.~Jacob, J.F. de~Oliveira, C.B. Herbert, E.E.N. Macau, and E.~Rosa.
\newblock Temperature effects on neuronal synchronization in seizures.
\newblock \emph{Chaos}, 34\penalty0 (8), 2024.

\bibitem[Franzke et~al.(2015)Franzke, Osprey, Davini, and Watkins]{FrOs15}
C.L.E. Franzke, S.M. Osprey, P.~Davini, and N.W. Watkins.
\newblock A dynamical systems explanation of the hurst effect and atmospheric low-frequency variability.
\newblock \emph{Sci. Rep.}, 5\penalty0 (1):\penalty0 9068, 2015.

\bibitem[Ghosh and Fatoyinbo(2025)]{GhFa25b}
I.~Ghosh and H.O. Fatoyinbo.
\newblock Fractional order induced bifurcations in {C}aputo-type denatured {M}orris--{L}ecar neurons.
\newblock \emph{Commun. Nonlinear Sci. Numer. Simul.}, page 108984, 2025.

\bibitem[Ghosh et~al.(2023)Ghosh, Muni, and Fatoyinbo]{GhMu23}
I.~Ghosh, S.S. Muni, and H.O. Fatoyinbo.
\newblock On the analysis of a heterogeneous coupled network of memristive {Chialvo} neurons.
\newblock \emph{Nonlinear Dyn.}, 111\penalty0 (18):\penalty0 17499--17518, 2023.

\bibitem[Ghosh et~al.(2024)Ghosh, Nair, Fatoyinbo, and Muni]{GhNa24}
I.~Ghosh, A.S. Nair, H.O. Fatoyinbo, and S.S. Muni.
\newblock Dynamical properties of a small heterogeneous chain network of neurons in discrete time.
\newblock \emph{Eur. Phys. J. Plus}, 139\penalty0 (6):\penalty0 545, 2024.

\bibitem[Ghosh et~al.(2025)Ghosh, Fatoyinbo, and Muni]{GhFa25}
I~Ghosh, H.O. Fatoyinbo, and S.S. Muni.
\newblock Comprehensive analysis of slow-fast denatured morris-lecar neurons.
\newblock \emph{Phys. Rev. E}, 111\penalty0 (4):\penalty0 044204, 2025.

\bibitem[Gon{\c{c}}alves et~al.(2025)Gon{\c{c}}alves, Labouriau, and Rodrigues]{GoLa25}
B.F.F. Gon{\c{c}}alves, I.S. Labouriau, and A.A.P. Rodrigues.
\newblock Bifurcations and canards in two coupled fitzhugh--nagumo equations.
\newblock \emph{arXiv preprint arXiv:2503.12596}, 2025.

\bibitem[Gottwald and Melbourne(2009{\natexlab{a}})]{GoMe09}
G.A Gottwald and I.~Melbourne.
\newblock On the implementation of the 0--1 test for chaos.
\newblock \emph{SIAM J. Appl. Dyn. Syst.}, 8\penalty0 (1):\penalty0 129--145, January 2009{\natexlab{a}}.

\bibitem[Gottwald and Melbourne(2009{\natexlab{b}})]{GoMe09b}
G.A. Gottwald and I.~Melbourne.
\newblock On the validity of the 0--1 test for chaos.
\newblock \emph{Nonlinearity}, 22\penalty0 (6):\penalty0 1367, 2009{\natexlab{b}}.

\bibitem[Gottwald and Melbourne(2016)]{GoMe16}
G.A. Gottwald and I.~Melbourne.
\newblock The 0--1 test for chaos: {A} review.
\newblock In \emph{Chaos Detection and Predictability}, Lecture notes in physics, pages 221--247. Springer Berlin Heidelberg, Berlin, Heidelberg, 2016.

\bibitem[Granger(1969)]{Gr69}
C.W.J. Granger.
\newblock Investigating causal relations by econometric models and cross-spectral methods.
\newblock \emph{Econometrica}, pages 424--438, 1969.

\bibitem[Hodgkin and Huxley(1952)]{HoHu52}
A.L. Hodgkin and A.F. Huxley.
\newblock A quantitative description of membrane current and its application to conduction and excitation in nerve.
\newblock \emph{J. Physiol.}, 117\penalty0 (4):\penalty0 500, 1952.

\bibitem[Hormuzdi et~al.(2004)Hormuzdi, Filippov, Mitropoulou, Monyer, and Bruzzone]{HoFi04}
S.G. Hormuzdi, M.A. Filippov, G.~Mitropoulou, H.~Monyer, and R.~Bruzzone.
\newblock Electrical synapses: a dynamic signaling system that shapes the activity of neuronal networks.
\newblock \emph{Biochim. Biophys. Acta, Biomembr.}, 1662\penalty0 (1-2):\penalty0 113--137, 2004.

\bibitem[Horn et~al.(2012)Horn, Memelli, and Solomon]{HoMe12}
K.G. Horn, H.~Memelli, and I.C. Solomon.
\newblock Emergent central pattern generator behavior in gap-junction-coupled {Hodgkin--Huxley} style neuron model.
\newblock \emph{Comput. Intell. Neurosci.}, 2012\penalty0 (1):\penalty0 173910, 2012.

\bibitem[Huang et~al.(2025)Huang, Wu, Ding, Jia, and Zhan]{HuWu25}
W.~Huang, Y.~Wu, Q.~Ding, Y.~Jia, and X.~Zhan.
\newblock Effects of external periodic stimuli and higher-order interactions on the synchronization of {Morris--Lecar} neurons.
\newblock \emph{Phys. A}, 669:\penalty0 130610, 2025.

\bibitem[Hurst(1951)]{Hu51}
H.E. Hurst.
\newblock Long-term storage of reservoirs.
\newblock \emph{Trans. Am. Soc. Civ. Eng.}, 116, 1951.

\bibitem[Ibarz et~al.(2011)Ibarz, Casado, and Sanju{\'a}n]{IbCa11}
B.~Ibarz, J.M. Casado, and M.A.F Sanju{\'a}n.
\newblock Map-based models in neuronal dynamics.
\newblock \emph{Phys. Rep.}, 501\penalty0 (1-2):\penalty0 1--74, April 2011.

\bibitem[Ikeda and Bekkers(2006)]{IkBe06}
K.~Ikeda and J.M. Bekkers.
\newblock Autapses.
\newblock \emph{Curr. Biol.}, 16\penalty0 (9):\penalty0 R308, 2006.

\bibitem[Izhikevich(2007)]{Iz07}
E.M. Izhikevich.
\newblock \emph{Dynamical systems in neuroscience}.
\newblock MIT press, 2007.

\bibitem[Izhikevich and FitzHugh(2006)]{IzFi06}
E.M. Izhikevich and R.~FitzHugh.
\newblock {F}itz{H}ugh-{N}agumo model.
\newblock \emph{Scholarpedia}, 1\penalty0 (9):\penalty0 1349, 2006.
\newblock \doi{10.4249/scholarpedia.1349}.
\newblock revision \#123664.

\bibitem[Kaizuka and Takumi(2018)]{KaTa18}
T.~Kaizuka and T.~Takumi.
\newblock Postsynaptic density proteins and their involvement in neurodevelopmental disorders.
\newblock \emph{J. Biochem.}, 163\penalty0 (6):\penalty0 447--455, 2018.

\bibitem[Korneev et~al.(2021)Korneev, Semenov, Slepnev, and Vadivasova]{KoSe21}
I.A. Korneev, V.V. Semenov, A.V. Slepnev, and T.E. Vadivasova.
\newblock The impact of memristive coupling initial states on travelling waves in an ensemble of the {FitzHugh--Nagumo} oscillators.
\newblock \emph{Chaos Soliton Fract.}, 147:\penalty0 110923, 2021.

\bibitem[Korneev et~al.(2024)Korneev, Ramazanov, Slepnev, Vadivasova, and Semenov]{KoRa24}
I.A. Korneev, I.R. Ramazanov, A.V. Slepnev, T.E. Vadivasova, and V.V. Semenov.
\newblock Traveling waves in an ensemble of excitable oscillators: The interplay of memristive coupling and noise.
\newblock \emph{Chaos}, 34\penalty0 (12), 2024.

\bibitem[Kuehn(2015)]{Ku15}
C.~Kuehn.
\newblock \emph{Multiple time scale dynamics}, volume 191.
\newblock Springer, 2015.

\bibitem[Kuramoto(1984)]{Ku84}
Y.~Kuramoto.
\newblock \emph{Chemical oscillations, waves, and turbulence}, volume~8.
\newblock Springer, 1984.

\bibitem[Kuramoto and Battogtokh(2002)]{KuBa02}
Y.~Kuramoto and D.~Battogtokh.
\newblock Coexistence of coherence and incoherence in nonlocally coupled phase oscillators.
\newblock \emph{Nonlinear Phenom. Complex Syst.}, 5\penalty0 (4):\penalty0 380--385, 2002.

\bibitem[Majhi et~al.(2017)Majhi, Perc, and Ghosh]{MaPe17}
S.~Majhi, M.~Perc, and D.~Ghosh.
\newblock Chimera states in a multilayer network of coupled and uncoupled neurons.
\newblock \emph{Chaos}, 27\penalty0 (7), 2017.

\bibitem[Majhi et~al.(2019)Majhi, Bera, Ghosh, and Perc]{MaBe19}
S.~Majhi, B.K. Bera, D.~Ghosh, and M.~Perc.
\newblock Chimera states in neuronal networks: A review.
\newblock \emph{Phys. Life Rev.}, 28:\penalty0 100--121, 2019.

\bibitem[Mandelbrot(1985)]{Ma85}
B.B. Mandelbrot.
\newblock Self-affine fractals and fractal dimension.
\newblock \emph{Phys. Scr.}, 32\penalty0 (4):\penalty0 257, 1985.

\bibitem[Mathur(2008)]{Ma08}
N.D. Mathur.
\newblock The fourth circuit element.
\newblock \emph{Nature}, 455\penalty0 (7217):\penalty0 E13--E13, 2008.

\bibitem[Mirzaei et~al.(2022)Mirzaei, Mehrabbeik, Rajagopal, Jafari, and Chen]{MiMe22}
S.~Mirzaei, M.~Mehrabbeik, K.~Rajagopal, S.~Jafari, and G.~Chen.
\newblock Synchronization of a higher-order network of {Rulkov} maps.
\newblock \emph{Chaos}, 32\penalty0 (12), 2022.

\bibitem[Mishra et~al.(2021)Mishra, Ghosh, Dana, Kapitaniak, and Hens]{MiGh21}
A.~Mishra, S.~Ghosh, S.K. Dana, T.~Kapitaniak, and C.~Hens.
\newblock Neuron-like spiking and bursting in josephson junctions: a review.
\newblock \emph{Chaos}, 31\penalty0 (5), 2021.

\bibitem[Morris and Lecar(1981)]{MoLe81}
C.~Morris and H.~Lecar.
\newblock Voltage oscillations in the barnacle giant muscle fiber.
\newblock \emph{Biophys. J.}, 35\penalty0 (1):\penalty0 193--213, 1981.

\bibitem[Muni and Provata(2020)]{MuPr20}
S.S. Muni and A.~Provata.
\newblock Chimera states in ring--star network of chua circuits.
\newblock \emph{Nonlinear Dyn.}, 101\penalty0 (4):\penalty0 2509--2521, 2020.

\bibitem[Muni et~al.(2022)Muni, Fatoyinbo, and Ghosh]{MuFa22}
S.S. Muni, H.O. Fatoyinbo, and I.~Ghosh.
\newblock Dynamical effects of electromagnetic flux on {Chialvo} neuron map: Nodal and network behaviors.
\newblock \emph{Int. J. Bifurcation Chaos}, 32\penalty0 (09), July 2022.

\bibitem[Nair et~al.(2024)Nair, Ghosh, Fatoyinbo, and Muni]{NaGh24}
A.S. Nair, I.~Ghosh, H.O. Fatoyinbo, and S.S. Muni.
\newblock On the higher-order smallest ring-star network of {Chialvo} neurons under diffusive couplings.
\newblock \emph{Chaos}, 34\penalty0 (7), 2024.

\bibitem[Njitacke et~al.(2022)Njitacke, Ramakrishnan, Rajagopal, Fozin, and Awrejcewicz]{NjRa22}
Z.T. Njitacke, B.~Ramakrishnan, K.~Rajagopal, T.F. Fozin, and J.~Awrejcewicz.
\newblock Extremely rich dynamics of coupled heterogeneous neurons through a josephson junction synapse.
\newblock \emph{Chaos Soliton Fract.}, 164:\penalty0 112717, 2022.

\bibitem[Nolte et~al.(2020)Nolte, Gal, Markram, and Reimann]{NoGa20}
M.~Nolte, E.~Gal, H.~Markram, and M.W. Reimann.
\newblock Impact of higher order network structure on emergent cortical activity.
\newblock \emph{Netw. Neurosci.}, 4\penalty0 (1):\penalty0 292--314, 2020.

\bibitem[Omelchenko et~al.(2015)Omelchenko, Provata, Hizanidis, Sch{\"o}ll, and H{\"o}vel]{OmPr15}
I.~Omelchenko, A.~Provata, J.~Hizanidis, E.~Sch{\"o}ll, and P.~H{\"o}vel.
\newblock Robustness of chimera states for coupled {FitzHugh-Nagumo} oscillators.
\newblock \emph{Phys. Rev. E}, 91\penalty0 (2):\penalty0 022917, 2015.

\bibitem[Parastesh et~al.(2022)Parastesh, Mehrabbeik, Rajagopal, Jafari, and Perc]{PaMe22}
F.~Parastesh, M.~Mehrabbeik, K.~Rajagopal, S.~Jafari, and M.~Perc.
\newblock Synchronization in {Hindmarsh--Rose} neurons subject to higher-order interactions.
\newblock \emph{Chaos}, 32\penalty0 (1), 2022.

\bibitem[Park and Friston(2013)]{PaFr13}
H.-J. Park and K.~Friston.
\newblock Structural and functional brain networks: from connections to cognition.
\newblock \emph{Science}, 342\penalty0 (6158):\penalty0 1238411, 2013.

\bibitem[Qian and Rasheed(2004)]{QiRa04}
B.~Qian and K.~Rasheed.
\newblock Hurst exponent and financial market predictability.
\newblock In \emph{IASTED conference on Financial Engineering and Applications}, pages 203--209. Proceedings of the IASTED International Conference. Chicago Cambridge, MA, 2004.

\bibitem[Richman and Moorman(2000)]{RiMo00}
J.S. Richman and J.R. Moorman.
\newblock Physiological time-series analysis using approximate entropy and sample entropy.
\newblock \emph{Am. J. Physiol. Heart Circ. Physiol.}, 278\penalty0 (6), 2000.

\bibitem[Rybalova et~al.(2019)Rybalova, Anishchenko, Strelkova, and Zakharova]{RyAn19}
E.~Rybalova, V.S. Anishchenko, G.I. Strelkova, and A.~Zakharova.
\newblock Solitary states and solitary state chimera in neural networks.
\newblock \emph{Chaos}, 29\penalty0 (7), 2019.

\bibitem[Rybalova et~al.(2021)Rybalova, Zakharova, and Strelkova]{RyZa21}
E.V. Rybalova, A.~Zakharova, and G.I. Strelkova.
\newblock Interplay between solitary states and chimeras in multiplex neural networks.
\newblock \emph{Chaos Soliton Fract.}, 148:\penalty0 111011, 2021.

\bibitem[Schaeffer and Cain(2018)]{ScCa18}
D.G. Schaeffer and J.W. Cain.
\newblock \emph{Ordinary differential equations: {B}asics and beyond}.
\newblock Springer, 2018.

\bibitem[Schölzel(2019)]{Sc19}
C.~Schölzel.
\newblock Nonlinear measures for dynamical systems.
\newblock \url{https://pypi.org/project/nolds/}, 2019.
\newblock URL \url{https://doi.org/10.5281/zenodo.3814723}.

\bibitem[Seabold and Perktold(2010)]{SePe10}
S.~Seabold and J.~Perktold.
\newblock statsmodels: Econometric and statistical modeling with {Python}.
\newblock In \emph{9th Python in Science Conference}, 2010.

\bibitem[Segall et~al.(2014)Segall, Guo, Crotty, Schult, and Miller]{SeKe14}
K.~Segall, S.~Guo, P.~Crotty, D.~Schult, and M.~Miller.
\newblock Phase-flip bifurcation in a coupled {Josephson} junction neuron system.
\newblock \emph{Phys. B}, 455:\penalty0 71--75, 2014.

\bibitem[Seth(2007)]{Se07}
A.~Seth.
\newblock {G}ranger causality.
\newblock \emph{Scholarpedia}, 2\penalty0 (7):\penalty0 1667, 2007.
\newblock \doi{10.4249/scholarpedia.1667}.
\newblock revision \#127333.

\bibitem[Seth(2005)]{Se05}
A.K. Seth.
\newblock Causal connectivity of evolved neural networks during behavior.
\newblock \emph{Netw. Comput. Neural Syst.}, 16\penalty0 (1):\penalty0 35--54, 2005.

\bibitem[Seth and Edelman(2007)]{SeEd07}
A.K. Seth and G.M. Edelman.
\newblock Distinguishing causal interactions in neural populations.
\newblock \emph{Neural Comput.}, 19\penalty0 (4):\penalty0 910--933, 2007.

\bibitem[Seung et~al.(2000)Seung, Lee, Reis, and Tank]{SeLe00}
H.S. Seung, D.D. Lee, B.Y. Reis, and D.W. Tank.
\newblock The autapse: a simple illustration of short-term analog memory storage by tuned synaptic feedback.
\newblock \emph{J. Comput. Neurosci.}, 9:\penalty0 171--185, 2000.

\bibitem[Shepelev et~al.(2020)Shepelev, Bukh, Muni, and Anishchenko]{ShBu20}
I.A. Shepelev, A.V. Bukh, S.S. Muni, and V.S. Anishchenko.
\newblock Role of solitary states in forming spatiotemporal patterns in a 2d lattice of van der pol oscillators.
\newblock \emph{Chaos Solit.}, 135:\penalty0 109725, 2020.

\bibitem[Shi et~al.(2024)Shi, Cao, Banerjee, Ahmad, and Mou]{ShCa24}
F.~Shi, Y.~Cao, S.~Banerjee, A.M. Ahmad, and J.~Mou.
\newblock A novel neural networks with memristor coupled memcapacitor-synapse neuron.
\newblock \emph{Chaos Soliton Fract.}, 189:\penalty0 115723, 2024.

\bibitem[Simo et~al.(2021)Simo, Louodop, Ghosh, Njougouo, Tchitnga, and Cerdeira]{SiLo21}
G.R. Simo, P.~Louodop, D.~Ghosh, T.~Njougouo, R.~Tchitnga, and H.A. Cerdeira.
\newblock Traveling chimera patterns in a two-dimensional neuronal network.
\newblock \emph{Phys. Lett. A}, 409:\penalty0 127519, 2021.

\bibitem[Somers and Kopell(1993)]{SoKo93}
D.~Somers and N.~Kopell.
\newblock Rapid synchronization through fast threshold modulation.
\newblock \emph{Biol. Cybernet.}, 68\penalty0 (5):\penalty0 393--407, 1993.

\bibitem[Stewart et~al.(2014)Stewart, Popov, Kraev, Medvedev, and Davies]{StPo14}
M.G. Stewart, V.I. Popov, I.V. Kraev, N.~Medvedev, and H.A. Davies.
\newblock Structure and complexity of the synapse and dendritic spine.
\newblock In \emph{The synapse}, pages 1--20. Elsevier, 2014.

\bibitem[Strogatz(2000)]{St00}
S.H. Strogatz.
\newblock From {K}uramoto to {C}rawford: exploring the onset of synchronization in populations of coupled oscillators.
\newblock \emph{Phys. D}, 143\penalty0 (1-4):\penalty0 1--20, 2000.

\bibitem[Strogatz(2012)]{St12}
S.H. Strogatz.
\newblock \emph{Sync: How order emerges from chaos in the universe, nature, and daily life}.
\newblock Hachette Books, 2012.

\bibitem[Sumpter(2006)]{Su06}
D.J.T. Sumpter.
\newblock The principles of collective animal behaviour.
\newblock \emph{Philos. Trans. R. Soc. Lond. B: Biol. Sci.}, 361\penalty0 (1465):\penalty0 5--22, 2006.

\bibitem[Timofeev et~al.(2012)Timofeev, Bazhenov, Seigneur, and Sejnowski]{TiBa12}
I.~Timofeev, M.~Bazhenov, J.~Seigneur, and T.~Sejnowski.
\newblock Neuronal synchronization and thalamocortical rhythms in sleep, wake and epilepsy.
\newblock \emph{Jasper's Basic Mechanisms of the Epilepsies [Internet]. 4th edition}, 2012.

\bibitem[Tlaie et~al.(2019)Tlaie, Leyva, and Sendi{\~n}a-Nadal]{TlLe19}
A.~Tlaie, I.~Leyva, and I.~Sendi{\~n}a-Nadal.
\newblock High-order couplings in geometric complex networks of neurons.
\newblock \emph{Phys. Rev. E}, 100\penalty0 (5):\penalty0 052305, 2019.

\bibitem[Vadivasova et~al.(2016)Vadivasova, Strelkova, Bogomolov, and Anishchenko]{VaSt16}
T.E. Vadivasova, G.I. Strelkova, S.A. Bogomolov, and V.S. Anishchenko.
\newblock Correlation analysis of the coherence-incoherence transition in a ring of nonlocally coupled logistic maps.
\newblock \emph{Chaos}, 26\penalty0 (9), 2016.

\bibitem[Van~Hook(2020)]{Va20}
M.J. Van~Hook.
\newblock Temperature effects on synaptic transmission and neuronal function in the visual thalamus.
\newblock \emph{PloS one}, 15\penalty0 (4):\penalty0 e0232451, 2020.

\bibitem[Vismaya et~al.(2025)Vismaya, Muni, Panda, and Mondal]{ViMu25}
V.S. Vismaya, S.S. Muni, A.K. Panda, and B.~Mondal.
\newblock {Degn–Harrison} map: Dynamical and network behaviours with applications in image encryption.
\newblock \emph{Chaos Soliton Fract.}, 192:\penalty0 115987, March 2025.

\bibitem[Wang et~al.(2008)Wang, Lu, and Li]{WaLu08}
J.~Wang, M.~Lu, and H.~Li.
\newblock Synchronization of coupled equations of {Morris--Lecar} model.
\newblock \emph{Commun. Nonlinear Sci. Numer. Simul.}, 13\penalty0 (6):\penalty0 1169--1179, 2008.

\bibitem[Wang et~al.(2024)Wang, Chen, Xi, Tian, and Yang]{WaCh24}
Z.~Wang, M.~Chen, X.~Xi, H.~Tian, and R.~Yang.
\newblock Multi-chimera states in a higher order network of {FitzHugh--Nagumo} oscillators.
\newblock \emph{Eur. Phys. J.: Spec. Top.}, 233\penalty0 (4):\penalty0 779--786, 2024.

\bibitem[Wei et~al.(2025)Wei, Li, and Zhang]{WeLi25}
L.~Wei, D.~Li, and J.~Zhang.
\newblock Dynamics and synchronization of the {Morris--Lecar} model with field coupling subject to electromagnetic excitation.
\newblock \emph{Commun. Nonlinear Sci. Numer. Simul.}, 140:\penalty0 108457, 2025.

\bibitem[Xu et~al.(2017)Xu, Jia, Ma, Alsaedi, and Ahmad]{XuJi17}
Y.~Xu, Y.~Jia, J.~Ma, A.~Alsaedi, and B.~Ahmad.
\newblock Synchronization between neurons coupled by memristor.
\newblock \emph{Chaos Soliton Fract.}, 104:\penalty0 435--442, 2017.

\bibitem[Yoshioka(2005)]{Yo05}
M.~Yoshioka.
\newblock Chaos synchronization in gap-junction-coupled neurons.
\newblock \emph{Phys. Rev. E}, 71\penalty0 (6):\penalty0 065203, 2005.

\bibitem[You et~al.(2023)You, Tian, and Tu]{YoTi23}
Y.~You, J.~Tian, and J.~Tu.
\newblock Synchronization of memristive {FitzHugh--Nagumo} neural networks.
\newblock \emph{Commun. Nonlinear Sci. Numer. Simul.}, 125:\penalty0 107405, 2023.

\bibitem[Zhang et~al.(2020)Zhang, Xu, Yao, and Ma]{ZhXu20}
Y.~Zhang, Y.~Xu, Z.~Yao, and J.~Ma.
\newblock A feasible neuron for estimating the magnetic field effect.
\newblock \emph{Nonlinear Dyn.}, 102\penalty0 (3):\penalty0 1849--1867, 2020.

\bibitem[Zhang et~al.(2023)Zhang, Lucas, and Battiston]{ZhLu23}
Y.~Zhang, M.~Lucas, and F.~Battiston.
\newblock Higher-order interactions shape collective dynamics differently in hypergraphs and simplicial complexes.
\newblock \emph{Nat. Commun.}, 14\penalty0 (1):\penalty0 1605, 2023.

\end{thebibliography}

\end{document}